\documentclass[amssymb,amsmath,aps,twocolumn,showpacs,10pt]{revtex4}

\usepackage{color,graphicx,bm}
\usepackage{overpic}
\usepackage{microtype}
\usepackage{upgreek}

\newcommand{\etal}{{\sl et\,al.}}

\newcommand{\ua}{\uparrow}
\newcommand{\da}{\downarrow}
\newcommand{\tj}{\hat J}

\renewcommand{\k}{{\bf k}}
\newcommand{\q}{{\bf q}}
\newcommand{\mf}{^{\scalebox{0.5}{M\!F}}}

\begin{document}

\title{\,
\vspace*{0 mm}
Modeling of Superconducting Stripe Phases in High-$\bm{T_{\rm c}}$ Cuprates
\vspace{0cm}}

\author{F.~Loder$^{1,2}$, S.~Graser$^{1,2}$, M.~Schmid$^1$, A.\,P.~Kampf$^1$, and T.~Kopp$^2$
\vspace{0,2cm}}

\affiliation{$^1$Theoretical Physics III, $^2$Experimental Physics VI,\\
Center for Electronic Correlations and Magnetism, Institute 
of Physics, \\ University of Augsburg,
D-86135 Augsburg, Germany
}
 
\date{\today}

\begin{abstract}
Even before the experimental discovery of spin- and charge-stripe order in La$_{2-x-y}$Nd$_y$Sr$_x$CuO$_4$ and La$_{2-x}$Ba$_x$CuO$_4$ at $x=1/8$, stripe formation was predicted from theoretical considerations. Nevertheless, a consistent description of the complex coexistence of stripe order with superconductivity has remained a challenge. Here we introduce a Hartree-Fock decoupling scheme which unifies previous approaches and allows for a detailed analysis of the competition between antiferromagnetism and superconductivity in real and momentum space. We identify two distinct parameter regimes, where spin-stripe order coexists with either one- or two-dimensional superconductivity; experiments on different striped cuprates are either compatible with the former or the latter regime. We argue that the cuprates at $x=1/8$ fall into an intermediate coupling regime with a crossover to long-range phase coherence between individual superconducting stripes.
\end{abstract}


\maketitle

%\color{blue}
\section{Introduction}\label{sec:intro}

Since the discovery of high-$T_{\rm c}$ superconductivity in La$_{2-x}$Ba$_x$CuO$_4$~\cite{Bernordz}, the analysis of many experiments has led to a complex phase diagram of cuprate materials, in part originating from the competition between antiferromagnetic (AF) and superconducting (SC) correlations. Both are induced by the Coulomb repulsion on the copper $d$-orbitals in the CuO$_2$ planes. At a critical hole-doping level antiferromagnetism and superconductivity are in balance, and details of the material or its environment determine which order is realized or whether a regime of coexistence or local phase separation exists. At very low temperatures most cuprates show a transition from an AF to a SC state, if the density of charge carriers exceeds a critical value, and superconductivity vanishes again beyond a higher carrier density, when the Coulomb interaction becomes less significant.

\begin{figure}[t!]
\centering
\vspace{2.5mm}
\begin{overpic}
[width=0.74\columnwidth]{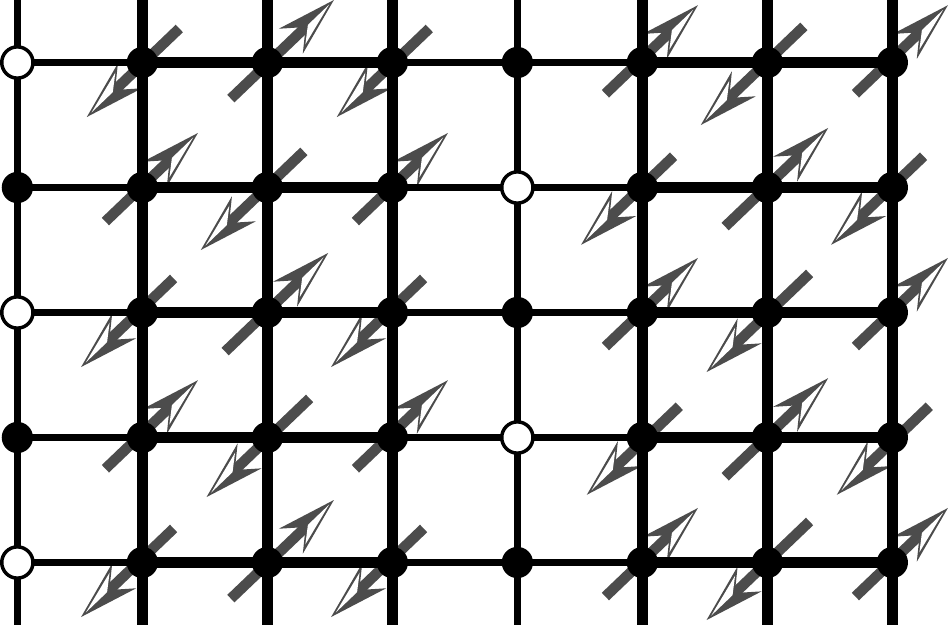}
\end{overpic}\hspace{3mm}
\caption{Three-legged spin-ladder structure with an anti-phase domain wall as suggested by Tranquada \etal~\cite{TranquadaLNdSCO} for striped cuprates at hole doping $x=1/8$. The ladders are half filled and antiferromagnetically ordered, whereas the domain walls are quarter filled and metallic.
}
\label{fig0}
\end{figure}

Above the critical hole doping level, AF correlations on reduced time and length scales persist and fluctuating antiferromagnetism may coexist with Cooper pairing. One prominent phase with a combination of both has \textquotedblleft stripe order\textquotedblright. An ordered phase with static stripes was first observed in 1994 by Tranquada {\it et al}. in the nickelate La$_2$NiO$_{4.125}$~\cite{TranquadaNickel} and later on in the rare-earth doped cuprate La$_{2-x-y}$Nd$_y$Sr$_x$CuO$_4$~\cite{TranquadaLNdSCO,Buchner94} and in the original high-$T_{\rm c}$ cuprate La$_{2-x}$Ba$_x$CuO$_4$~\cite{TranquadaLSCO,fujita,abbamonte}. The cuprate systems evolve from a Mott insulator into a homogeneous SC phase as $x$ is increased, but as $x$ approaches $1/8$, antiferromagnetism returns in a characteristic spin-density-wave (SDW) pattern and superconductivity is suppressed~\cite{Buchner94,Tranquada97,Hucker10}. Based on his neutron scattering data, Tranquada suggested the existence of antiferromagnetically ordered spin ladders separated by metallic lines, thus forming stripes with a width of four lattice constants (figure~\ref{fig0}). The metallic lines form anti-phase domain-wall boundaries between the AF stripes, which optimizes the kinetic energy through virtual electronic hopping processes into the AF stripes~\cite{KivelsonRMP}. The appearance of static stripe order seems to be tied to the structural transition into the anisotropic low-temperature tetragonal phase (LTT) in which at low temperatures superconductivity possibly coexists with stripe order~\cite{berg07,Tranquada08,Hucker10}. The spin dynamics of such systems, has been explained quite successfully using models of coupled spin ladders~\cite{vojta,uhrig,greiter}, although their microscopic origin has remained unresolved within this ansatz, and superconductivity was not incorporated. In this framework both, site and bond centered (i.e. three- or two-legged) ladders are compatible with the experimental results for the dynamical spin susceptibility, although distinct theoretical considerations suggest that site centered stripes are favored~\cite{greiter}.

Unidirectional charge- and spin-stripe order was indeed predicted for cuprate systems before their experimental discovery. In 1989, Zaanen and Gunnarson~\cite{ZaanenHF} and Machida~\cite{machida} found AF stripe formation in the mean-field solution of a two-band Hubbard model. After stripes had been discovered experimentally, the $t$--$J$ model was identified as a more appropriate approach towards stripe formation in hole doped cuprates. Extensive numerical studies were performed using exact diagonalization~\cite{Prelovsek}, DMRG methods~\cite{white,white09} or Monte Carlo simulations~\cite{himeda,capello}, all indicating that stripes indeed form in the hole doped $t$--$J$ model. Although these numerically exact results yielded a consistent picture, their range of validity is limited by small system sizes and boundary effects, and the inclusion of superconductivity has only recently become possible~\cite{capello,white09}.

Meanwhile, new approaches have been developed for a phenomenological characterization of the SC state that possibly coexists with spin- and charge-stripe order. Berg {\it et al}. suggested a new type of SC state, termed \textquotedblleft pair density wave\textquotedblright\ (PDW), in which Cooper pairs with center-of-mass momenta $\q$ and $-\q$ coexist and the pair density oscillates with wave vector $\q$~\cite{Agterberg08,Baruch:2008,Berg09,Berg:2009,Loder:2010}. The latter oscillation is analogous to the FFLO solution of a superconductor in magnetic fields presented by Fulde and Ferrell~\cite{fulde} and by Larkin and Ovchinnikov~\cite{larkin}, and is typically accompanied by a CDW with wave vector $2\q$. Houzet and Buzdin introduced a description of the FFLO state on the Ginzburg-Landau level~\cite{Houzet}, which was later used as a model for the PDW by Agterberg and Tsunetsugu~\cite{Agterberg08} and by Berg \textit{et al}.~\cite{Berg09}. Subsequently, the PDW state without antiferromagnetism was shown to be the energetically stable solution for a microscopic nearest-neighbor pairing model with sufficiently strong pairing interaction~\cite{Loder:2010}. This state indeed explains qualitatively some of the observed properties of striped cuprates~\cite{Berg:2009}.

To examine the nature of the SC state in coexistence with spin- and charge-stripe order, a simple microscopic model is desirable which allows to adequately describe magnetic and SC order simultaneously, merging the strong-coupling ansatz of the $t$--$J$ model with the weak-coupling framework of BCS theory. Work in this direction was recently performed using a Gutzwiller renormalized mean-field ansatz for the $t$--$J$ model~\cite{Raczkowski07,Yang:2009}. However, the PDW  did not prove to be the groundstate, but rather a \textquotedblleft modulated $d$-wave\textquotedblright\ state without phase shift of the SC order parameter between neighboring stripes was found as the lowest energy solution. In contrast to the PDW, this \textquotedblleft modulated $d$-wave\textquotedblright\ has a pair density that oscillates around a finite uniform $\q=\bf 0$ component, and pair and charge densities oscillate with the same wave vector.

In this paper we introduce a modified $t$--$\hat J$ model as an extension of the standard $t$--$J$  model which allows for a straightforward mean-field decoupling. This method of relaxing the no-double-occupancy constraint and extending the $t$--$J$ model by an additional on-site repulsion was originally introduced by Kagan and Rice~\cite{kagan} to demonstrate the possibility of $d$-wave superconductivity in the $t$--$J$ model. This ansatz enables us to study stripe formation and superconductivity in a wide range of parameters, and it merges two previously distinct approaches to theoretically describe cuprates in different regimes of the renormalized exchange coupling $\hat J$.

In section~\ref{sec:tj-bcs} we introduce the $t$--$\hat J$ model and identify two parameter regimes where a mean-field description of magnetic correlations is appropriate for hole-doped systems: $i)$~$\tj\ll t$. 
In this limit superconductivity dominates and antiferromagnetism is controlled by the on-site Coulomb repulsion rather than by the superexchange interaction. For small $\tj$ the $t$--$\tj$ model therefore maps onto a simple nearest-neighbor pairing Hamiltonian with an additional on-site repulsion $U$; this model is derived in section~\ref{sec:bcs} as the \textquotedblleft $U$-model\textquotedblright.
$ii)$~$\tj\sim t$. A strong exchange interaction tends to phase separate the system into a hole-free antiferromagnet and a hole-rich part~\cite{emery, kivelson.emery,white00}. We show that for $\hat J\sim t$ an intermediate regime exists in which the charge is separated periodically in half-filled AF regions and quarter-filled metallic regions. In this regime the Coulomb repulsion is less significant and the $t$--$\tj$ model approaches the so called \textquotedblleft$V$-model\textquotedblright, which was investigated in detail in~\cite{loder11} in particular with respect to the breakdown of superconductivity in La$_{2-x-y}$Nd$_y$Sr$_x$CuO$_4$ near $x=1/8$. A similar regime has also been found in the $t$--$J$ model itself using DMRG calculations~\cite{white09} and variational Monte Carlo simulations~\cite{himeda}.

Section~\ref{sec:bdg} describes the numerical strategy to obtain striped solutions within the framework of the Bogoliubov - de Gennes (BdG) equations.
Although charge- and spin-stripe order coexisting with superconductivity is found for $\tj\ll t$ as well as for $\tj\sim t$, there are important qualitative differences in the two regimes. We provide a detailed comparison for the origin of stripe formation in both regimes in section~\ref{sec:U-V}. The striped groundstate solution presented in this section is energetically favored among a considerable number of other possible solutions. Many of these energetically unfavorable solutions exist only in certain temperature ranges. We therefore present a survey of the temperature dependences of the most regular solutions in section~\ref{sec:T}. In section~\ref{sec:conc} we draw conclusions and discuss our findings in relation to striped cuprate materials.

\section{$\bm t$--$\bm J$ versus BCS Model}\label{sec:tj-bcs}

Shortly after the discovery of high-$T_{\rm c}$-cuprates it was agreed upon that the electronic structure of the CuO$_2$ planes is well described by a three-band Hubbard (Emery) model~\cite{unger}, which can be further reduced to a one-band repulsive Hubbard model for electron densities above $2/3$~\cite{anderson87}. Apart from numerically exact solutions, two different approaches have commonly been followed to treat the Hubbard interaction $U\sum_{i,s}n_{i,s}n_{i,-s}$: $i)$~In the strong-coupling regime ($U\sim10\,t$), the Hubbard model is approximated by the $t$--$J$ model with a finite superexchange coupling $J$~\cite{anderson59,zhang}. This effective interaction is non-local and responsible for superconductivity as well as for antiferromagnetism. $ii)$~Alternatively, effective non-local interactions in the finite $U$ Hubbard model can be computed numerically using, for example, QMC simulations~\cite{scalapino} or determined diagrammatically, e.g. in the RPA approximation. One thereby obtains a non-local attractive interaction between electrons on nearest-neighbor sites which may serve as a basis for a BCS type mean-field description.

In this section, we derive mean-field Hamiltonians for both of the above approaches, $i)$ and $ii)$, and show that they indeed describe the same physics and are formally equivalent.

\subsection{$\bm t$--$\bm\tj$ Model}\label{sec:tj}

We start from the $t$--$J$ Hamiltonian:
\begin{align}
{\cal H}_{t\text{-}J}=-\sum_{i,j,s}t_{ij}\hat c^\dag_{is}\hat c_{js}+J\sum_{\langle i,j\rangle}\left({\bf S}_i\cdot{\bf S}_j-\frac{n_in_j}{4}\right),
\label{tj1}
\end{align}
where $n_{is}=c^\dag_{is}c_{is}$, $n_{i}=n_{i\ua}+n_{i\da}$, and $\hat c_{i,s}=c_{i,s}(1-n_{i,-s})$ is the projected annihilation operator which enforces the constraint that no lattice site is doubly occupied. The spin operator is ${\bf S}_i=\sum_{s,s'}c^\dag_{is}\bm\upsigma_{ss'}c_{is'}/2$ with $\bm\upsigma=(\sigma^x,\sigma^y,\sigma^z)$ containing the Pauli matrices. Here we use the matrix elements $t_{ij}=\{t,t'\}$ for nearest and next-nearest neighbor hopping with $t'=-0.4\,t$, and $\langle i,j\rangle$ denotes all pairs of nearest-neighbor sites $i$ and $j$. Note that the three-site terms are omitted in the Hamiltonian~(\ref{tj1}).

The constraint on the site occupation is the main obstacle in solving the $t$--$J$ model and also prevents a straightforward mean-field approximation. An ansatz to solve this problem is the slave-boson technique (see~\cite{Fancini} for a review). Here we follow alternatively a method by Kagan and Rice~\cite{kagan} who introduced an extension of ${\cal H}_{t\text-J}$, in which an additional Hubbard repulsion $\hat U$ replaces the local constraints:
\begin{multline}
{\cal H}_{t\text{-}\tj}=-\sum_{i,j,s}t_{ij}c^\dag_{i,s}c_{j,s}+\hat J\sum_{\langle i,j\rangle}\left({\bf S}_i\cdot{\bf S}_j-\frac{n_in_j}{4}\right)\\
+\frac{\hat U}{2}\sum_{i,s}n_{i,s}n_{i,-s}.
\label{tj2}
\end{multline}
In the limit $\hat U\rightarrow\infty$, one recovers ${\cal H}_{t\text -J}$ from~(\ref{tj2}) with $J= \hat J-4t^2/\hat U$, if the three-sites terms are omitted. 
Generally, mean-field solutions of ${\cal H}_{t\text{-}\tj}$ overestimate antiferromagnetism as compared to superconductivity at large interaction strength. Therefore more physical results are expected for smaller values of $\hat U$ (c.f. the discussion in section~\ref{sec:U-V}).

The Hamiltonian ${\cal H}_{t\text{-}\tj}$ is readily decoupled on the Hartree-Fock level. With
\begin{multline}
{\bf S}_i\cdot{\bf S}_j-\frac{n_in_j}{4}\\=
\frac{1}{2}\big[-n_{i\ua}n_{j\da}-n_{i\da}n_{j\ua}+c^\dag_{i\ua}c_{i\da}c^\dag_{j\da}c_{j\ua}+c^\dag_{i\da}c_{i\ua}c^\dag_{j\ua}c_{j\da}\big]
\label{tj3}
\end{multline}
the decoupled Hamiltonian becomes
\begin{align}
{\cal H}\mf_{t\text{-}\tj}=&-\sum_{i,j,s}t_{ij}c^\dag_{i,s}c_{j,s}+\frac{\hat U}{2}\sum_{i,s}\bar n_{i,s}c^\dag_{i,-s}c_{i,-s}\nonumber\\
&+\sum_{\langle i,j\rangle}\Big[\Delta^{\!s*}_{ij}c_{j\da}c_{i\ua}+\Delta^{\!s}_{ji}c^\dag_{i\ua}c^\dag_{j\da}\label{tj4}\\
&-\sum_s\left(\hat J\bar n_{i,-s}c^\dag_{j,s}c_{j,s}-{\cal X}_{ij,-s}c^\dag_{i,s}c_{j,s}\right)\Big]+C_1.\nonumber
\end{align}
with
\begin{align}
\Delta^{\!s}_{ij}&=-\frac{\hat J}{2}\left(\langle c_{j\da}c_{i\ua}\rangle-\langle c_{j\ua}c_{i\da}\rangle\right),
\label{tj5}\\
\bar n_{is}&=\langle c^\dag_{is}c_{is}\rangle,
\label{tj6}\\
{\cal X}_{ijs}&=\hat J\langle c^\dag_{is}c_{js}\rangle.
\end{align}
The constant $C_1$ is given by
\begin{align}
\scalebox{0.95}[1]{$\displaystyle
C_1\!=\!\sum_{\langle i,j\rangle}\!\left[\!\frac{\displaystyle\Delta^{\!s*}_{ji}\Delta^{\!s}_{ij}}{\hat J}-\frac{{\cal X}^*_{ji\da}{\cal X}_{ij\ua}}{2\hat J}+\frac{\hat J}{2}\bar n_{i\ua}\bar n_{j\da}\!\right]\!-\frac{\hat U}{2}\sum_{i,s}\bar n_{i,s}\bar n_{i,-s}$}.
\label{tj4.1}
\end{align}
$C_1$ is essential for determining the free energy correctly.
The superconducting order parameter $\Delta^{\!s}_{ij}$
represents nearest-neighbor electron pairing in the spin-singlet channel, while possible spin triplet contributions cancel within the Hartree-Fock decoupling of the $t$--$\hat J$ model. $\bar n_{is}$ is the spin resolved thermal average of the local charge density $n_i$ and ${\cal X}_{ijs}$ renormalizes the nearest-neighbor hopping amplitude $t$.

The pairing term in ${\cal H}\mf_{t\text{-}\tj}$ drives superconductivity with a maximum energy gain for an order parameter $\Delta^{\!s}_{ij}$ with $d$-wave symmetry, i.e. $\Delta_{i,i\pm\hat x}=-\Delta_{i,i\pm\hat y}$. Furthermore, ${\cal H}\mf_{t\text{-}\tj}$ contains two terms which favor antiferromagnetic order. The Hubbard term $\hat U\bar n_{i,s}c^\dag_{i,-s}c_{i,-s}$ costs the energy $\hat U$ for each doubly occupied lattice site and thus polarizes the spin on each site. The kinetic energy in turn is optimized by ordering the spins antiferromagnetically. The term $-\hat J\bar n_{i,-s}c^\dag_{j,s}c_{j,s}$ directly supports an opposite spin alignment on nearest-neighbor sites by the energy $-\tj|m_i|(\bar n_{j,s}+\bar n_{j,-s})/2$ per bond, where $m_i=n_{i\ua}-n_{i\da}$ is the local magnetic moment. For large values of $\hat J$ the system approaches perfect AF order and the Hubbard term looses relevance. On the other hand, even if all lattice sites are fully spin polarized by a large $\hat U$, a finite $\hat J$ term in~(\ref{tj4}) leads to AF correlations.

The above reasoning justifies to use $\hat{\cal H}_{t\text{-}\tj}$ with a finite $\hat U$ instead of the original $t$--$J$ Hamiltonian for the following physical reason: In the limit $U\rightarrow\infty$, the half-filled one-band Hubbard model decouples into independent localized electrons without any magnetic order. In the $t$--$J$ model, this is reflected by $J=t^2/U\rightarrow0$. If a multi-band Hubbard model is mapped onto the $t$--$J$ model, the superexchange terms lead to a finite and sizable $J$ and thus to antiferromagnetic order even in the limit of a large intra-orbital $U$~\cite{anderson59}. Antiferromagnetism is in this case controlled by $J$ rather than by $U$.

\subsection{$\bm U$- and $\bm V$-Model}\label{sec:bcs}

The dominant effective interactions derived from the QMC calculations in~\cite{scalapino} are the on-site repulsion $U$ and an attractive interaction $V$ for electrons on nearest-neighbor sites. If only these two dominant interactions are kept, we arrive at the real-space BCS type Hamiltonian 
\begin{multline}
{\cal H}_{UV}=-\sum_{i,j,s}t_{ij}c^\dag_{i,s}c_{j,s}
-\frac{V}{2}\!\sum_{\langle i,j\rangle,s}\!c^\dag_{i,s}c^\dag_{j,-s}c_{j,-s}c_{i,s}\\
+\frac{U}{2}\sum_{i,s}n_{i,s}n_{i,-s},
\label{tj7}
\end{multline}
with $V>0$ and  $U>0$. 
The mean-field decoupling of ${\cal H}_{UV}$ generates the bilinear Hamiltonian
\begin{multline}
{\cal H}\mf_{UV}=-\sum_{i,j,s}t_{ij}c^\dag_{i,s}c_{j,s}+\frac{U}{2}\sum_{i,s}\bar n_{i,s}c^\dag_{i,-s}c_{i,-s}\\
+\frac{1}{2}\sum_{\langle i,j\rangle}\Big[\Delta_{ij}c^\dag_{i\ua}c^\dag_{j\da}+\Delta^{\!*}_{ji}c_{j\da}c_{i\ua}-V\sum_s\bar n_{i,-s}c^\dag_{j,s}c_{j,s}\Big]+C_2
\label{tj8}
\end{multline}
with
\begin{align}
\Delta_{ij}=-V\langle c_{j\da}c_{i\ua}\rangle
\label{tj9}
\end{align}
and
\begin{align}
C_2=-\frac{1}{2}\sum_{\langle i,j\rangle}\left[\frac{\displaystyle\Delta^{\!*}_{ji}\Delta_{ij}}{V}+V\bar n_{i\ua}\bar n_{j\da}\right]-\frac{U}{2}\sum_{i,s}\bar n_{i,s}\bar n_{i,-s}.
\label{tj8.1}
\end{align}

Comparing ${\cal H}_{t\text{-}\tj}$ to ${\cal H}_{UV}$ reveals that the mean-field decoupled Hamiltonians are almost identical, if one identifies the interaction parameters $\{\hat J,\hat U\}\longleftrightarrow\{V,U\}$ in ${\cal H}\mf_{t\text{-}\tj}$ and ${\cal H}\mf_{UV}$, respectively. In ${\cal H}\mf_{UV}$, the term ${\cal X}_{ijs}$ is missing, which renormalizes the nearest-neighbor hopping $t$ in the $t$--$\hat J$ model. Since it effectively decreases $t$, it tends to stabilize superconductivity; its qualitative influence on the groundstate solution is however marginal (c.f.~\cite{ghosal}). More relevant are the differences in the definitions of the SC order parameters in ${\cal H}\mf_{t\text{-}\tj}$ and ${\cal H}\mf_{UV}$: whereas $\Delta^{\!s}_{ij}$ contains only the spin-singlet channel, $\Delta_{ij}$ includes also the $S_z=0$ triplet channel. In a non-magnetic system, the triplet component generally vanishes for a nearest-neighbor interaction, because the pair wave-function has even parity. In the presence of antiferromagnetism however, a finite triplet admixture appears (c.f.~\cite{zhang97,ghosal}). In ${\cal H}\mf_{UV}$, this \textquotedblleft $\pi$-triplet\textquotedblright\ has its own order parameter $\Delta^{\!d}_{ij}=-V(\langle c_{j\da}c_{i\ua}\rangle+\langle c_{j\ua}c_{i\da}\rangle)/2$, the spin-triplet component of $\Delta_{ij}$. However, $\Delta^{\!d}_{ij}$ remains small compared to the spin-singlet component, as we have verified. Therefore we neglect the triplet component in the following calculations and discussions and use the spin-singlet order parameter $\Delta^{\!s}_{ij}=-V(\langle c_{j\da}c_{i\ua}\rangle-\langle c_{j\ua}c_{i\da}\rangle)/2$ alone with the coupling constant $V$ [instead of $\hat J$ as in~(\ref{tj5})]. Furthermore, ${\cal H}\mf_{UV}$ has an additional prefactor $1/2$ in the SC term. Therefore SC order is weighted stronger in ${\cal H}\mf_{t\text{-}\tj}$ than in ${\cal H}\mf_{UV}$ and consequently $\hat J$ has to be larger than $V$ in order to stabilize antiferromagnetism (see the discussion in Sec.~\ref{sec:U}). 

Notice that the last term $V\sum_s\bar n_{i,-s}c^\dag_{j,s}c_{j,s}$ in~(\ref{tj8}), which appears identically also in ${\cal H}\mf_{t\text{-}\tj}$, is absent in the classical BCS theory, because in a homogeneous system it only renormalizes the chemical potential $\mu$ used to control the particle number (see section~\ref{sec:bdg}). It is however a source of AF order in the nearest-neighbor pairing model ${\cal H}_{UV}$.

A mean-field approach does not necessarily capture all the physics contained in the underlying model Hamiltonians. The above decoupling scheme lacks e.g. the proper physics of the Mott insulator to metal transition. There are however two specific cases in which the mean-field solutions of ${\cal H}_{UV}$ or ${\cal H}_{t\text-\tj}$ are indeed self-consistent: 

\noindent
1. If $V\ll t$, the term $-V\bar n_{i,-s}c^\dag_{j,s}c_{j,s}$ contributes little to the emergence of antiferromagnetism, which in this case is controlled mainly by $U$. If also $U$ is small, superconductivity is the dominating order and the SC gap is larger than the AF gap. Therefore the failure of the mean-field theory to describe the quasiparticle peak at the Fermi energy of a doped Mott insulator is not relevant. Thus, for small $V$,
${\cal H}\mf_{UV}$ reduces to the \textquotedblleft$U$-model\textquotedblright:
\begin{multline}
{\cal H}\mf_{U}=-\sum_{i,j,s}t_{ij}c^\dag_{i,s}c_{j,s}+\frac{U}{2}\sum_{i,s}\bar n_{i,s}c^\dag_{i,-s}c_{i,-s}\\
+\frac{1}{2}\sum_{\langle i,j\rangle}\Big[\Delta^{\!s}_{ij}c^\dag_{i\ua}c^\dag_{j\da}+\Delta^{\!s*}_{ji}c_{j\da}c_{i\ua}\Big]
\label{j11}
\end{multline}
discussed in detail in section~\ref{sec:U}.

\noindent
2. If $V$ is close to or larger than $t$, the term $V\bar n_{i,-s}c^\dag_{j,s}c_{j,s}$ is the dominant source of antiferromagnetism. The system separates into nearly half-filled AF regions (stripes) where superconductivity is suppressed, and into empty or quarter-filled metallic (or superconducting) regions, depending on the type of solution. Since commensurate antiferromagnetism appears only in the half-filled regions, a mean-field approach is consistent also in this case, and ${\cal H}\mf_{UV}$ reduces to the \textquotedblleft $V$-model\textquotedblright:
\begin{multline}
{\cal H}\mf_{V}=-\sum_{i,j,s}t_{ij}c^\dag_{i,s}c_{j,s}\\
+\frac{1}{2}\sum_{\langle i,j\rangle}\Big[\Delta^{\!s}_{ij}c^\dag_{i\ua}c^\dag_{j\da}+\Delta^{\!s*}_{ji}c_{j\da}c_{i\ua}-V\sum_s\bar n_{i,-s}c^\dag_{j,s}c_{j,s}\Big]
\label{j12}
\end{multline}
discussed in~\cite{loder11} and in section~\ref{sec:V}. Within the $V$-model, a finite $U$-term can be added without significant influence on the groundstate solution, since the $V$-term alone is sufficient to drive the AF order parameter close to its maximum.

\section{Striped Solutions}\label{sec:bdg}

\subsection{The Bogoliubov - de Gennes Equations}

Here we briefly summarize the basic steps to solve the Bogoliubov - de Gennes (BdG) equations for ${\cal H}\mf_U$ and ${\cal H}\mf_V$ and explain  how striped (or other inhomogeneous) solutions are identified. 

The Bogoliubov transformation which diagonalizes the Hamiltonians~(\ref{j11}) and~(\ref{j12}), respectively, is given by
\begin{align}
c_{i\ua}&=\sum_n\left[u_{ni\ua}a_{n\ua}-v^*_{ni\da}a^\dag_{n\da}\right],
\label{j4.1}\\
c_{i\da}&=\sum_n\left[u_{ni\da}a_{n\da}+v^*_{ni\ua}a^\dag_{n\ua}\right],
\label{j4}
\end{align}
where the coefficients $u_{nis}$ and $v_{nis}$ are obtained from the eigenvalue equation
\begin{align}
\begin{pmatrix}\hat t_s&\hat\Delta\cr\displaystyle\hat\Delta^{\!*}&-\hat t^*_{-s}\end{pmatrix}\begin{pmatrix}{\bf u}_{n,s}\cr{\bf v}_{n,-s}\end{pmatrix}=E_n\begin{pmatrix}{\bf u}_{n,s}\cr{\bf v}_{n,-s}\end{pmatrix}
\label{j5}
\end{align}
with ${\bf u}_{n,s}=(u_{n1s},\dots,u_{nis},\dots)$ and the corresponding vector ${\bf v}_{n,s}$.
The sum over $n$ in~(\ref{j4.1}) and~(\ref{j4}) extends over all positive eigenvalues $E_n$ and the corresponding eigenvectors.
The operators $\hat t_s$ and $\hat\Delta$ act on ${\bf u}_{n,s}$ and ${\bf v}_{n,s}$ as
\begin{align}
\hat t_su_{nis}&=-\sum_{l}t_{il}u_{nls}-\sum_{j}V\bar n_{j,-s}u_{nis}+U\bar n_{i,-s}u_{nis},\\
\hat\Delta v_{nis}&=\sum_{j}\Delta^{\!s}_{ij}v_{njs},
\label{j6}
\end{align}
where $j$ labels all nearest-neighbor sites of $i$ and $l$ labels all sites for which $t_{il}$ is finite, i.e. nearest and next-nearest neighbors.

Using the symmetry of the energy spectrum $E_n$ around $E=0$, it is sufficient to diagonalize~(\ref{j5}) for a single spin component, say $s=\ \ua$.
Inserting the transformation (\ref{j4}) into~(\ref{tj6}) and~(\ref{tj9}) leads to the self-consistency equations for the order parameter and the spin resolved densities $\bar n_{i\ua}$ and $\bar n_{i\da}$:
\begin{align}
\scalebox{0.98}[1]{$\displaystyle
\Delta^{\!s}_{ij}=\frac{V}{2}\sum_n\left[u_{ni\ua}v^*_{nj\da}f(E_n-\mu)+u_{nj\ua}v^*_{ni\da}f(-E_n+\mu)\right],$}
\label{j7}
\end{align}\vspace{-5mm}
\begin{align}
\bar n_{i\ua}&=\sum_nu^2_{ni\ua}f(E_n-\mu),\label{j8}\\
\bar n_{i\da}&=1-\sum_nv^2_{ni\da}f(E_n-\mu).
\label{j9}
\end{align}
Here we used that $\sum_nv^2_{ni\da}f(E_n-\mu)$ equals the number of holes with spin $s=\ \da$, and $f(E-\mu)$ denotes the Fermi function. The chemical potential $\mu$ fixes the total particle number $N$ to the desired value. Note that in the presence of antiferromagnetism, $\Delta^{\!s}_{ij}$ has a finite spin-triplet component.

The BdG equations~(\ref{j5}) and (\ref{j7} -- \ref{j9}) are solved iteratively until self-consistency in $\Delta^{\!s}_{ij}$ and $\bar n_{is}$ is achieved. In each loop $\mu$ is adjusted to keep $N$ constant. To obtain a higher momentum and energy resolution, we use the super cell method to block-diagonalize the eigenvalue problem~(\ref{j5}). This procedure was introduced by Wang and MacDonald~\cite{Wang} for magnetic super cells (for more details, see~\cite{ghosal} and~\cite{Schmid}). Here the size of a super cell must be chosen commensurate with the wavelength of the striped solution we attempt to find.

The BdG equations typically have more than one solution, into which the self-consistency cycle may converge. An anticipated solution can usually be selected by choosing appropriate initial values for $\Delta^{\!s}_{ij}$ and $\bar n_{is}$ (and for $\mu$), the search for new solutions is however always a demanding task. Typically, two types of SC solutions compete in the presence of AF stripes: the \textquotedblleft modulated $d$-wave\textquotedblright\ and the PDW solution. Both are characterized by an order parameter with local $d$-wave symmetry, i.e. $\Delta^{\!s}_{i,i\pm\hat x}$ has the opposite sign of $\Delta^{\!s}_{i,i\pm\hat y}$, but the absolute values are different, if an extended $s$-wave component exists. In the PDW solution, $\Delta^{\!s}_{ij}$ also has opposite signs on neighboring stripes. Such sign changes in $\Delta^{\!s}_{ij}$ or in the antiferromagnetic order parameter $M_i=(-1)^im_i$ do not evolve continuously from a uniform initial state. In order to trace the anti phase striped solutions discussed in section~\ref{sec:U-V}, the initial values of $\Delta^{\!s}_{ij}$ and $M_i$ must therefore have the same anti phase stripe pattern with equal wavelength.

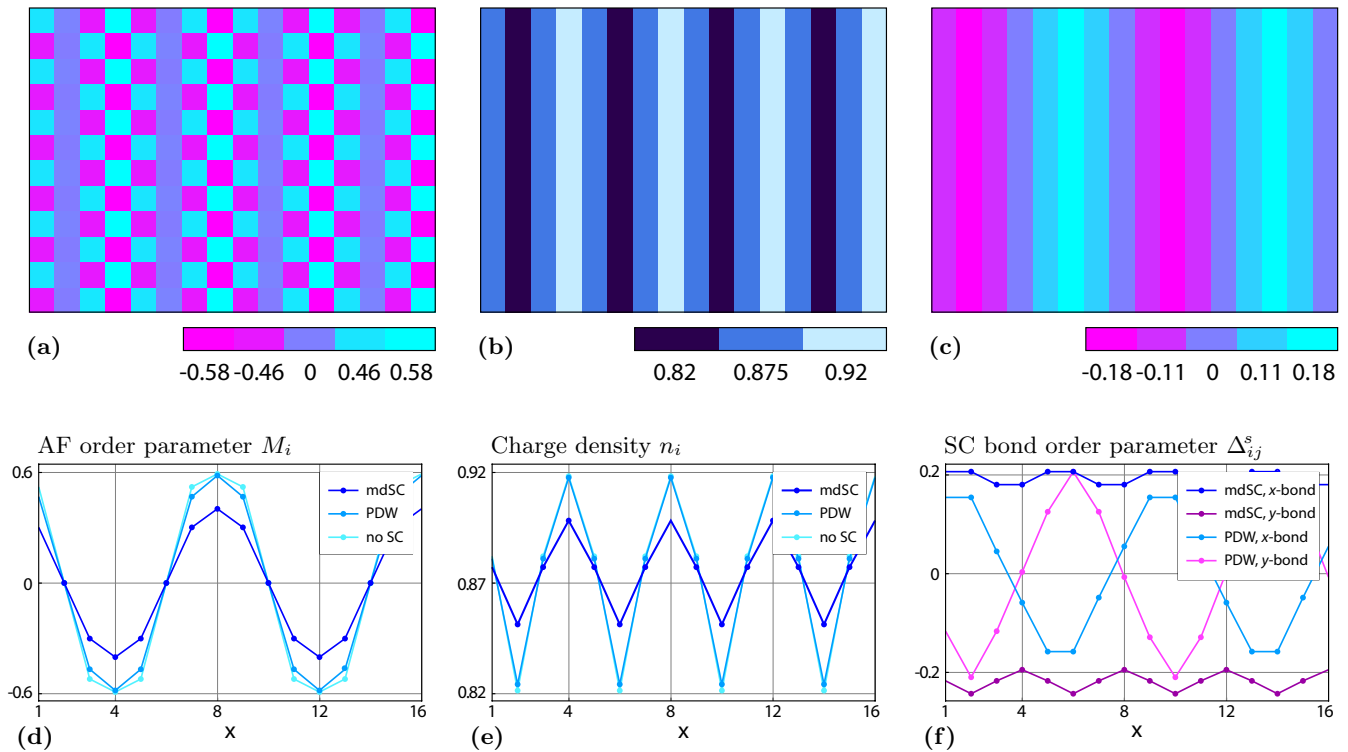
\begin{figure*}[t!]
\centering
\vspace{2.5mm}
\hspace{3mm}
\begin{overpic}
[width=0.635\columnwidth]{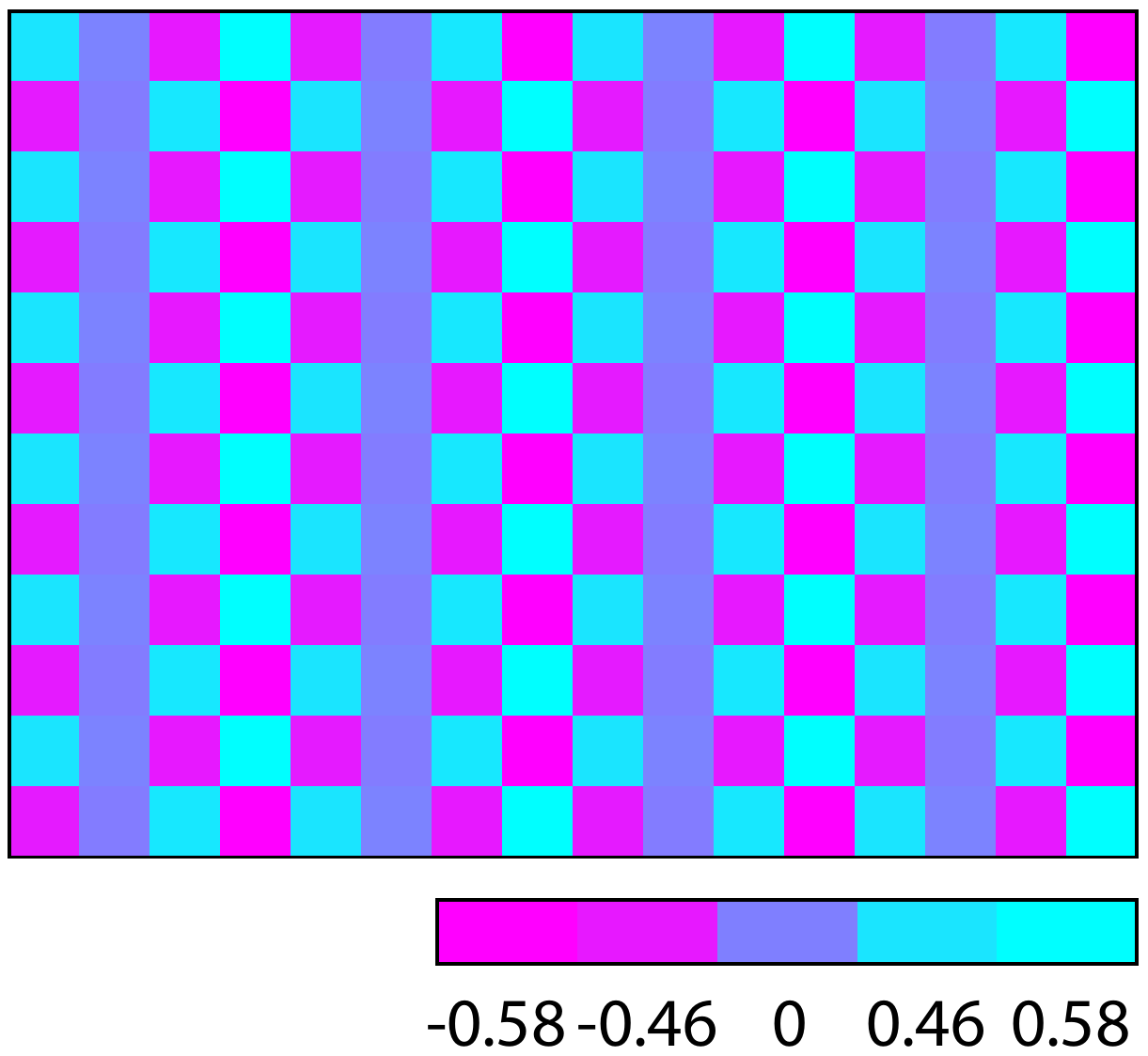}
\put(0,7){\bf(a)}
\end{overpic}\hspace{4mm}
\begin{overpic}
[width=0.635\columnwidth]{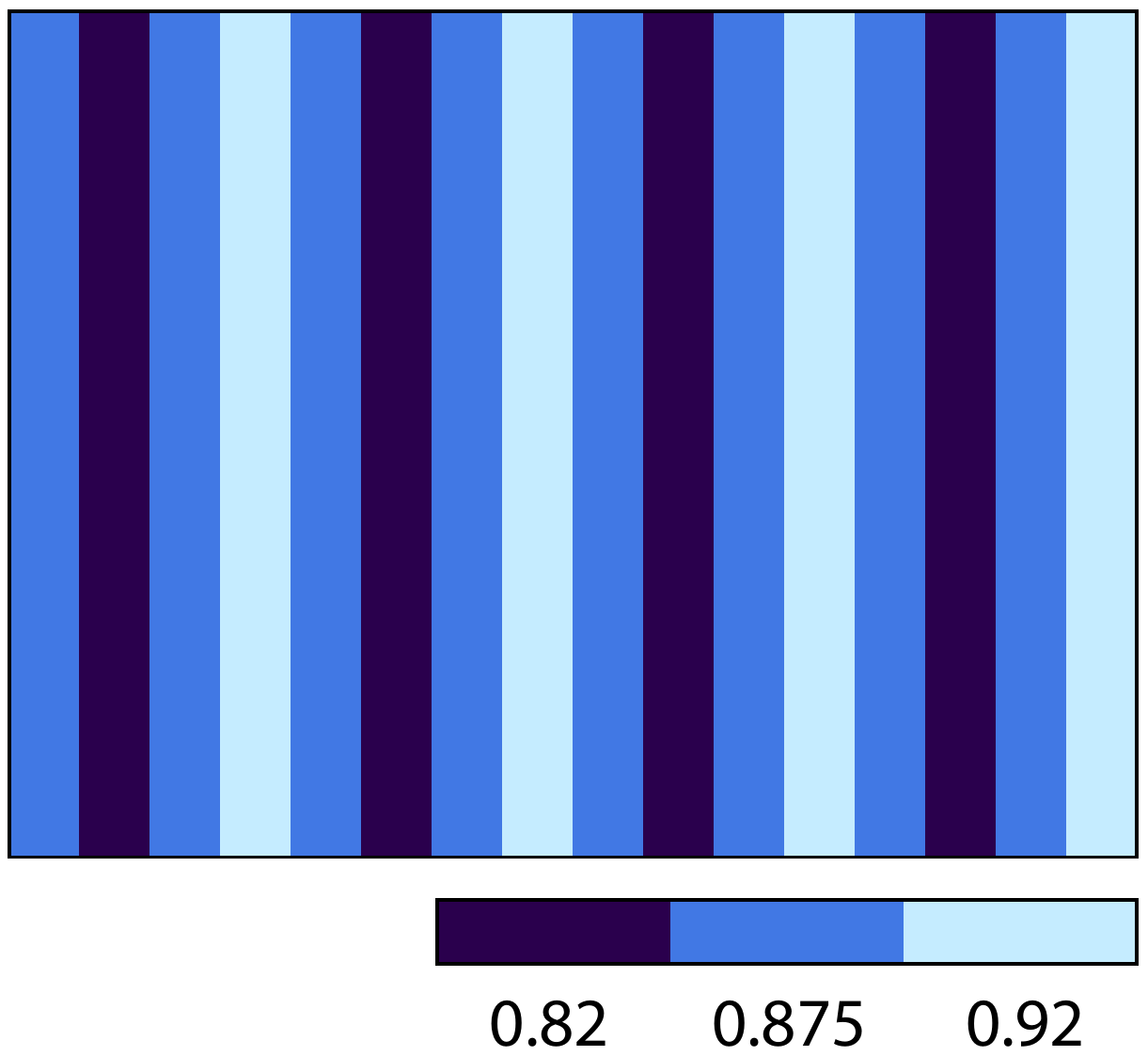}
\put(0,7){\bf(b)}
\end{overpic}\hspace{4mm}
\begin{overpic}
[width=0.635\columnwidth]{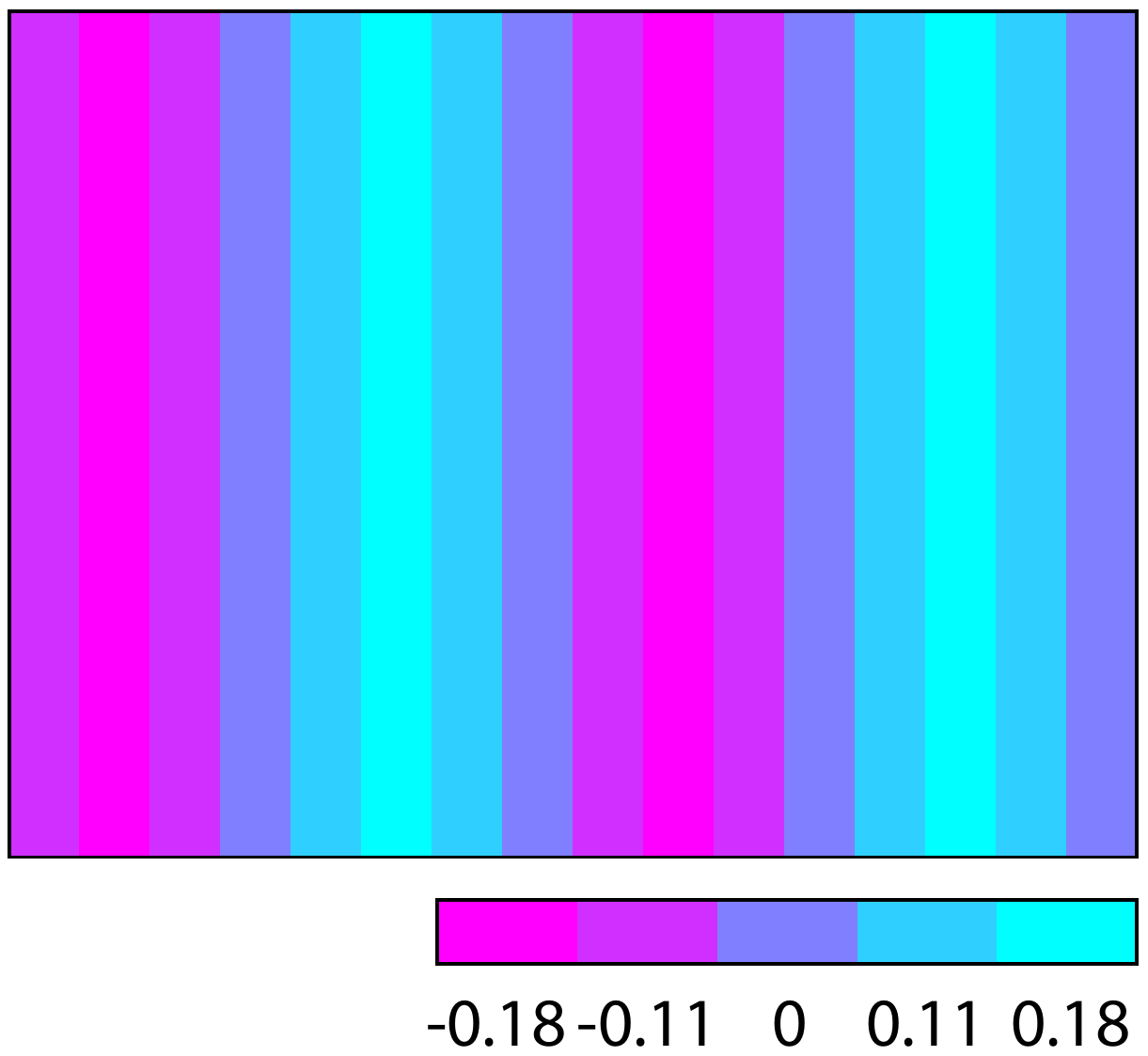}
\put(0,7){\bf(c)}
\end{overpic}\\[10mm]
\begin{overpic}
[width=0.65\columnwidth]{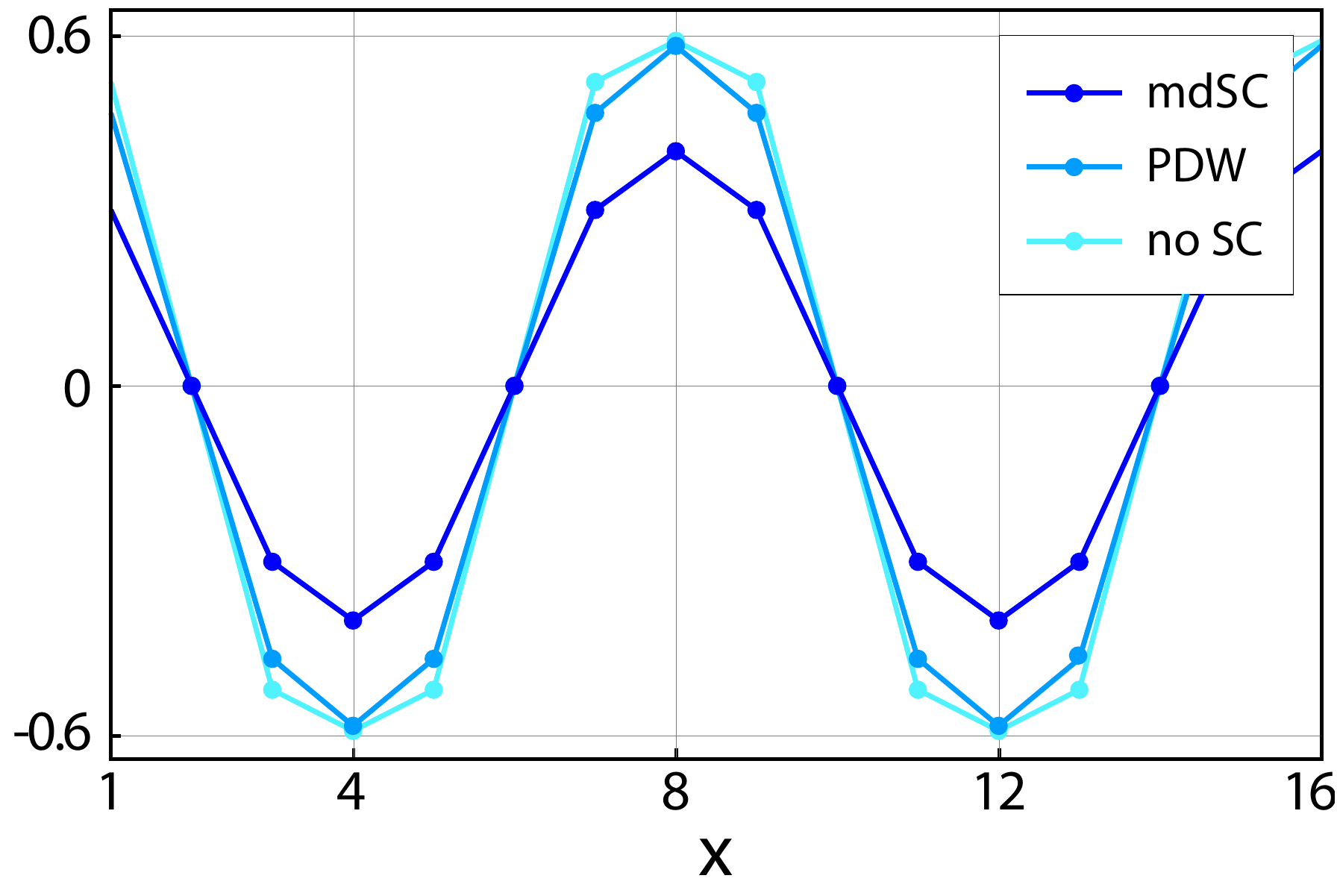}
\put(8,69){AF order parameter $M_i$}
\put(3,0){\bf(d)}
\end{overpic}\hspace{3mm}
\begin{overpic}
[width=0.65\columnwidth]{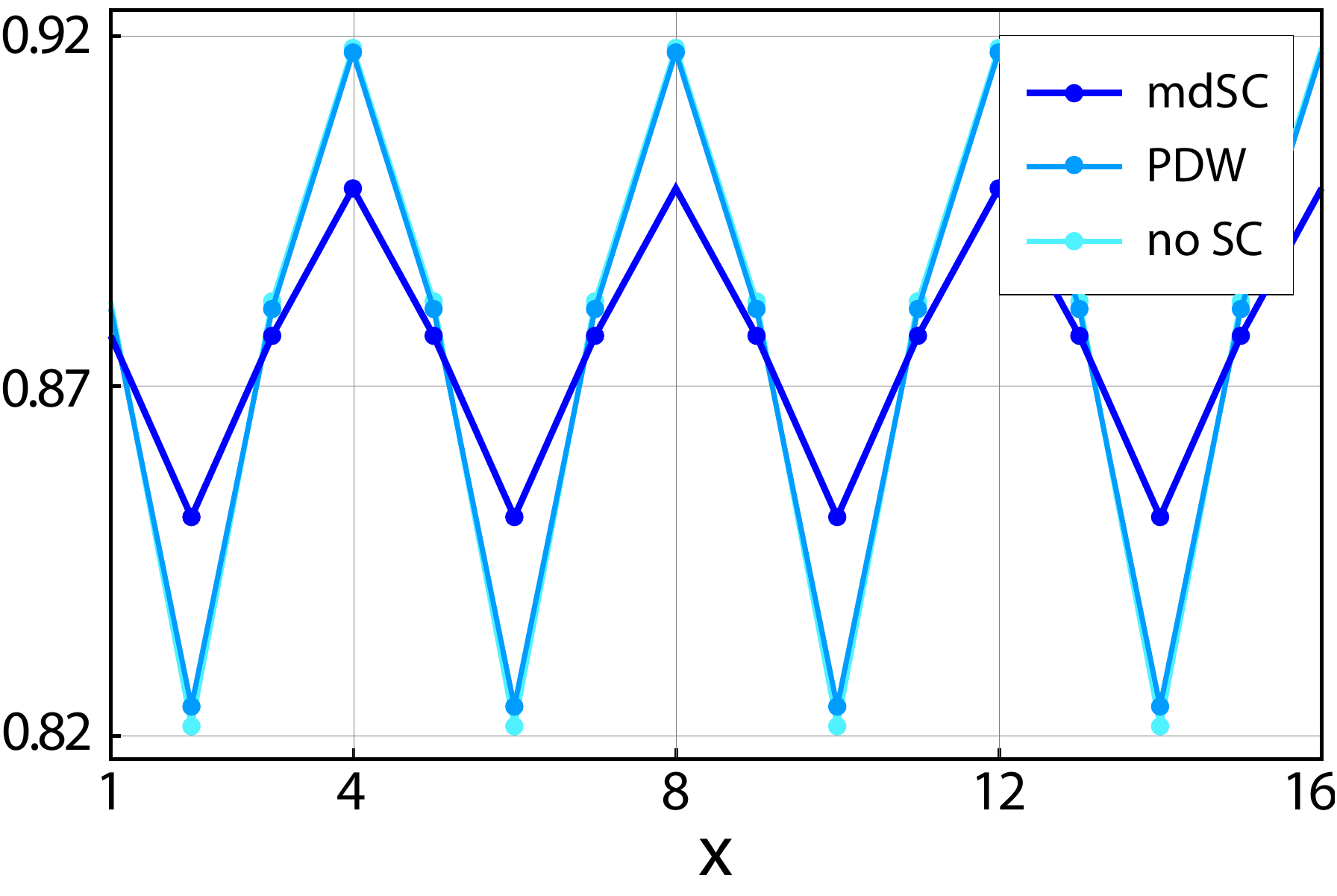}
\put(8,69){Charge density $n_i$}
\put(3,0){\bf(e)}
\end{overpic}\hspace{3mm}
\begin{overpic}
[width=0.65\columnwidth]{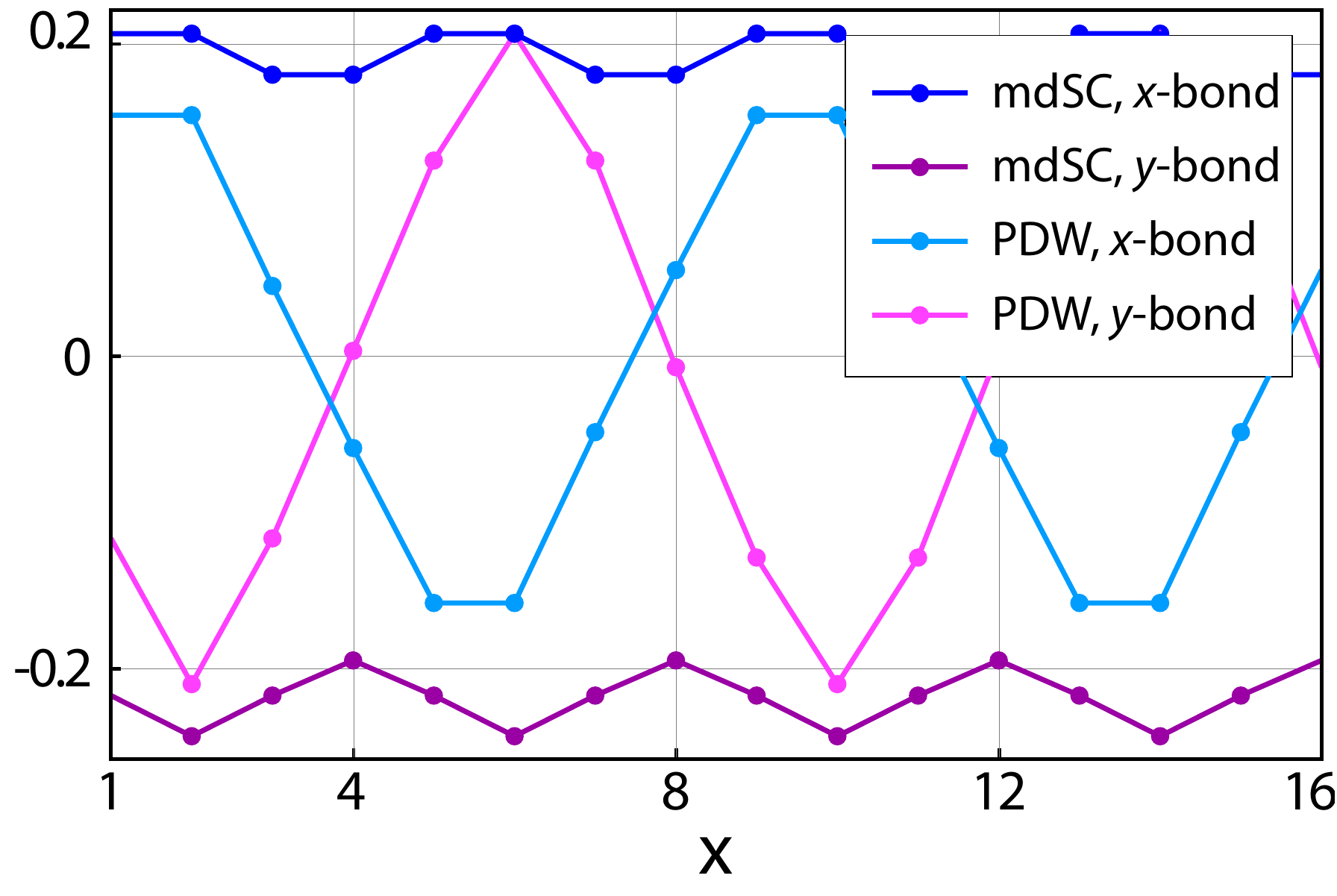}
\put(8,69){SC bond order parameter $\Delta^{\!s}_{ij}$}
\put(3,0){\bf(f)}
\end{overpic}
\caption{Real-space characterization of the \textquotedblleft modulated $d$-wave\textquotedblright (mdSC), PDW, and non-SC solutions of the $U$-model. (a) Local magnetization $m_i=n_{i\ua}-n_{i\da}$. (b) Charge density $n_i=n_{i\ua}+n_{i\da}$. (c) $d$-wave projection $\Delta^{\!d}_i$ of the SC order parameter. (a - c) correspond to the PDW solution.
(d) AF order parameter $M_i=(-1)^im_i$. (e) Charge density $n_i$. (f) SC bond order-parameter $\Delta^{\!s}_{i,i+\hat x}$ and $\Delta^{\!s}_{i,i+\hat y}$. All results were obtained on a 16$\times$12 lattice for ${V=2\,t}$, and $U=3.4\,t$.
}
\label{fig1}
\end{figure*}

The groundstate for each set of parameters is found by minimizing the system's free energy $F=\langle{\cal H}\rangle-TS$ over all self-consistent solutions. This means minimizing $F$ over stripe patterns with different wavelength, but also over bond- and site-centered stripes and different types of SC states, i.e. a \textquotedblleft modulated $d$-wave\textquotedblright\ or a PDW state. Three general observations are made:

1. For those values of $V$ and $U$ for which striped solutions exist at all, solutions of the BdG equations exist for all stripe widths that are commensurate with the finite-size lattice, with AF stripes separated by an anti phase hole-rich domain wall.

2. For $x$ close to $1/8$ the striped groundstate has a stripe wavelength $\lambda=8\,a$, where $a$ is the lattice constant. For small values of $V$ and $U$, the groundstate is a homogeneous non-magnetic $d$-wave superconductor, whereas phase separation occurs, if $V$ or $U$ exceeds a certain limit.

3. A SC state of PDW type occurs only, if the pairing interaction $V$ is large enough ($V\gtrsim t$). Its energy however is \textit{always} larger than the energy of a \textquotedblleft modulated $d$-wave\textquotedblright\ state, although the energy difference becomes vanishingly small for large $V$. This result is in agreement to earlier work using either a Gutzwiller approximation~\cite{Yang:2009} or DMRG calculations~\cite{white09} for the $t$--$J$ model. In contrast, the pure PDW in non-magnetic systems was shown to be the SC groundstate for a sufficiently strong nearest-neighbor pairing interaction~\cite{Loder:2010}.

A thorough analysis of the groundstate properties in different parameter regimes follows in section~\ref{sec:U-V}. All calculations, except for section~\ref{sec:T}, were performed at temperature $k_{\rm B}T=0.01\,t$.

\subsection{Comment on Energy Minimization}\label{sec:energy}

In the grand canonical framework of BCS-type mean-field theories, the thermodynamically stable groundstate is defined by the global minimum of the grand canonical potential $\Omega=\langle{\cal H}\rangle-TS-\mu N$, where $\mu$ is the chemical potential and the entropy $S$ of the system is given by
\begin{align}
S=-k_{\rm B}\sum_n\left[f(E_n)\ln f(E_n)+f(-E_n)\ln f(-E_n)\right].
\label{j10}
\end{align}
Self-consistent solutions of the BdG equations correspond to local minima in $\Omega$. To determine the groundstate or the thermodynamically stable state at finite temperatures, it is therefore necessary to compare different self-consistent solutions of the BdG equations for each chosen set of parameters (e.g., with or without sign changes in $\Delta^{\!s}_{ij}$, site-centered or bond-centered stripe patterns, etc.), and to select the solution with the lowest $\Omega$.

This minimization procedure is valid for the grand-canonical ensemble with a fixed chemical potential $\mu$ and a variable particle number $N$. However, our calculations aim at solutions with fixed $N$ by adjusting $\mu$, although all expectation values are evaluated grand canonically. Therefore the thermodynamically stable state is determined by the minimum of $F=\Omega+\mu N$.

It is instructive to compare $\Omega$ and $F$ for solutions of the BdG equations with fixed $\mu$ or with fixed $N$, respectively. For fixed $N$, one can prepare the system initially in a state where $\Omega$ is minimal, but not $F$. Specifically this would be a state where antiferromagnetism is absent. For fixed $N$, $\Omega$ has a global minimum without AF order, whereas $F$ has two degenerate minima with AF order.
As $F$ is not minimal for this state, the non-magnetic solution is unstable and the BdG equations will eventually converge into one of the two minima of $F$. By contrast, the roles of $\Omega$ and $F$ are interchanged, if $\mu$ is fixed and $N$ varied. Therefore fixing $N$ instead of $\mu$ is numerically equivalent to a Legendre transformation from the grand canonical back to the canonical ensemble.

\section{Stripe patterns in the $\bm U$- and the $\bm V$-model}\label{sec:U-V}

\subsection{$\bm U$-Model}\label{sec:U}

\begin{figure}[t!]
\centering
\centering
\vspace{2.5mm}
\begin{overpic}
[width=0.44\columnwidth]{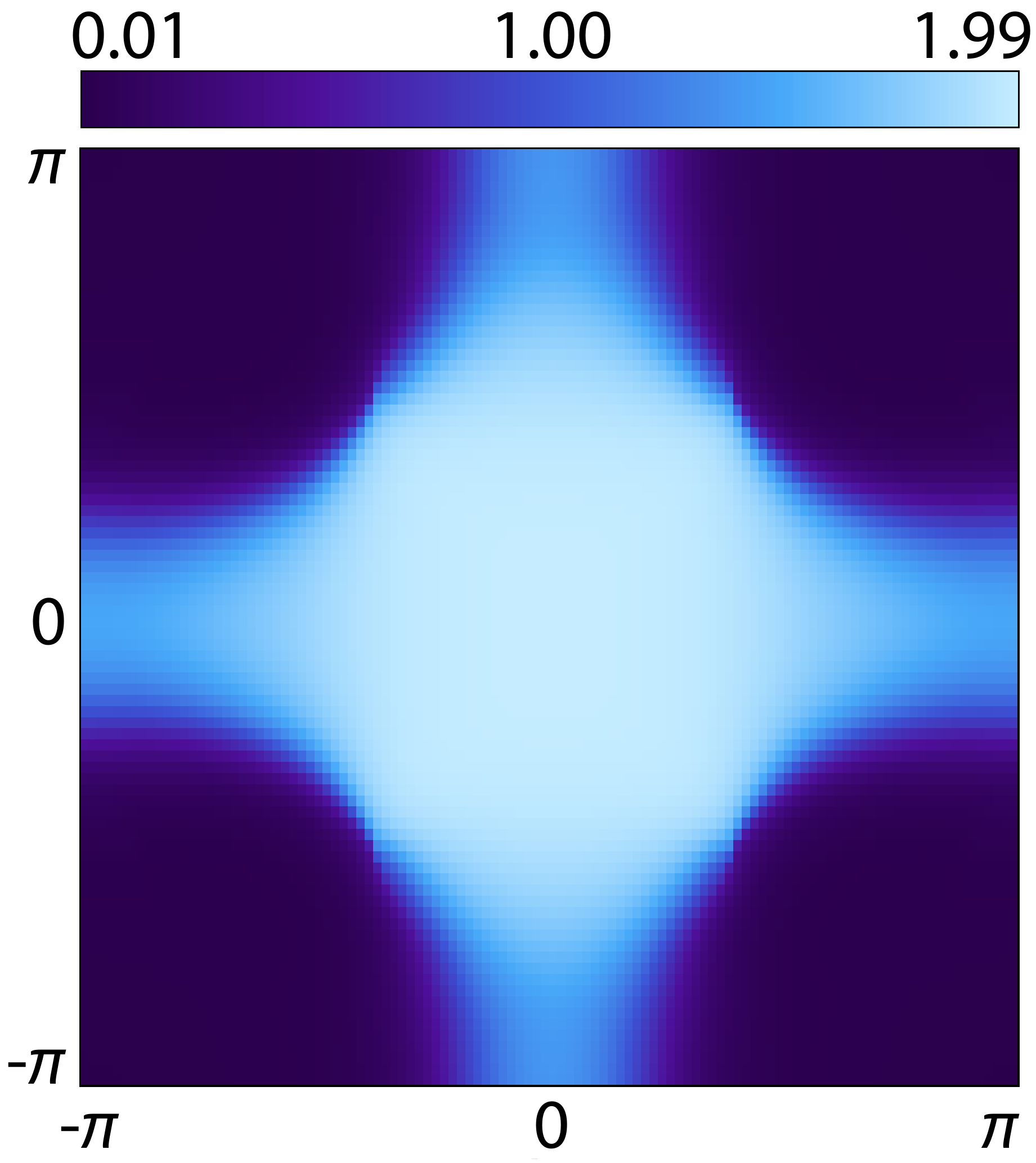}
\put(-8,92){\bf(a)}
\end{overpic}\hspace{5mm}
\begin{overpic}
[width=0.44\columnwidth]{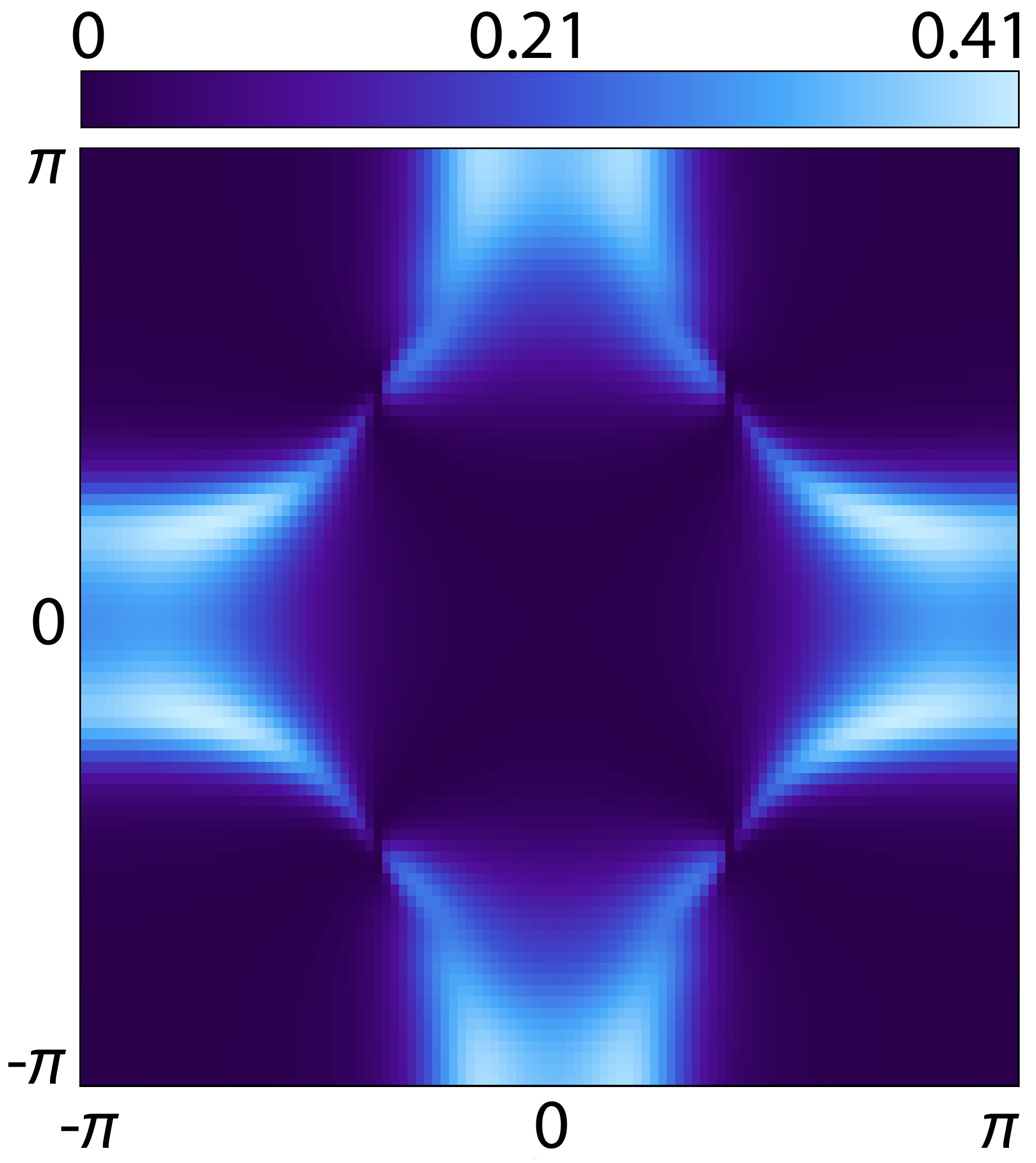}
\put(-8,92){\bf(b)}
\end{overpic}\\\vspace{3mm}
\begin{overpic}
[width=0.44\columnwidth]{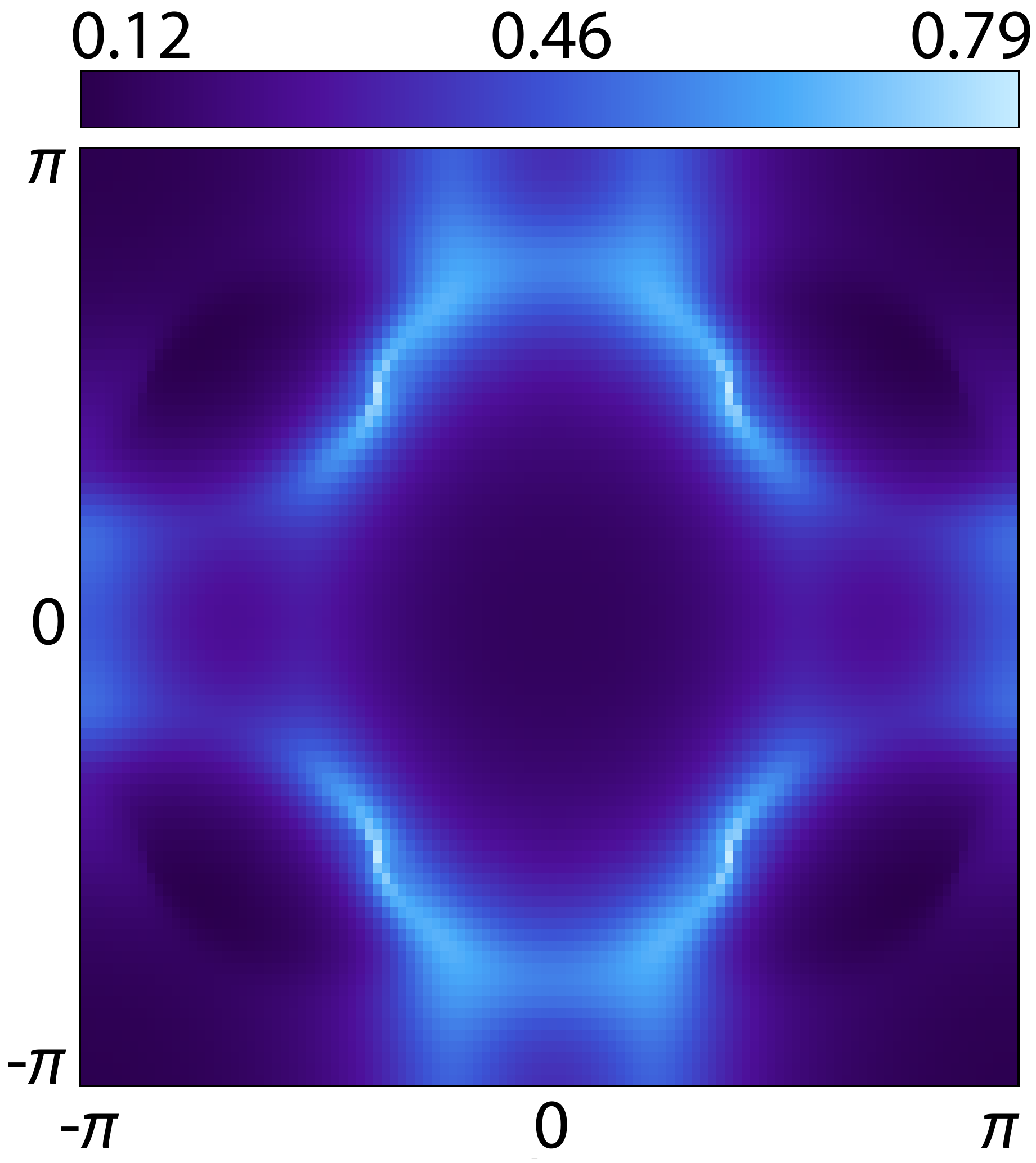}
\put(-8,92){\bf(c)}
\end{overpic}\hspace{5mm}
\begin{overpic}
[width=0.44\columnwidth]{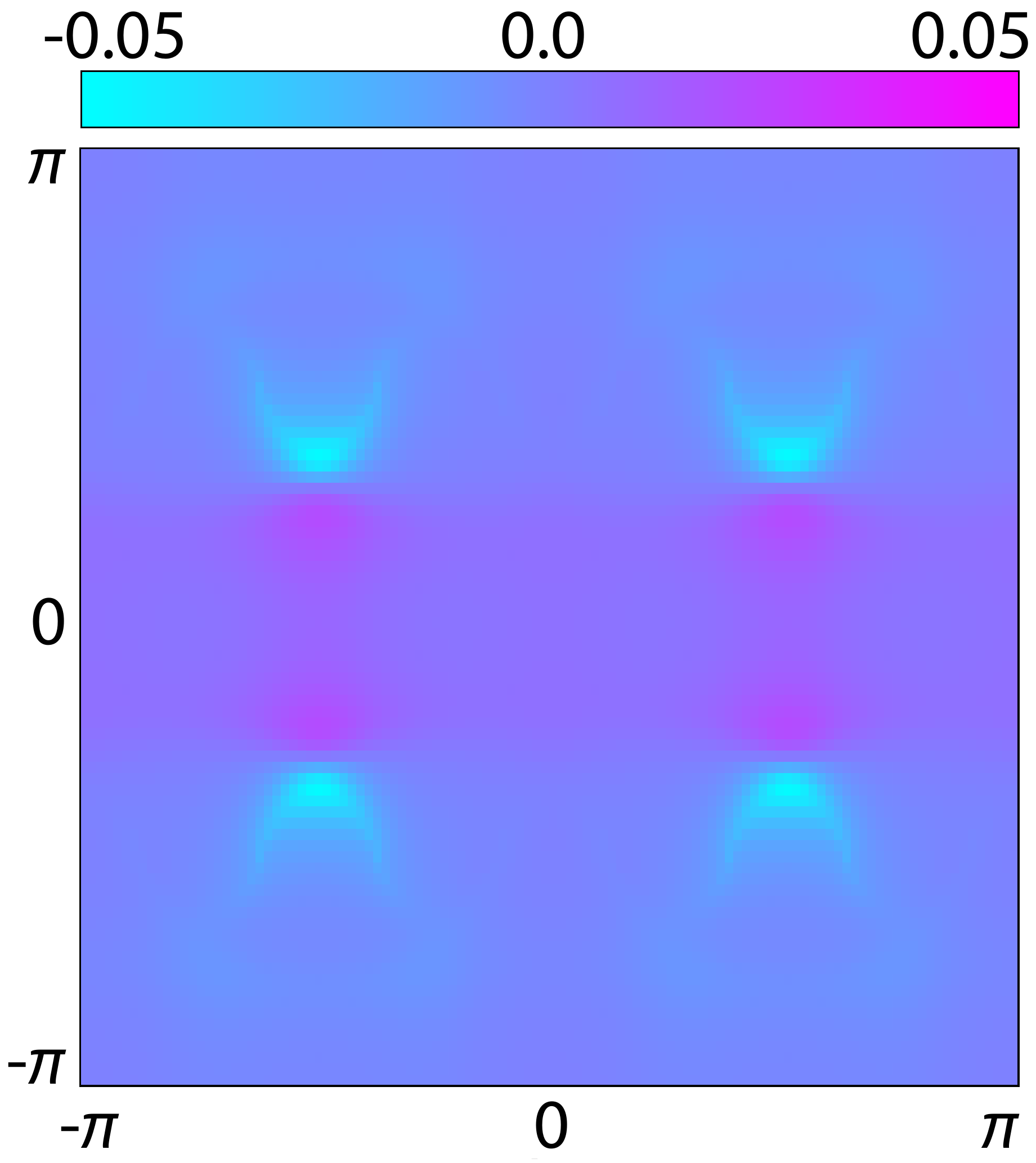}
\put(-8,92){\bf(d)}
\end{overpic}
\caption{Momentum-space characterization of the \textquotedblleft modulated $d$-wave\textquotedblright\ solution of the $U$ model for $x=1/8$, ${V=2\,t}$, and $U=3.4\,t$ on a 16$\times$12 lattice with 7$\times$7 supercells. (a) Occupation probability $n({\bf k})$, (b) pair density $P({\bf k})$, (c) spin order $\rho_S({\bf k})$, and (d) charge order $\rho_C({\bf k})$.
}
\label{fig3}
\end{figure}

If $V\ll t$ (say $V\approx 0.2\,t$), the energy gain originating from $-V\sum_s\bar n_{i,-s}c^\dag_{j,s}c_{j,s}$ upon ordering antiferromagnetically is smaller than the accompanying cost in kinetic energy. If also $U\lesssim t$, the groundstate will be a homogeneous, non-magnetic $d$-wave superconductor, whereas larger values of $U$ induce AF order. For moderate values of $U$, $d$-wave superconductivity is still the dominating order, which coexists with weak antiferromagnetism above a critical value of $U$ depending on the pairing interaction strength. This regime is described by the $U$-model given by the simplified mean-field Hamiltonian ${\cal H}\mf_U$ in~(\ref{j11}).

The $U$-model supplemented with non.magnetic impurity potentials has been widely and successfully used to describe AF correlations in disordered cuprate superconductors~\cite{andersen07,Schmid} or around vortex cores in magnetic fields~\cite{Wang,zhu,ghosal,Schmid}; Andersen and Hedeg\aa rd showed the existence of striped solutions for this model~\cite{Andersen05}. In particular disorder induced antiferromagnetism appears above a critical value of $U$. Here we show that this model also allows for striped groundstate solutions in clean systems for $U_{{\rm c}1}<U<U_{{\rm c}2}$, where the critical value $U_{{\rm c}1}\approx3\,V$ is slightly larger than for anitferromagnetism in disordered systems. For values of $U>U_{{\rm c}2}\approx 6\,V$, we obtain a homogeneous AF solution without superconductivity, i.e. a doped AF Mott insulator, for which the mean-field treatment is not adequate.

\begin{figure}[t!]
\centering
\vspace{2.5mm}
\begin{overpic}
[width=0.44\columnwidth]{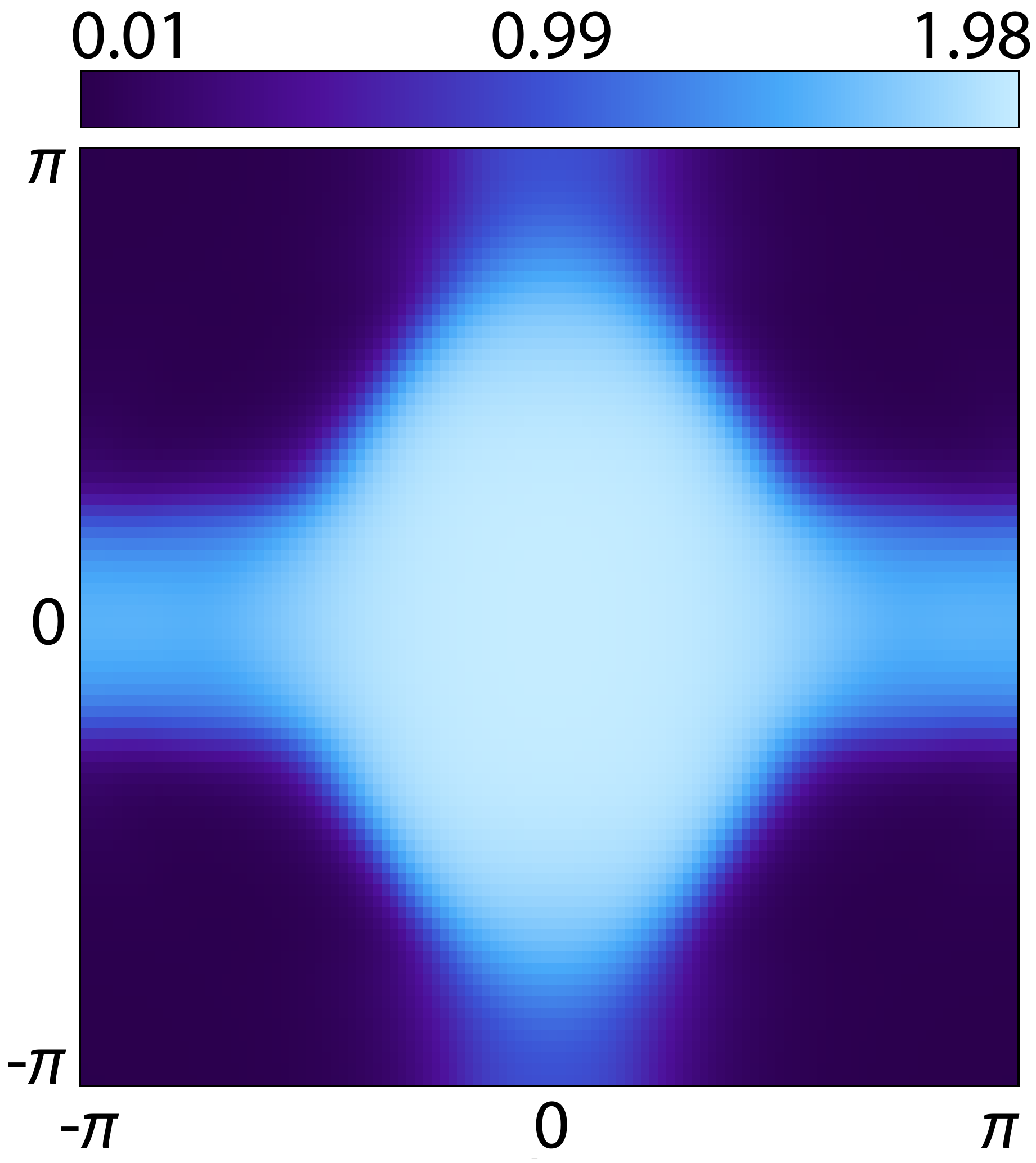}
\put(-8,92){\bf(a)}
\end{overpic}\hspace{5mm}
\begin{overpic}
[width=0.44\columnwidth]{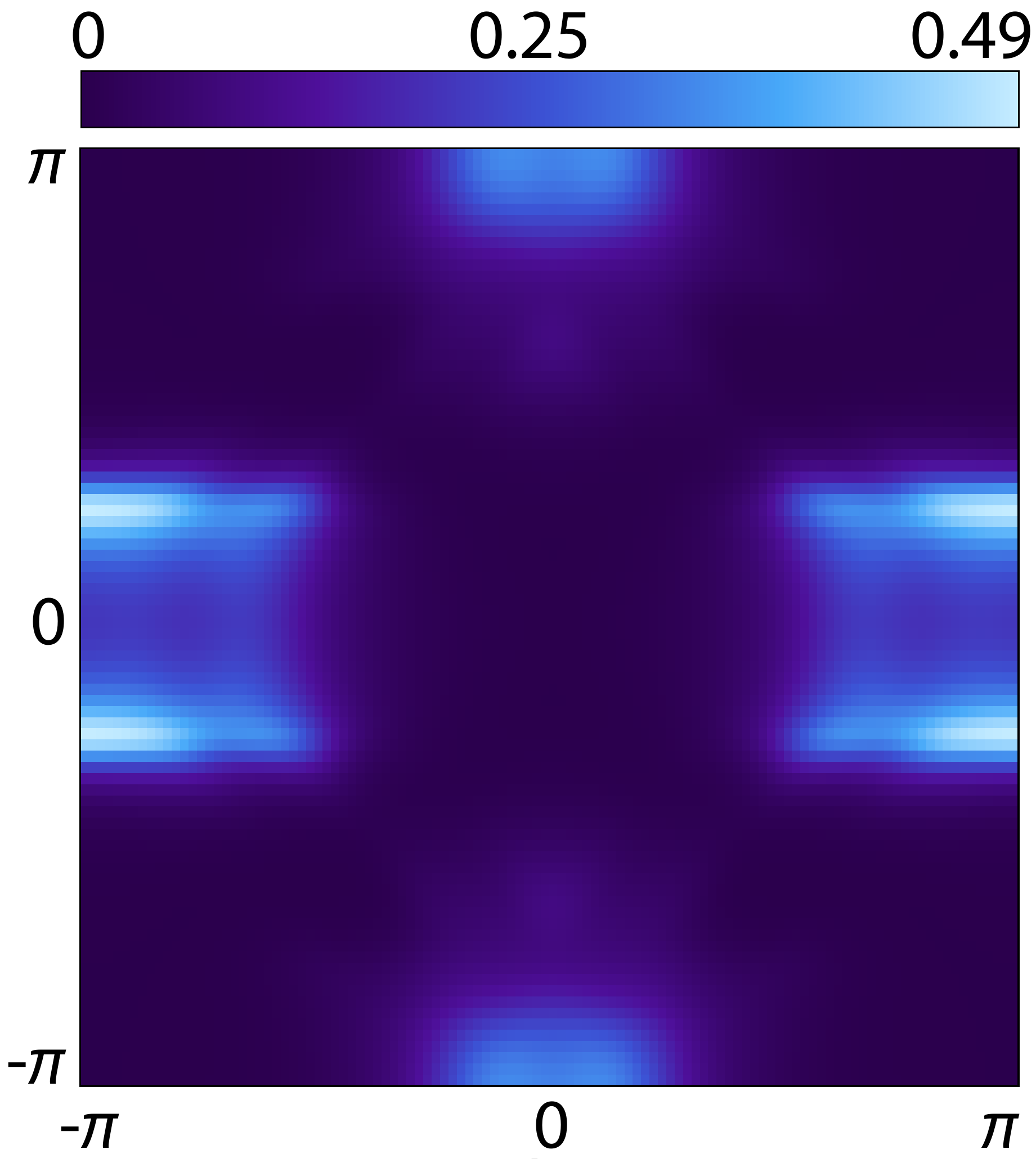}
\put(-8,92){\bf(b)}
\end{overpic}\\\vspace{3mm}
\begin{overpic}
[width=0.44\columnwidth]{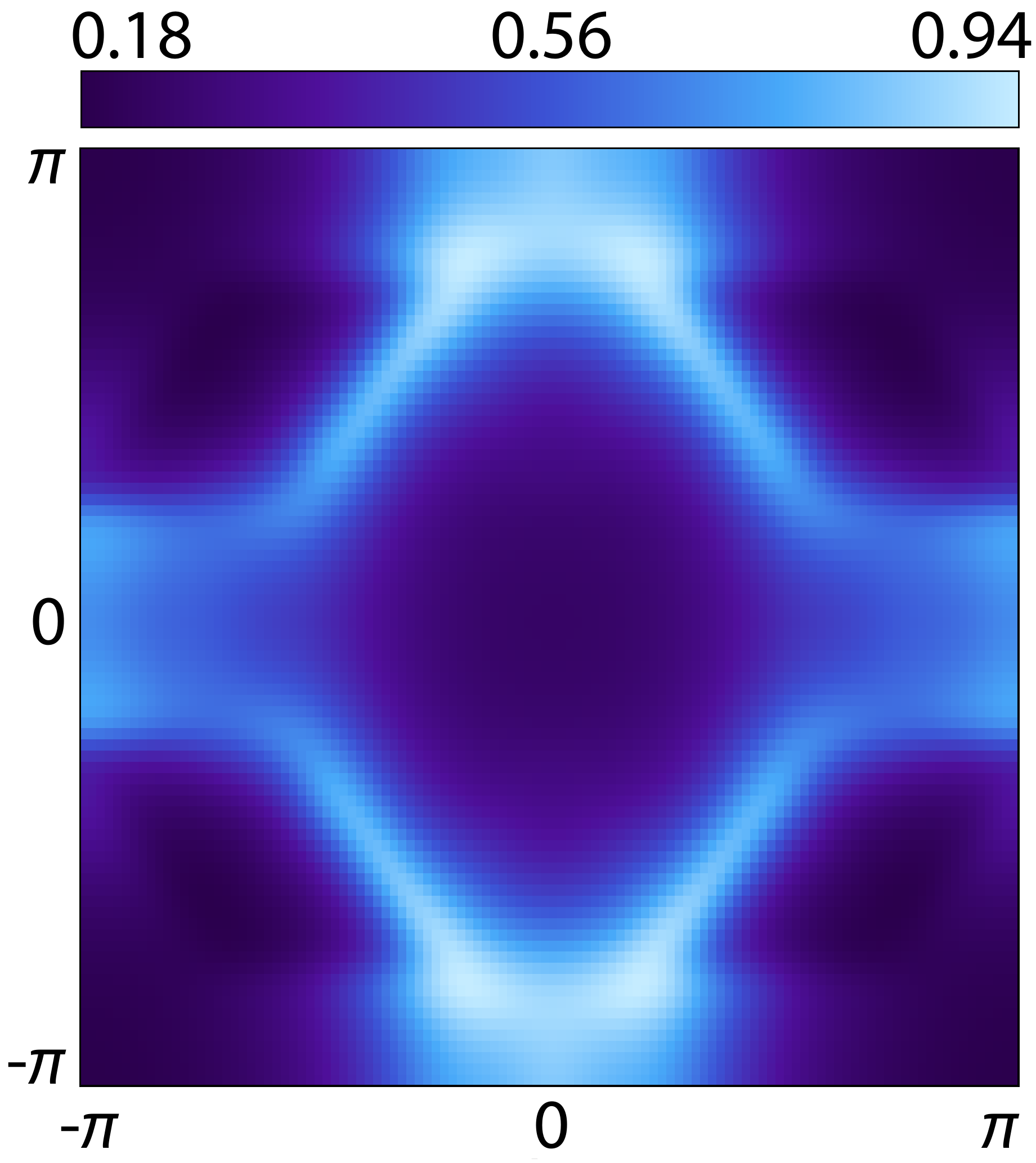}
\put(-8,92){\bf(c)}
\end{overpic}\hspace{5mm}
\begin{overpic}
[width=0.44\columnwidth]{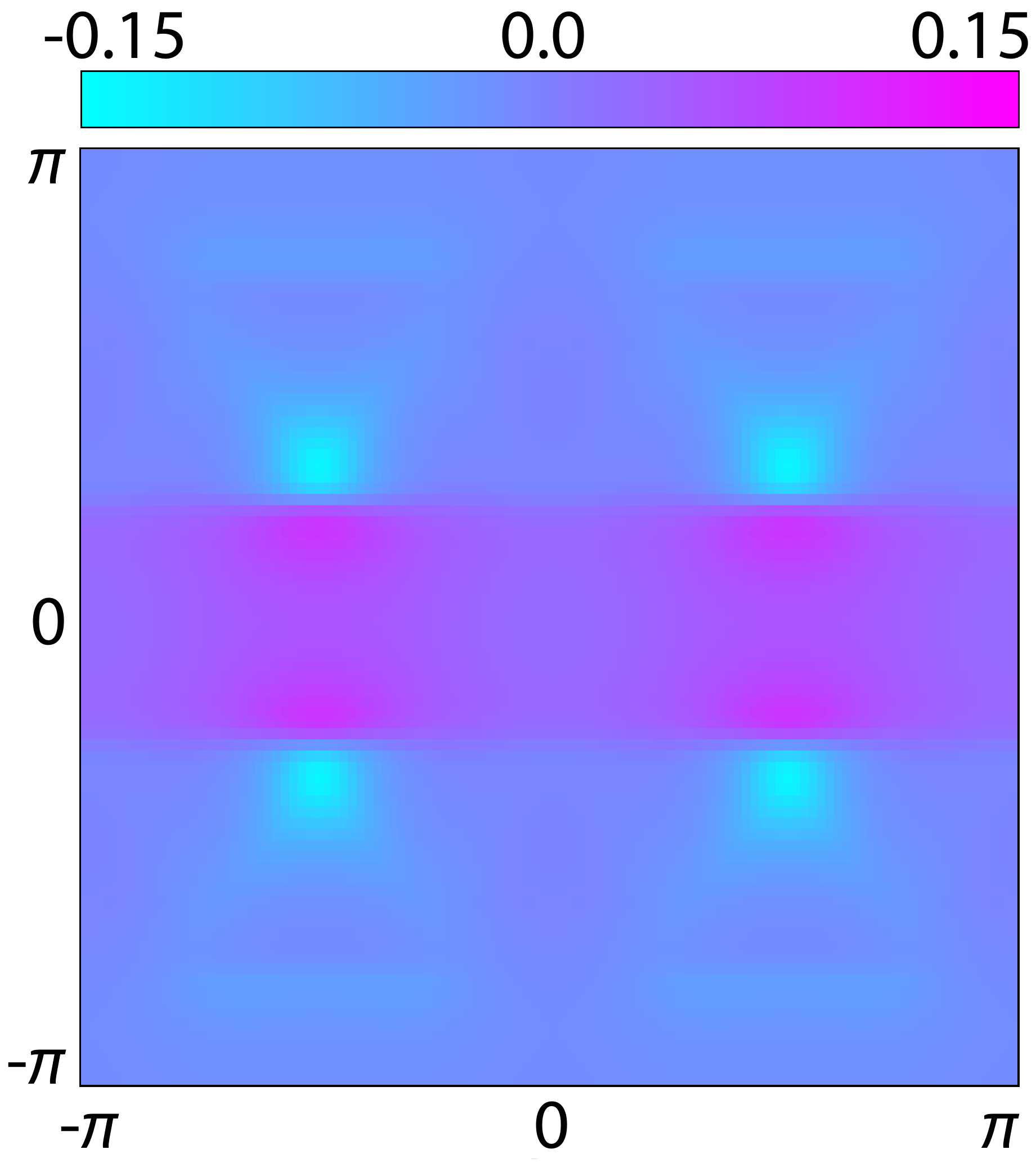}
\put(-8,92){\bf(d)}
\end{overpic}
\caption{The same quantities as in figure~\ref{fig3} but for the PDW solution for the same parameter set.
}
\label{fig4}
\end{figure}

Figure~\ref{fig1} shows the typical groundstate solution for hole doping $x=1/8$, $V=2\,t$, and $U=3.4\,t$. The local magnetization $m_i=n_{i\ua}-n_{i\da}$ (a) forms the well known site-centered spin-stripe structure with an anti-phase domain wall between the stripes and an overall periodicity of $8\,a$. Upon increasing $U$ through $U_{{\rm c}1}$, the maximum polarization $\max_i|m_i|$ abruptly increases to $\sim$$0.5$, and rises further to $\sim$$0.75$ towards $U_{{\rm c}2}$. Thus the system is quite far from the ideal spin-ladder structure with fully polarized AF stripes (in contrast to the $V$-model, c.f. section~\ref{sec:V}). Holes are expelled from the AF stripes, leading to a charge density modulation with a relative amplitude of $5$--$10\%$ [see figure~\ref{fig1}~(b) and~(e)].

A solution with bond-centered stripes in a two-legged spin-ladder structure also exists, but it is higher in energy for the parameters chosen above. If however $U$ is increased towards $U_{{\rm c}2}$, the groundstate changes to the bond-centered spin structure.

The groundstate for $U_{{\rm c}1}<U<U_{{\rm c}2}$ is a \textquotedblleft modulated $d$-wave\textquotedblright (mdSC) state. The $d$-wave projection of the SC order parameter
\begin{align}
\Delta^{\!d}_i=(\Delta^{\!s}_{i,i+\hat x}-\Delta^{\!s}_{i,i+\hat y}+\Delta^{\!s}_{i,i-\hat x}-\Delta^{\!s}_{i,i-\hat y})/4
\label{j22}
\end{align}
[see figure~\ref{fig1}~(c)] shows a similar stripe pattern as the charge density. Because of the $x$ - $y$ asymmetry of the striped system, a finite extended $s$-wave component 
\begin{align}
\Delta^{\!s}_i=(\Delta^{\!s}_{i,i+\hat x}+\Delta^{\!s}_{i,i+\hat y}+\Delta^{\!s}_{i,i-\hat x}+\Delta^{\!s}_{i,i-\hat y})/4
\label{j23}
\end{align}
is also induced. To fully characterize the SC state, $x$ and $y$ bond-order parameters are shown in figure~\ref{fig1}~(f) over a cross section of two stripe periods. For comparison, we also include a solution of the PDW type with a sign change in $\Delta^s_{ij}$ from one AF stripe to the next. The PDW solution has a reduced SC order parameter as compared to the mdSC state and a slightly higher free energy. Although the maximum AF order parameter $M_i$ is larger for the PDW solution, the gain in magnetic energy is not sufficient to compensate for the smaller SC condensation energy caused by the zeros in the SC order parameter in the center of the AF stripes [see figure~\ref{fig1}~(f)].

\begin{figure}[t!]
\centering
\vspace{2.5mm}
\begin{overpic}
[width=0.44\columnwidth]{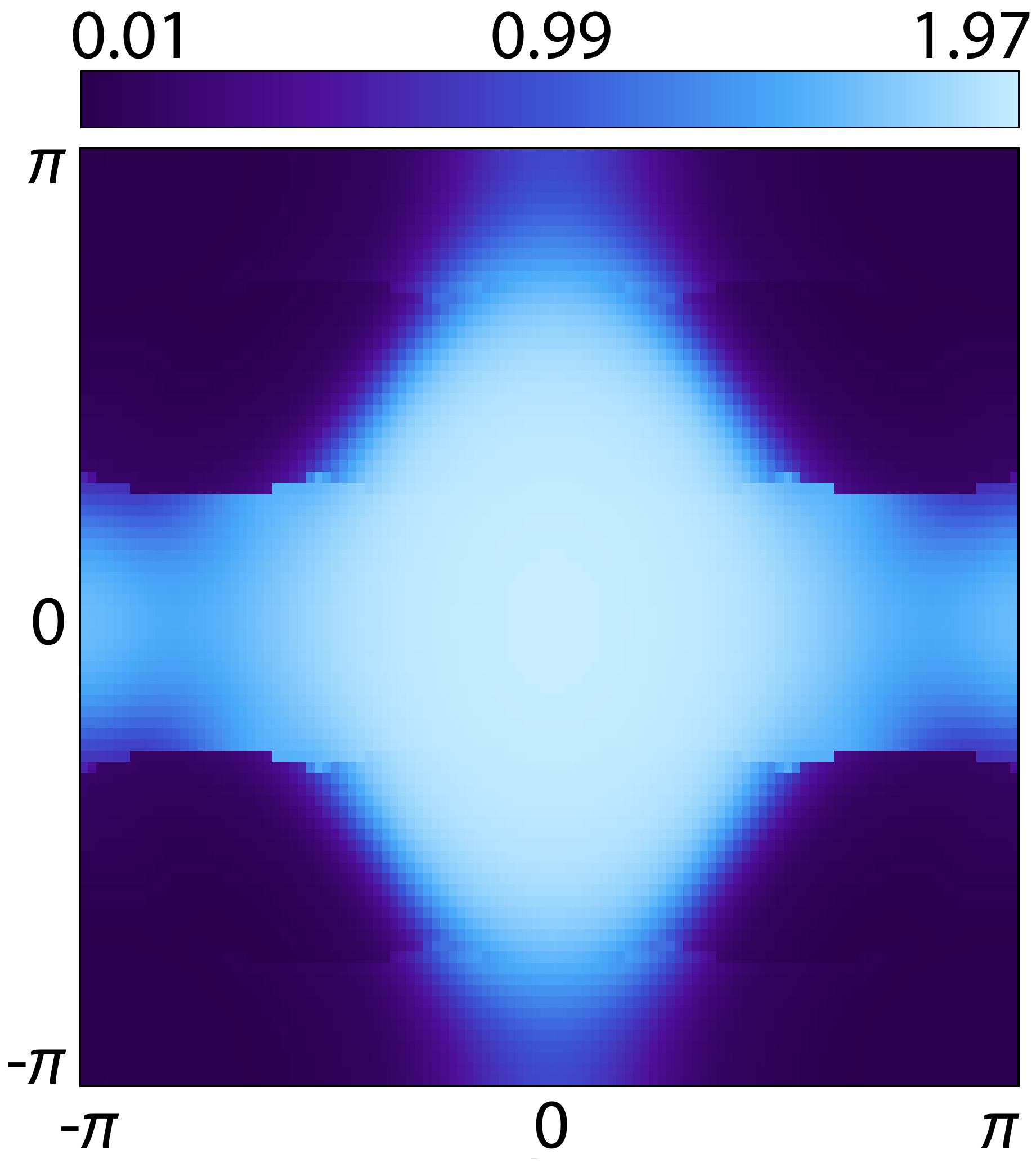}
\put(-8,92){\bf(a)}
\end{overpic}\hspace{5mm}
\begin{overpic}
[width=0.44\columnwidth]{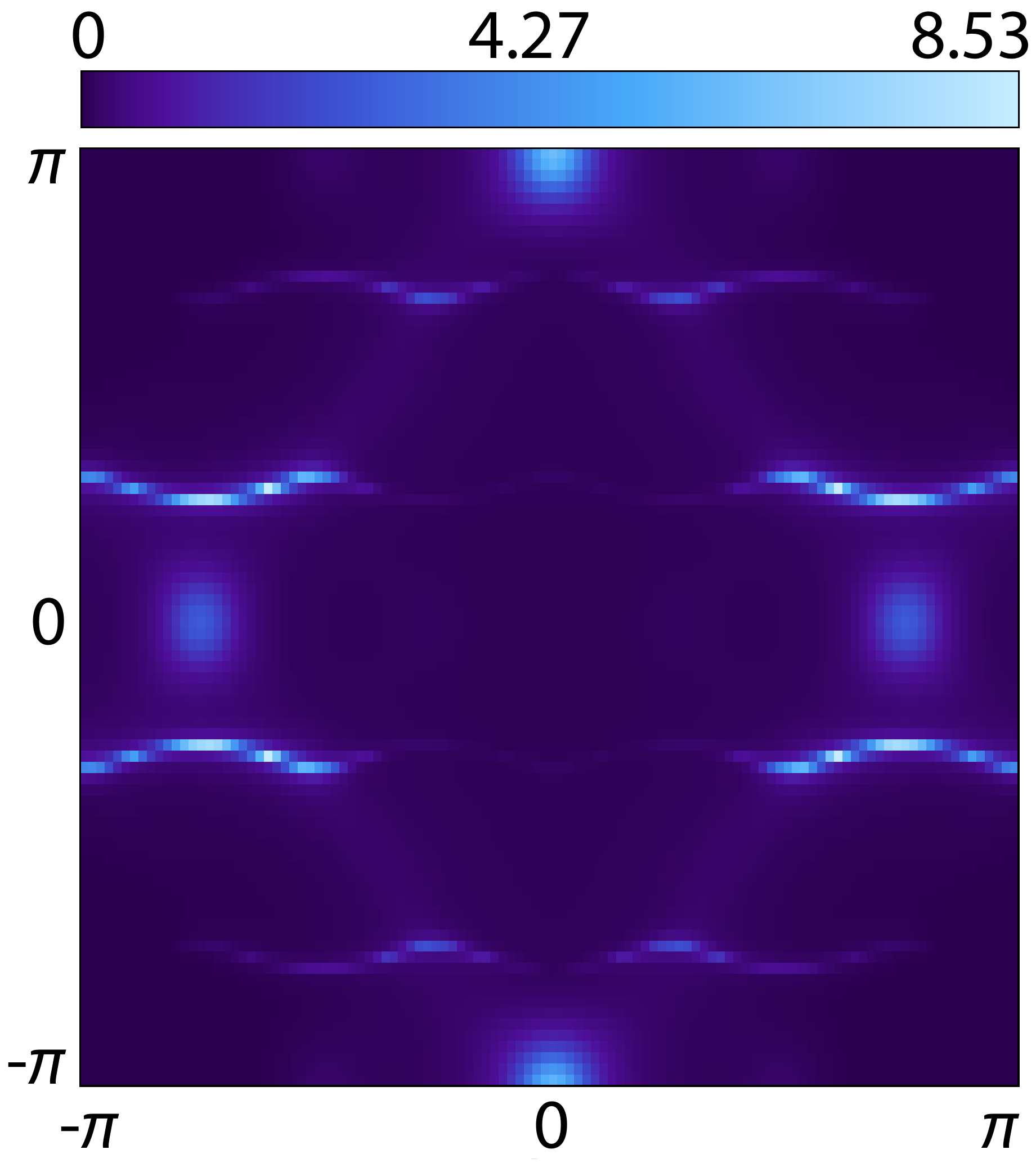}
\put(-8,92){\bf(b)}
\end{overpic}
\caption{Momentum-space characterization of the non-SC solution of the $U$ model for $x=1/8$, ${V=2\,t}$, and $U=3.4\,t$ on a 16$\times$12 lattice with 7$\times$7 supercells. (a) Occupation probability $n({\bf k})$, (b) spectral function $A({\bf k},\omega=0)$.
}
\label{fig4.0}
\end{figure}

If superconductivity is completely suppressed (by setting $\Delta^s_{ij}=0$ artificially), the modulations in $m_i$ and $n_i$ are barely larger than in the mdSC state and almost identical to the PDW solution [see figure~\ref{fig1}~(d) and~(e)]. This indicates that the spin polarization is limited mostly by the kinetic energy, which energetically favors a homogeneous charge and spin density, and less limited by the competition with superconductivity.
Indeed, antiferromagnetism and superconductivity rather avoid each other, as becomes evident from the momentum-space characterization discussed below.

A complementary description of the intertwined SC and AF orders, together with the resulting anisotropy of the system, is possible in momentum space. Note that although we measure all length in real space in units of the lattice constant $a$, we will set $a=1$ for the following discussions in momentum space.
Figure~\ref{fig3} characterizes the groundstate solution of the mdSC state. The occupation probability $n({\bf k})$ in figure~\ref{fig3}~(a) and the pair density $P({\bf k})$ in figure~\ref{fig3}~(b) are almost identical to the known distributions of a pure $d$-wave state without magnetism (c.f.~\cite{Loder:2010}), except for a small $x$-$y$ asymmetry. The pair density squared
\begin{multline}
P^2({\bf k})=\sum_\q\big[\langle c^\dag_{\k\ua}c_{\k\ua}c^\dag_{-\k+\q\da}c_{-\k+\q\da}\rangle\\
-\langle c^\dag_{\k\ua}c_{\k\ua}\rangle\langle c^\dag_{-\k+\q\da}c_{-\k+\q\da}\rangle\big]
\label{PD}
\end{multline}
measures the average correlated occupation of the electron states with momenta $\k$ and $-\k+\q$. Thus $P^2(\k)$ is largest at those momenta in the Brillouin zone where electron pairs predominately form. Within the mean-field decoupling scheme, equation~(\ref{PD}) reduces to $P^2(\k)=\sum_\q\langle c_{-\k+\q\da}c_{\k\ua}\rangle^2$. The pair density is concentrated around the Fermi surface of the normal state and vanishes at four nodal points near the zone diagonal.

Around these nodal points, AF correlations emerge. They are quantified by the spin order in momentum space, that is by $\rho_S({\bf k})=\sum_{\bf q}\sum_s\langle sc^\dag_{{\bf k}+{\bf q}s}c_{{\bf k}s}\rangle$ [see figure~\ref{fig3}~(c)]. $\rho_S$ indeed has its maxima at the nodal points, reflecting the competition between, or rather avoidance of superconductivity and antiferromagnetism, i.e. antiferromagnetism is strong where pair formation is weak. $\rho_C({\bf k})=\sum_{\bf q\neq\bm0}\sum_s\langle c^\dag_{{\bf k}+{\bf q}s}c_{{\bf k}s}\rangle$ [see figure~\ref{fig3}~(d)] is the analog for the charge order, and $\rho_C(\k)$ is non zero only in the presence of charge modulations.

It is instructive to compare figure~\ref{fig3} to the corresponding results for the PDW solution in figure~\ref{fig4} or to a state where superconductivity is suppressed artificially (figure~\ref{fig4.0}). If superconductivity is absent, the spin polarization is slightly stronger in the AF stripes which leads to the characteristic occupation probability $n(\k)$, shown in figure~\ref{fig4.0}~(a): most of the Fermi surface is gapped by the AF order with a continuous $n(\k)$ across, but discontinuities in $n(\k)$ remain on the border of a horizontal bar between $k_y=\pi/4$ and $k_y=-\pi/4$. These discontinuities constitute disconnected Fermi surface arcs which are clearly visible in the spectral function $A(\k,\omega=0)=-(1/\pi)\,\text{Im}\,G(\k,\k,\omega=0)$ shown in figure~\ref{fig4.0}~(b).

In the superconducting PDW solution (see figure~\ref{fig4}), antiferromagnetism is almost as strong as in the absence of superconductivity [c.f. figure~\ref{fig1}~(d)], thus $n(\k)$, shown in figure~\ref{fig4}~(a), has the same structure as in figure~\ref{fig4.0}~(a), but the entire Fermi surface is now gapped due to superconductivity.
The pair density $P({\bf k})$ of the PDW state shown in figure~\ref{fig4}~(b) is strongly anisotropic because of the line zeros of the order parameter enforced by the $\pi$ phase shift between neighboring stripes. $P({\bf k})$ is similar to the pure PDW without antiferromagntism described in~\cite{Loder:2010}, the latter solution is however not gapless. The pairing free pieces of the reconstructed Fermi surface of the pure PDW~\cite{Baruch:2008,Berg:2009,Loder:2010} are now gapped due to antiferromagnetism and extend towards the $(0,\pm\pi)$ points where superconductivity is weak for vertically oriented stripes [see figure~\ref{fig4}~(c)].
Although the pairing contributions around $(0,\pm\pi)$ are weak compared to the vicinity of $(\pm\pi,0)$, they are essential for a globally phase coherent SC state. 

\begin{figure}[t!]
\centering
\vspace{2.5mm}
\begin{overpic}
[width=0.8\columnwidth]{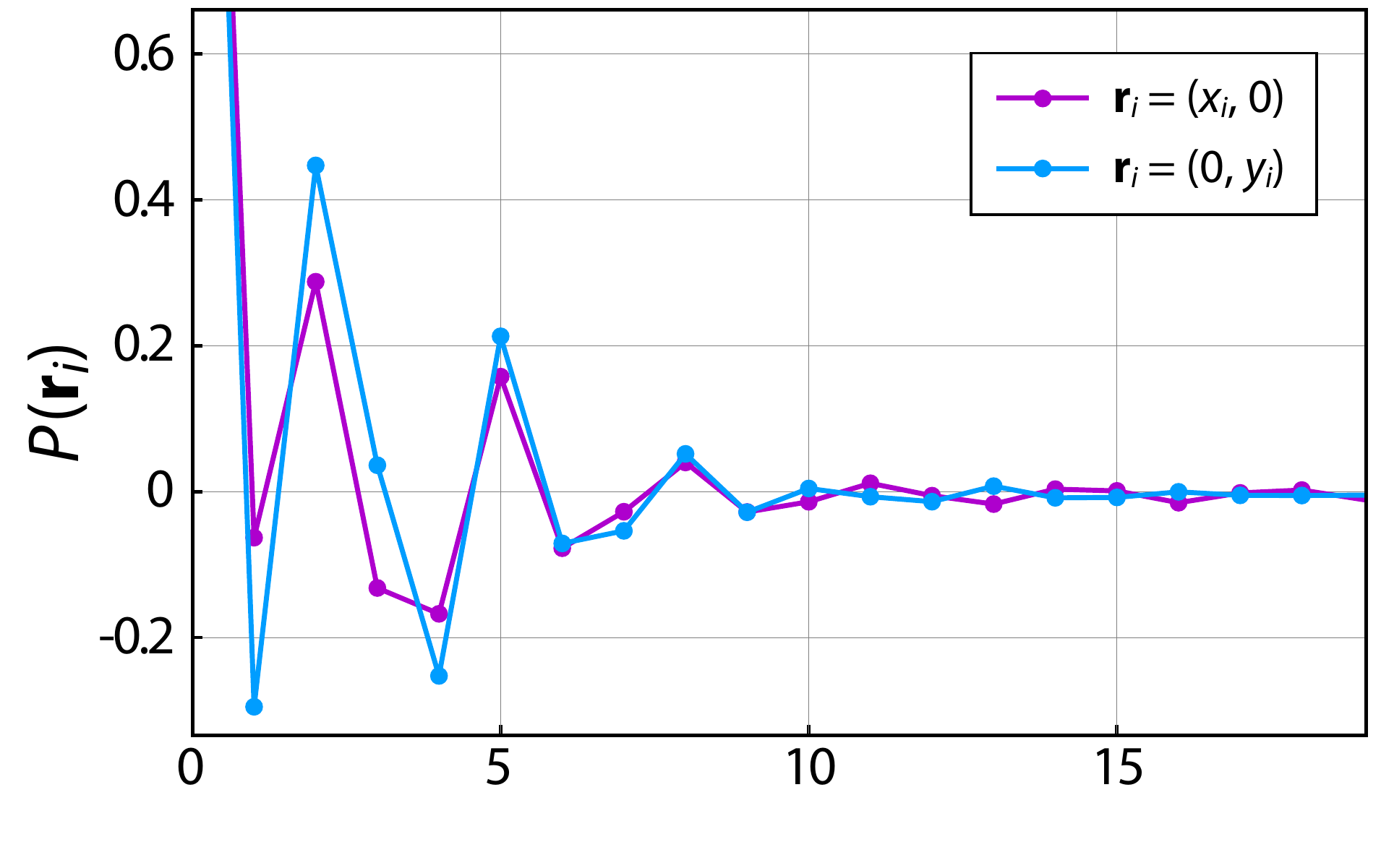}
\put(-0,59){\bf(a)}
\end{overpic}
\\\vspace{3mm}
\begin{overpic}
[width=0.8\columnwidth]{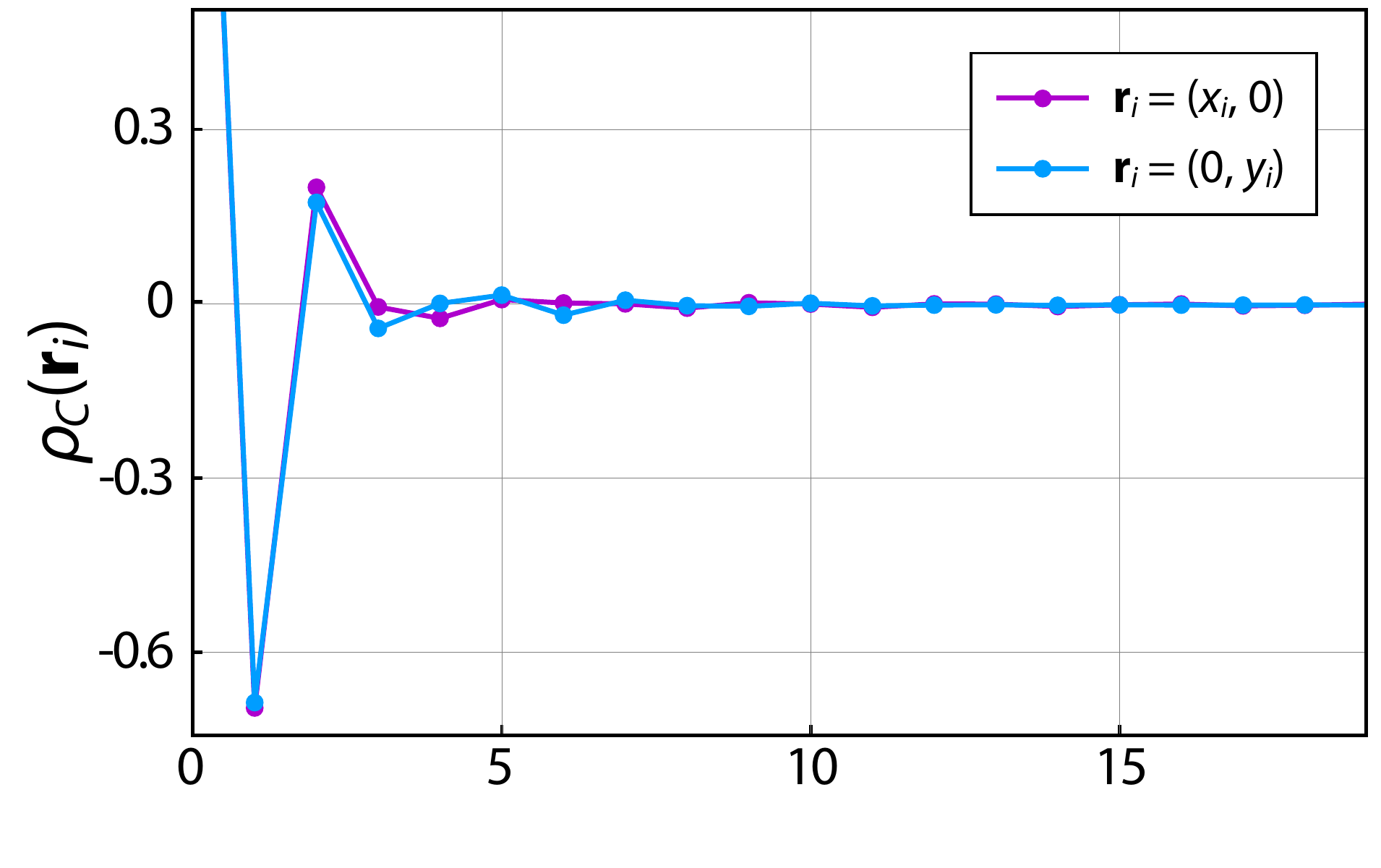}
\put(-0,59){\bf(b)}
\end{overpic}
\caption{Pairing and effective hopping amplitudes in real space for the $U$ model and the same parameters as in figure~\ref{fig4}. (a) Pairing amplitude $P({\bf r}_i)=\langle c_{0\da}c_{i\ua}\rangle$. The value of the on-site term is $P(0)=1.88$. (b) Effective hopping amplitude $\rho_C({\bf r}_i)=\sum_s\langle c^\dag_{0s}c_{is}\rangle$. The value of the on-site term is $\rho_C(0)=1.73$.
}
\label{fig4.5}
\end{figure}

It is instructive to analyze also charge and pair correlations in real space. The effective hopping amplitude $\rho_C({\bf r}_i)=\sum_s\langle c_{0s}c^\dag_{is}\rangle$ for hopping from site $0$ to site $i$ is obtained by Fourier transforming $n(\k)+\rho_C(\k)$.
The pairing amplitude in real space is $P({\bf r}_i)=\langle c_{0\da}c_{i\ua}\rangle$. Figure~\ref{fig4.5} shows $P({\bf r}_i)$ and $\rho_C({\bf r}_i)$ for the $U$-model. Both quantities exhibit an oscillating behavior in which $P({\bf r}_i)$ decays over the distance of the SC coherence length $\xi_0$, which is roughly 10 lattice constants for the chosen parameters. The oscillations in $P({\bf r}_i)$, as well as the Friedel type of oscillations in $\rho_C({\bf r}_i)$, extend over almost the same length in $x$ and $y$ direction, which verifies the 2D character of the solution of the $U$-model.

\begin{figure}[t!]
\centering
\vspace{2.5mm}
\begin{overpic}
[width=0.8\columnwidth]{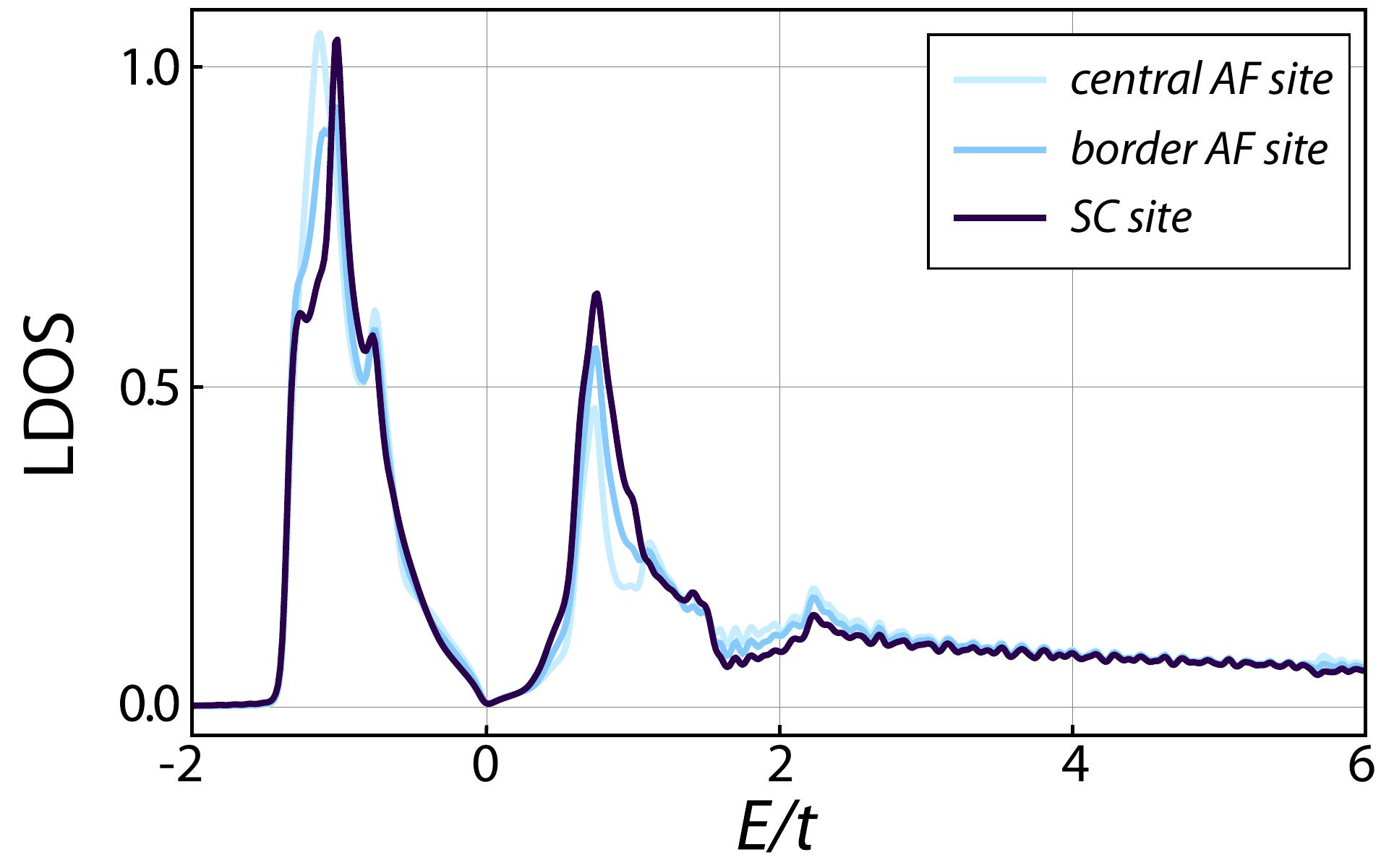}
\put(-0,59){\bf(a)}
\end{overpic}
\\\vspace{3mm}
\begin{overpic}
[width=0.8\columnwidth]{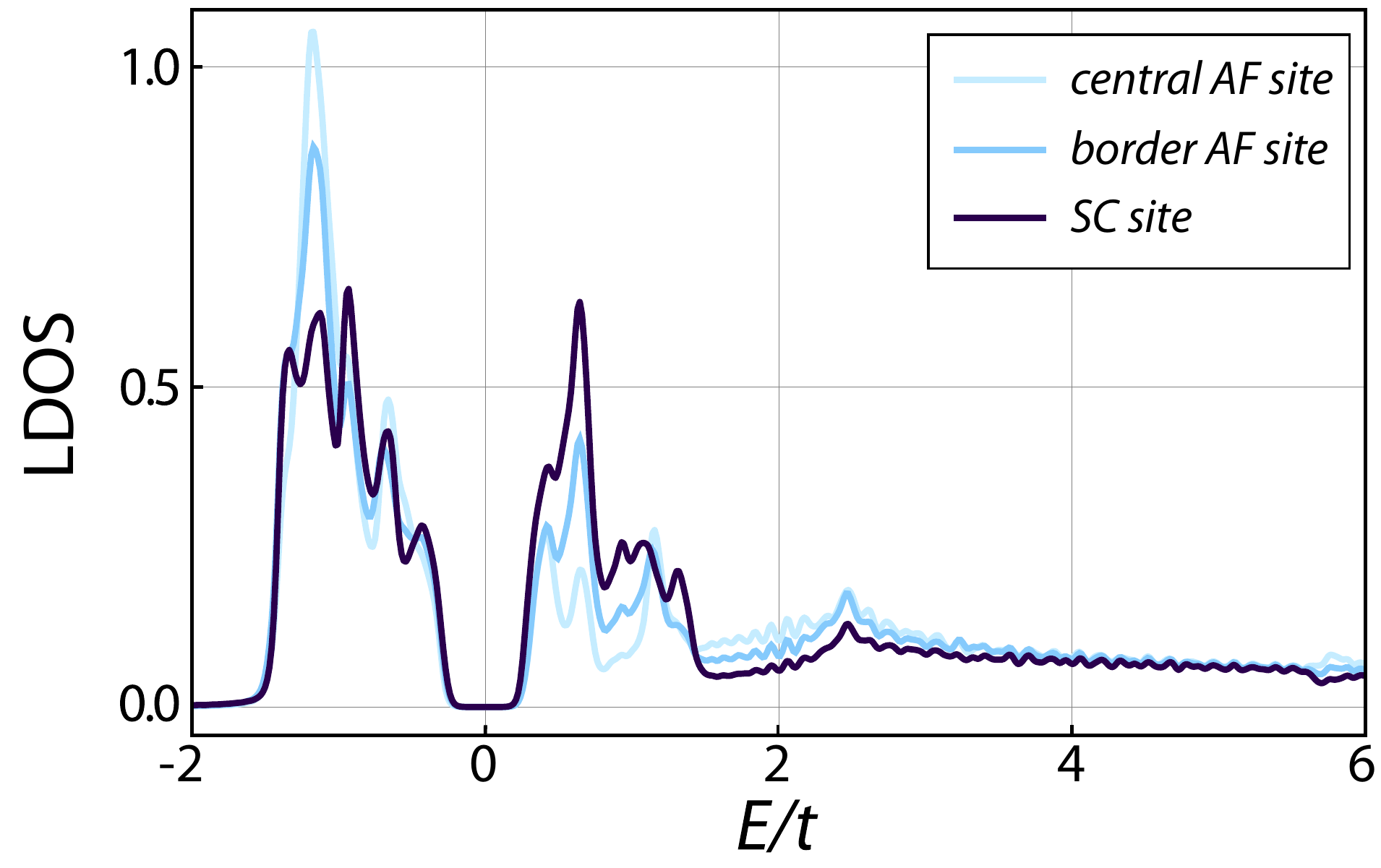}
\put(-0,59){\bf(b)}
\end{overpic}
\caption{LDOS for the $U$ model and the same parameters as in figure~\ref{fig4}. (a) mdSC solution, (b) PDW solution.
}
\label{fig5}
\end{figure}

The dominance of the SC order in the $U$-model is most apparent in the local density of states (LDOS) on magnetic and non-magnetic lattice sites as shown in figure~\ref{fig5} for the same parameters as above. For the \textquotedblleft modulated $d$-wave\textquotedblright\ solution [see figure~\ref{fig5}~(a)] the LDOS vanishes linearly upon approaching the Fermi energy, which is the characteristic of $d$-wave superconductivity, but it has a slightly asymmetric shape, caused by the particle-hole asymmetry induced by the presence of a charge density modulation. The LDOS of the PDW solution~[figure~\ref{fig5}~(b)] has a smaller but $s$-wave like gap, since it has a strong extended $s$-wave contribution. In both solutions the gap scales with the SC order parameter, and its site independence is characteristic for the $U$-model. If $U$ is further increased to values above $U_{{\rm c}2}$, the magnetism induced gap starts to dominate over the SC gap and superconductivity breaks down. The $U$-model therefore does only allow for SC solutions with superconductivity dominating over AF order. This behavior is typical for the $U$-model and quite different to the $V$-model, as we show below.

\subsection{$\bm V$-Model}\label{sec:V}

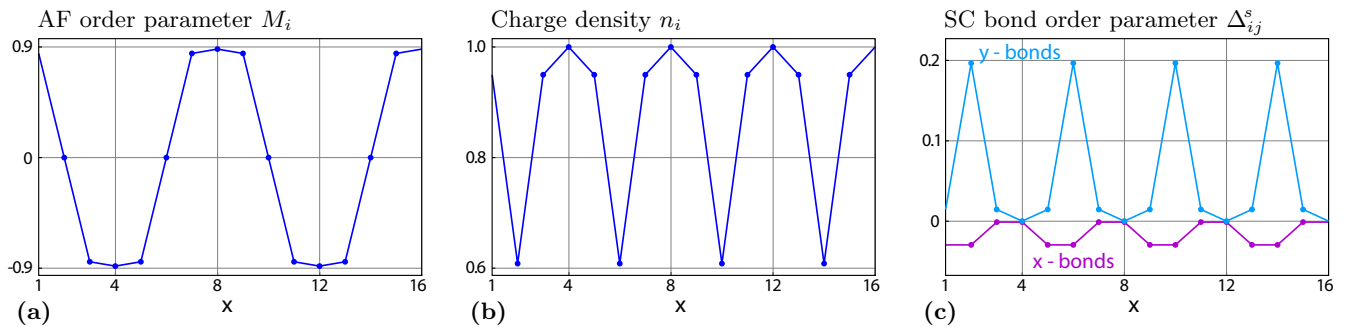
\begin{figure*}[t!]
\centering
\vspace{5mm}
\begin{overpic}
[width=0.65\columnwidth]{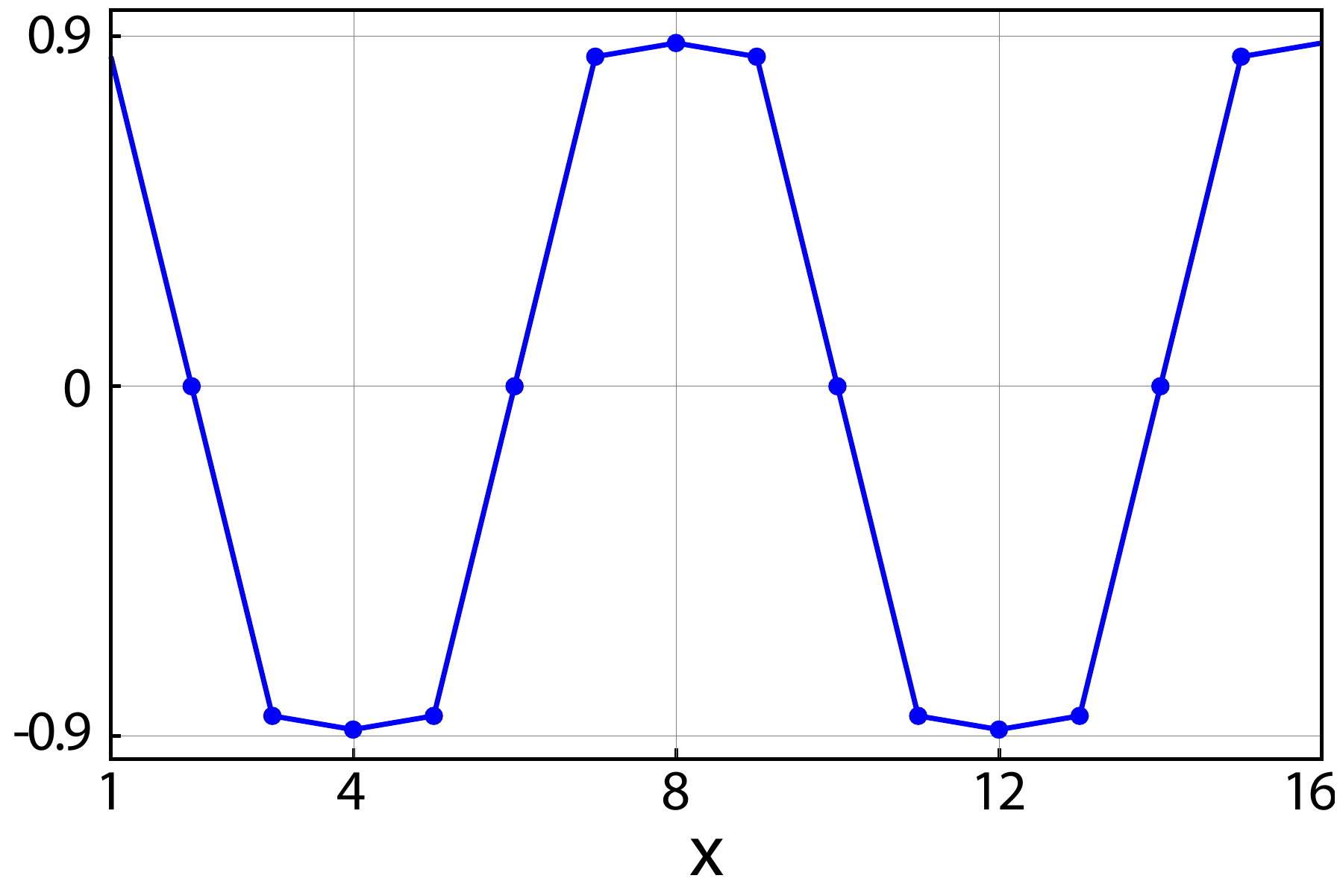}
\put(8,69){AF order parameter $M_i$}
\put(3,0){\bf(a)}
\end{overpic}\hspace{3mm}
\begin{overpic}
[width=0.65\columnwidth]{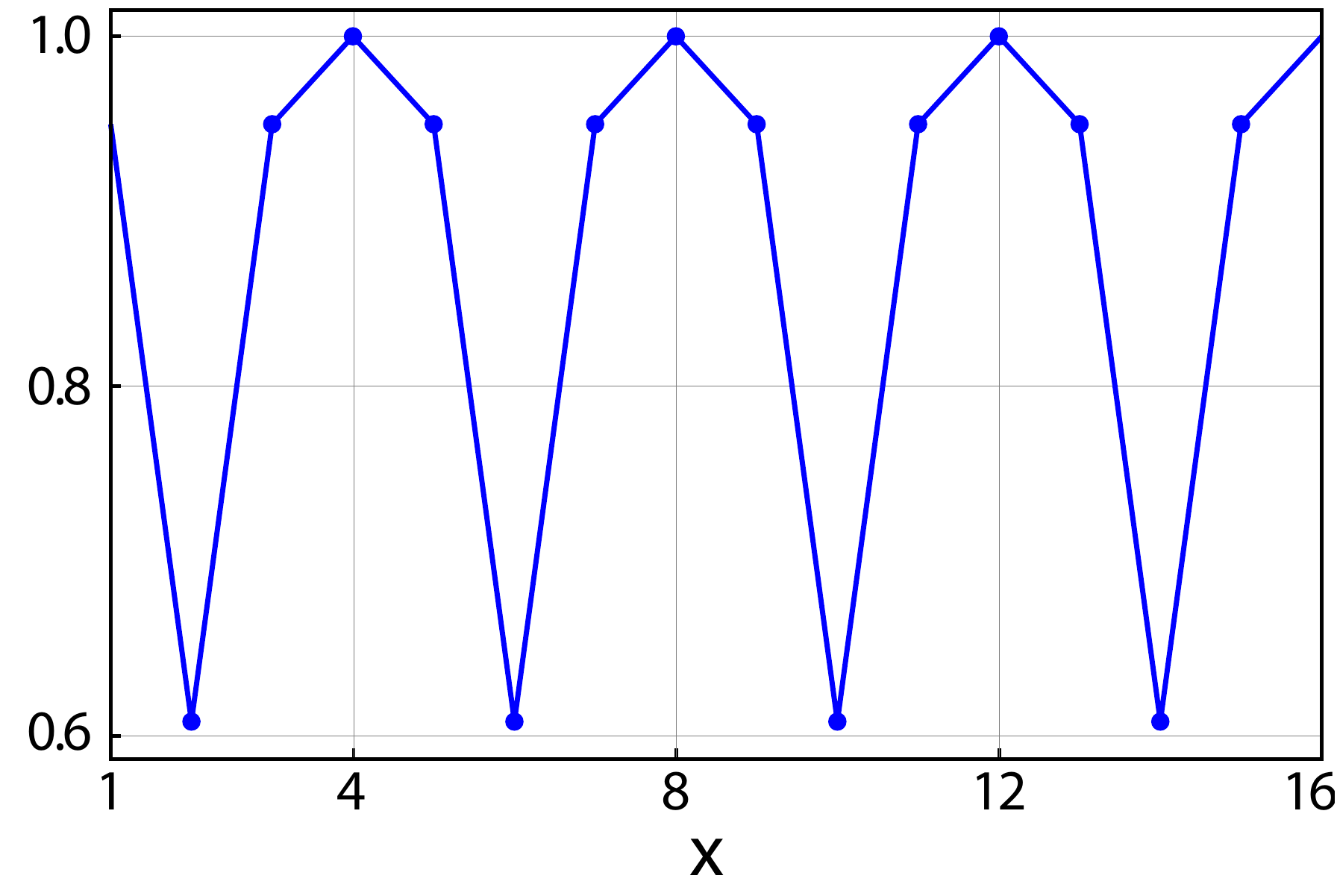}
\put(8,69){Charge density $n_i$}
\put(3,0){\bf(b)}
\end{overpic}\hspace{3mm}
\begin{overpic}
[width=0.65\columnwidth]{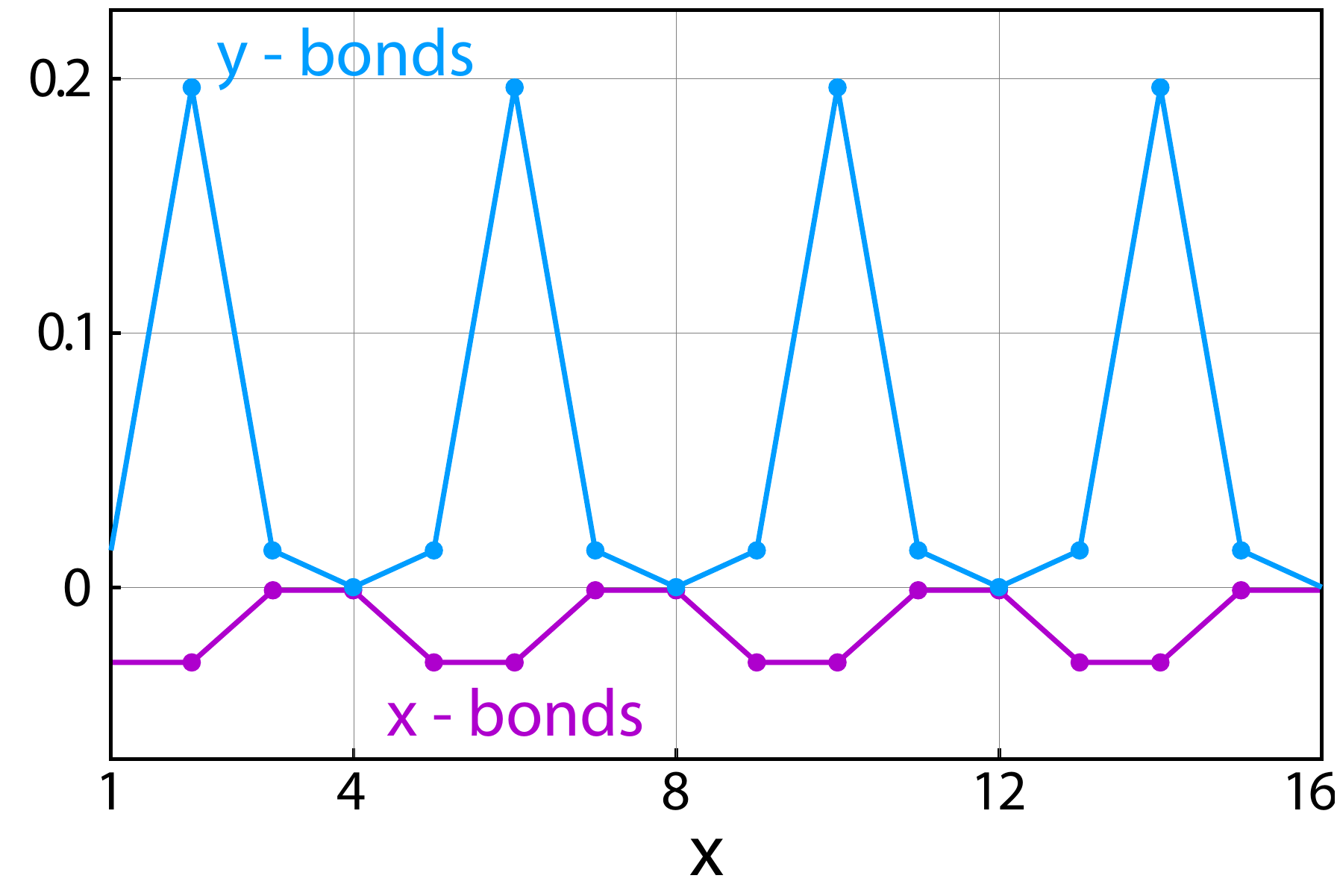}
\put(8,69){SC bond order parameter $\Delta^{\!s}_{ij}$}
\put(3,0){\bf(c)}
\end{overpic}
\caption{Real space characterization of the \textquotedblleft modulated $d$-wave\textquotedblright\ solution of the $V$-model for $x=1/8$ and ${V=2\,t}$ on a 16$\times$12 lattice. (a) AF order parameter $M_i=(-1)^im_i$, (b) charge density $n_i$, and (c) SC bond order parameters $\Delta^{\!s}_{i,i+\hat x}$ and $\Delta^{\!s}_{i,i+\hat y}$.}
\label{fig6}
\end{figure*}

A detailed description of the $V$-model ${\cal H}\mf_V$ at $x=1/8$ was given in~\cite{loder11}. Here we extend this analysis to clarify its relation to the $U$-model and the characteristic differences. In the $V$-model, the interaction strength $V$ controls both the SC and the AF correlations. Upon turning on and increasing $V$, we find a similar sequence of phases as in the $U$-model with increasing $U$ at $x=1/8$: For small $V$, the groundstate is a homogeneous, non-magnetic $d$-wave superconductor. At a critical value $V_{{\rm c}1}\approx0.8\,t$, AF stripe order with wavelength $8\,a$ sets in, similar to the solution of the $U$ model; this solution is characterized in real space in figure~\ref{fig6} for $V=2\,t$ at $x=1/8$. Above a second critical value $V_{{\rm c}2}\approx2.6\,t$, superconductivity disappears and the system becomes susceptible to phase separation.

Figure~\ref{fig6} shows that the AF stripes are almost half filled with a staggered spin polarization that is close to maximal, whereas superconductivity vanishes in the center of the AF stripes for both the $x$ and $y$ bonds. The system can therefore be described as locally phase separated into half-filled AF stripes and 1D metallic lines acting as anti phase domain walls. The crossover from the homogeneous $d$-wave superconductor to the striped solution with amplitudes as in figure~\ref{fig6} is rather sharp.

\begin{figure}[t!]
\centering
\vspace{2.5mm}
\begin{overpic}
[width=0.44\columnwidth]{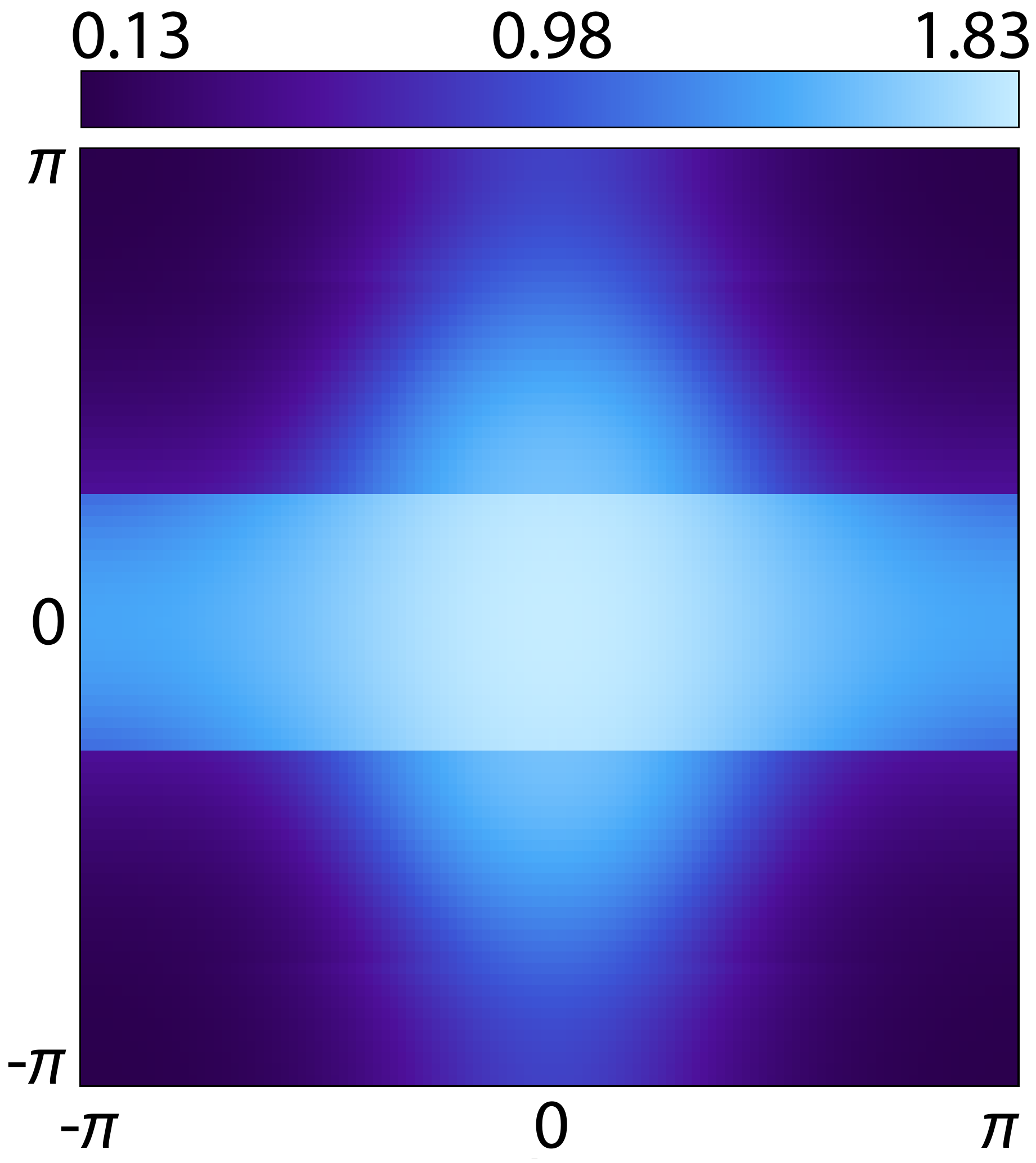}
\put(-8,92){\bf(a)}
\end{overpic}\hspace{5mm}
\begin{overpic}
[width=0.44\columnwidth]{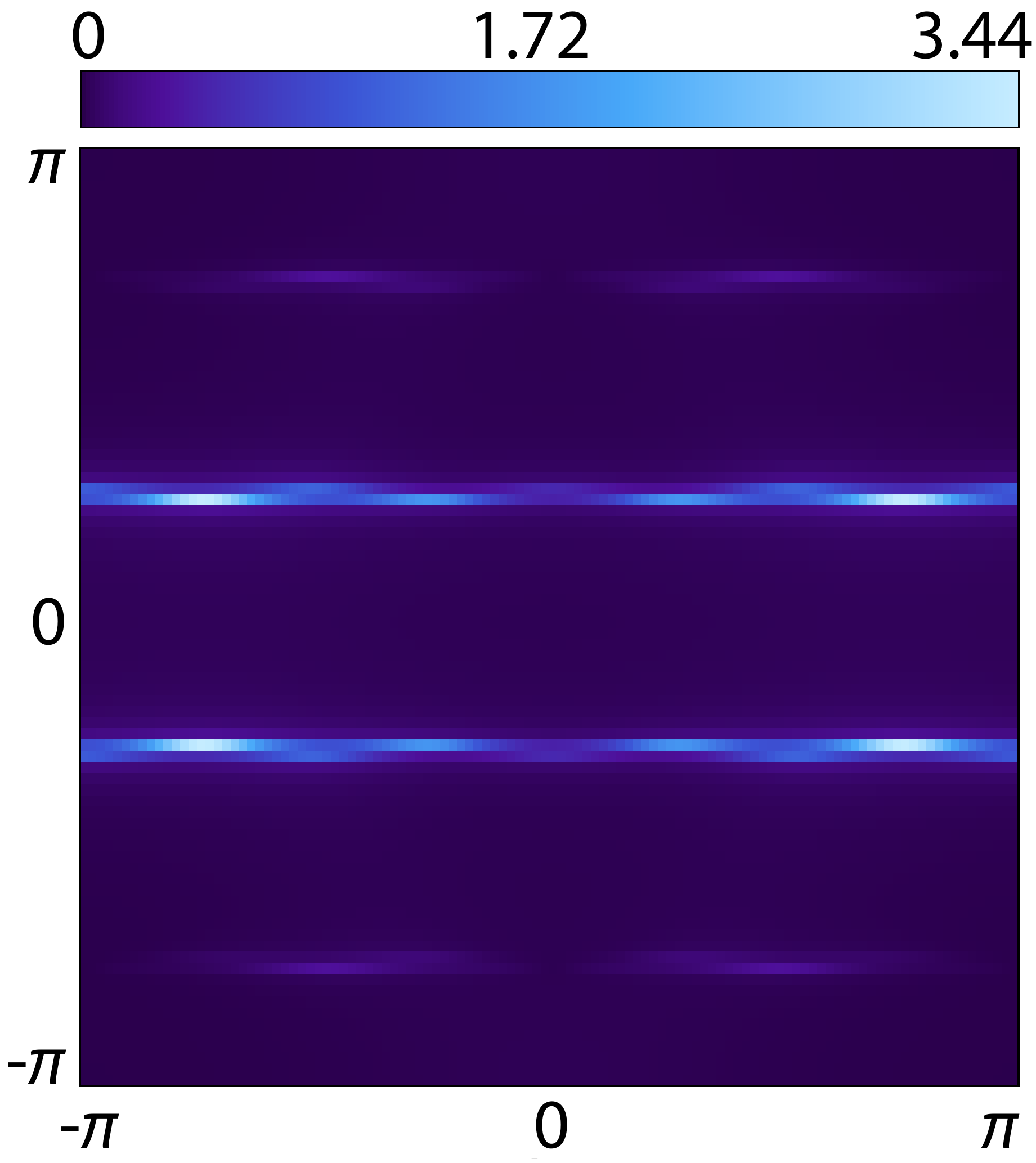}
\put(-8,92){\bf(b)}
\end{overpic}
\caption{Momentum-space characterization of the non-SC solution of the $V$ model for $x=1/8$ and ${V=2\,t}$ on a 16$\times$12 lattice with 7$\times$7 supercells. (a) Momentum occupation probability $n({\bf k})$, (b) spectral function $A({\bf k},\omega=0)$.
}
\label{fig7}
\end{figure}

The 1D character of the $V$-model solution is most apparent in momentum space, if superconductivity is again artificially suppressed. The momentum distribution $n({\bf k})$ [see figure~\ref{fig7}~(a)] consists of a horizontal bar with occupied states for $-\pi/4\leq k_y\leq\pi/4$, and a diffuse cloud of occupied states in the center of the Brillouin zone far below $E_{\rm F}$. The states with momenta inside the bar arise from the quasi 1D metallic lines between the AF stripes, whereas the cloud near the zone center originates from the AF stripes themselves. Both structures are tied to separate energy windows in the LDOS [c.f. figure~\ref{fig8}~(a)]. The borders of the bar form two 1D Fermi surfaces at $k_y=\pm\pi/4$, visible in the spectral function $A({\bf k},\omega=0)$ in figure~\ref{fig7}~(b). Pairing can occur exclusively around these two Fermi surfaces and is therefore restricted to the metallic hole-rich lines.

\begin{figure}[t!]
\centering
\vspace{2.5mm}
\begin{overpic}
[width=0.8\columnwidth]{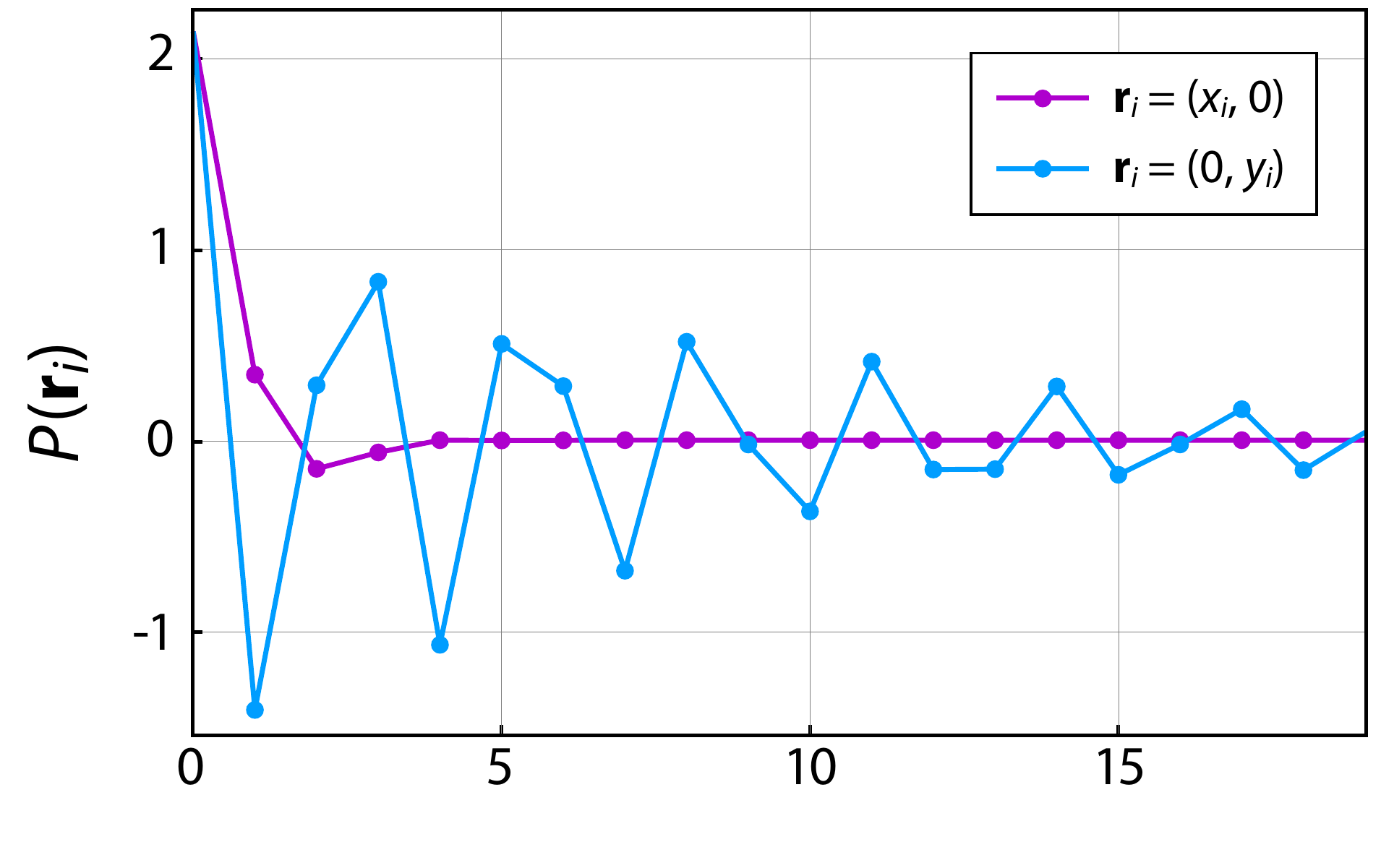}
\put(-0,59){\bf(a)}
\end{overpic}
\\\vspace{3mm}
\begin{overpic}
[width=0.8\columnwidth]{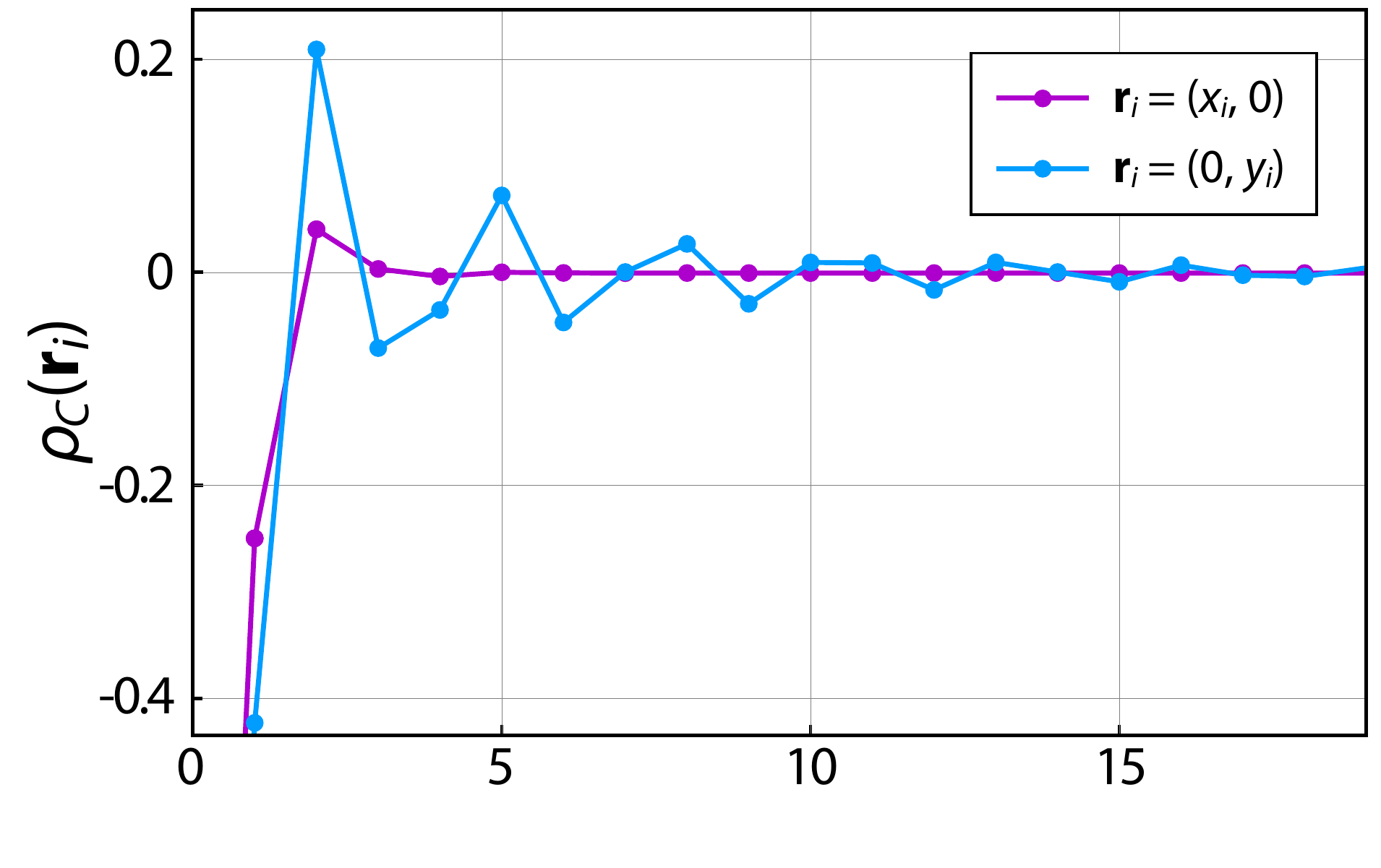}
\put(-0,59){\bf(b)}
\end{overpic}
\caption{Pairing and effective hopping amplitudes for the $V$ model and the same parameters as in figure~\ref{fig7}. (a) Pairing amplitude $P({\bf r}_i)=\langle c_{0\da}c_{i\ua}\rangle$, (b) effective hopping amplitude $\rho_C({\bf r}_i)=\sum_s\langle c^\dag_{0s}c_{is}\rangle$. The value of the on-site term is $\rho_C(0)=1.58$.
}
\label{fig7.5}
\end{figure}

\begin{figure}[t!]
\centering
\vspace{2.5mm}
\begin{overpic}
[width=0.8\columnwidth]{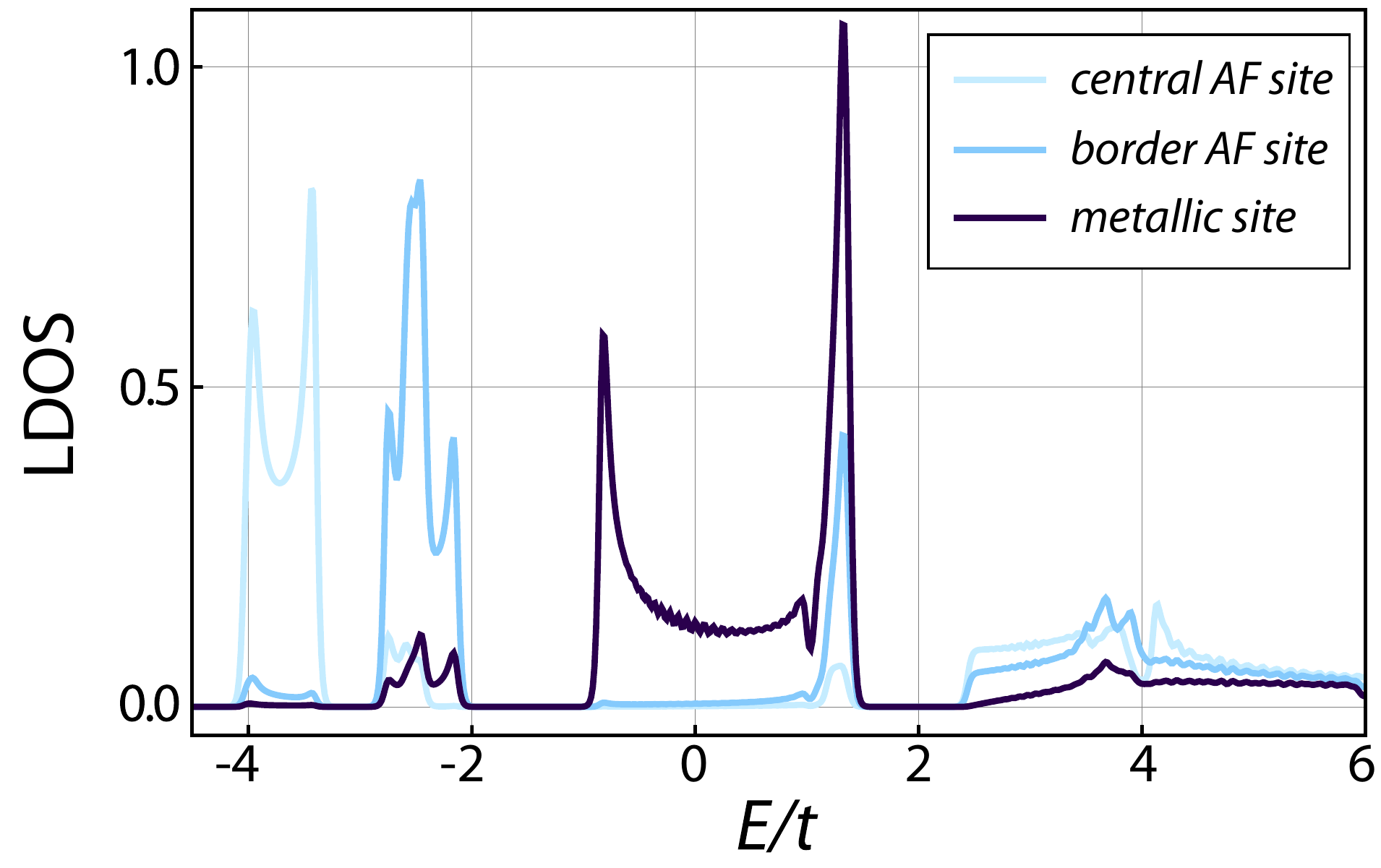}
\put(-0,59){\bf(a)}
\end{overpic}
\\\vspace{3mm}
\begin{overpic}
[width=0.8\columnwidth]{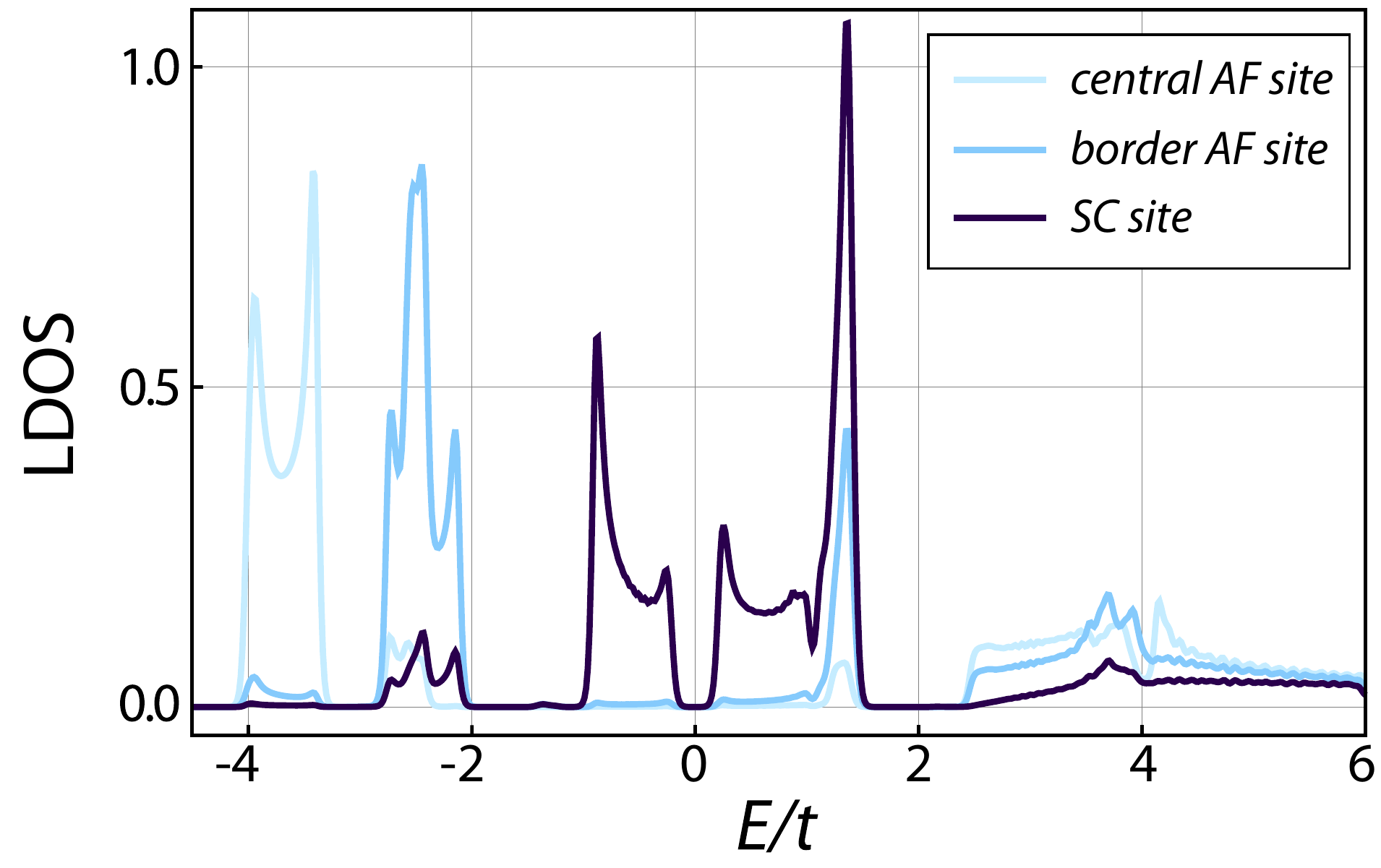}
\put(-0,59){\bf(b)}
\end{overpic}
\caption{LDOS for the $V$ model and the same parameters as in figure~\ref{fig7}. (a) Non-SC state and (b) SC state.
}
\label{fig8}
\end{figure}

As for the $U$-model, we also calculated the pairing and effective hopping amplitudes $P({\bf r}_i)$ and $\rho_C({\bf r}_i)$ in real space, shown in figure~\ref{fig7.5}. The pairing amplitude $P({\bf r}_i)$ exhibits a SC correlation length which is highly anisotropic with $\xi_0^x\approx2\,a$. The Cooper pair motion is therefore restricted to 1D channels along the stripes and consequently
the phases of the SC order parameters on adjacent non-magnetic lines become decoupled. In other words, the 2D phase coherent state is degenerate with uncorrelated 1D superconducting stripes. A globally phase coherent state may still be favorable by a Josephson coupling of the 1D SC stripes, however on the Hartree-Fock level Josephson coupling is not included. Since charge transport in the $x$-direction transverse to the stripes is absent, also the charge correlations do not extend from one stripe to the next, as is visible from $\rho_C({\bf r}_i)$ in figure~\ref{fig7.5}~(b). The dimensionality of the SC state therefore constitutes a qualitative disparity between the $U$- and $V$-model.

The formation of energetically separated states on metallic (or SC) and AF sites is the essential distinction between the $V$ model and the $U$ model. The LDOS of the non-SC system in figure~\ref{fig8}~(a) shows a broad metallic band on the metallic sites around $E_{\rm F}$, whereas on the AF sites all states are far from $E_{\rm F}$. A SC gap therefore appears only on the non-magnetic sites [see figure~\ref{fig8}~(b)]. If $V$ is lowered  below $V_{{\rm c}1}$, the magnetism induced gap, which reaches from $\sim-2\,t$ to $t$ in figure~\ref{fig8}, becomes smaller than the SC gap, as in the $U$-model. Here however magnetism vanishes below $V_{{\rm c}1}$. The $V$-model behaves therefore contrarily to the $U$-model, where superconductivity is the dominant order until it vanishes when the AF gap becomes too large.

\section{Temperature Dependence}\label{sec:T}

In section~\ref{sec:bdg} we mentioned already the existence of different self-consistent solutions of the BdG equations in both the $U$- and the $V$-model. They correspond to different local minima in the parameter space of the free energy $F$ above the global minimum representing the states described above. Some of these additional solutions can only be obtained in special parameter ranges, or within a specific temperature regime. In this section we present the temperature evolution of a selection of these states. The list of states presented here is incomplete, but contains the most stable and regular patterns that are obtainable in both the $U$- and the $V$-model

Whereas in the $V$-model the contribution to the free energy $F$ from magnetism is by far larger than the condensation energy of superconductivity, in the $U$-model both contributions are of similar size in the range of $U$-values where superconductivity and AF coexist. Therefore local minima in $F$ with distinct magnetic structures are much closer in energy and thus a large variety of solutions is obtained. On the other hand, in the $V$-model the dominant AF order allows only for a few self-consistent solutions.

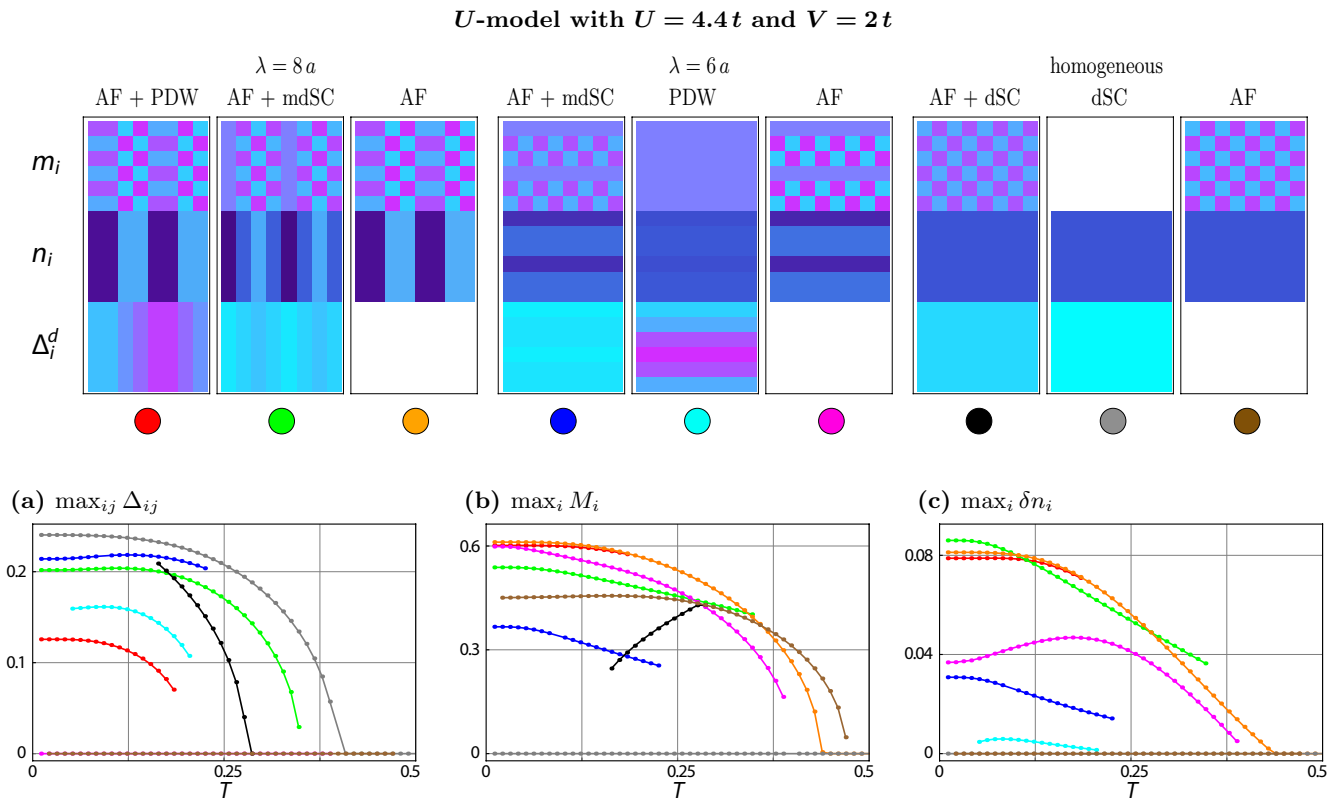
\begin{figure*}[t!]
\centering
\vspace{18mm}
\begin{overpic}
[width=2\columnwidth]{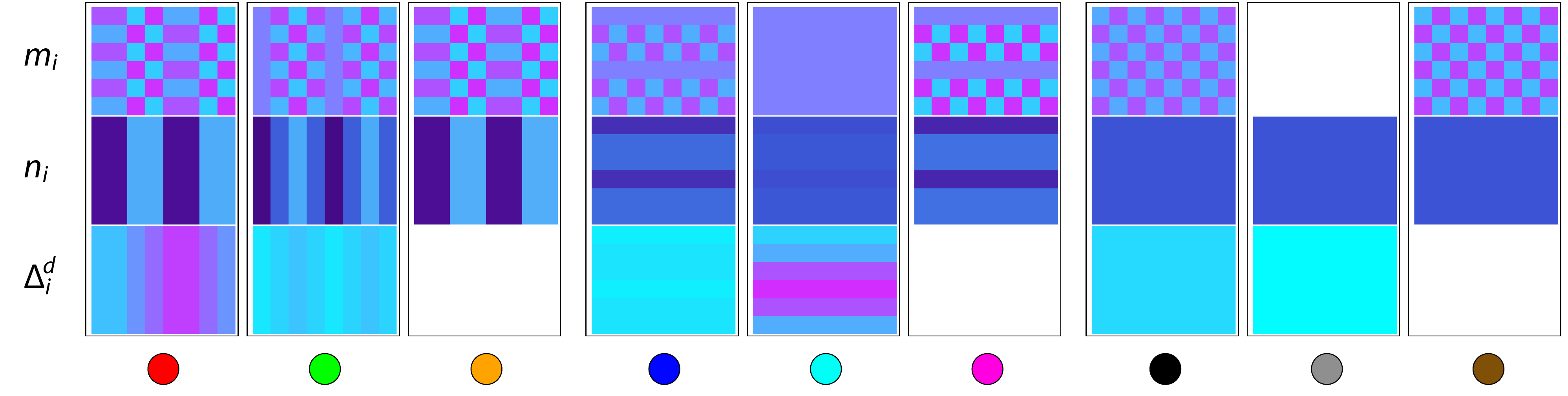}
\put(34,32){\bf $\bm U$-model with $\bm{U=4.4\,t}$ and $\bm{V=2\,t}$}
\put(18.7,28.5){\scalebox{0.8}[1]{$\lambda=8\,a$}}
\put(6.4,26){\scalebox{0.8}[1]{AF + PDW}}
\put(16.5,26){\scalebox{0.8}[1]{AF + mdSC}}
\put(29.8,26){\scalebox{0.8}[1]{AF}}
\put(50.5,28.5){\scalebox{0.8}[1]{$\lambda=6\,a$}}
\put(38.0,26){\scalebox{0.8}[1]{AF + mdSC}}
\put(50.5,26){\scalebox{0.8}[1]{PDW}}
\put(61.8,26){\scalebox{0.8}[1]{AF}}
\put(79.8,28.5){\scalebox{0.8}[1]{homogeneous}}
\put(70.5,26){\scalebox{0.8}[1]{AF + dSC}}
\put(83,26){\scalebox{0.8}[1]{dSC}}
\put(93.6,26){\scalebox{0.8}[1]{AF}}
\end{overpic}\vspace{10mm}
\begin{overpic}
[width=0.65\columnwidth]{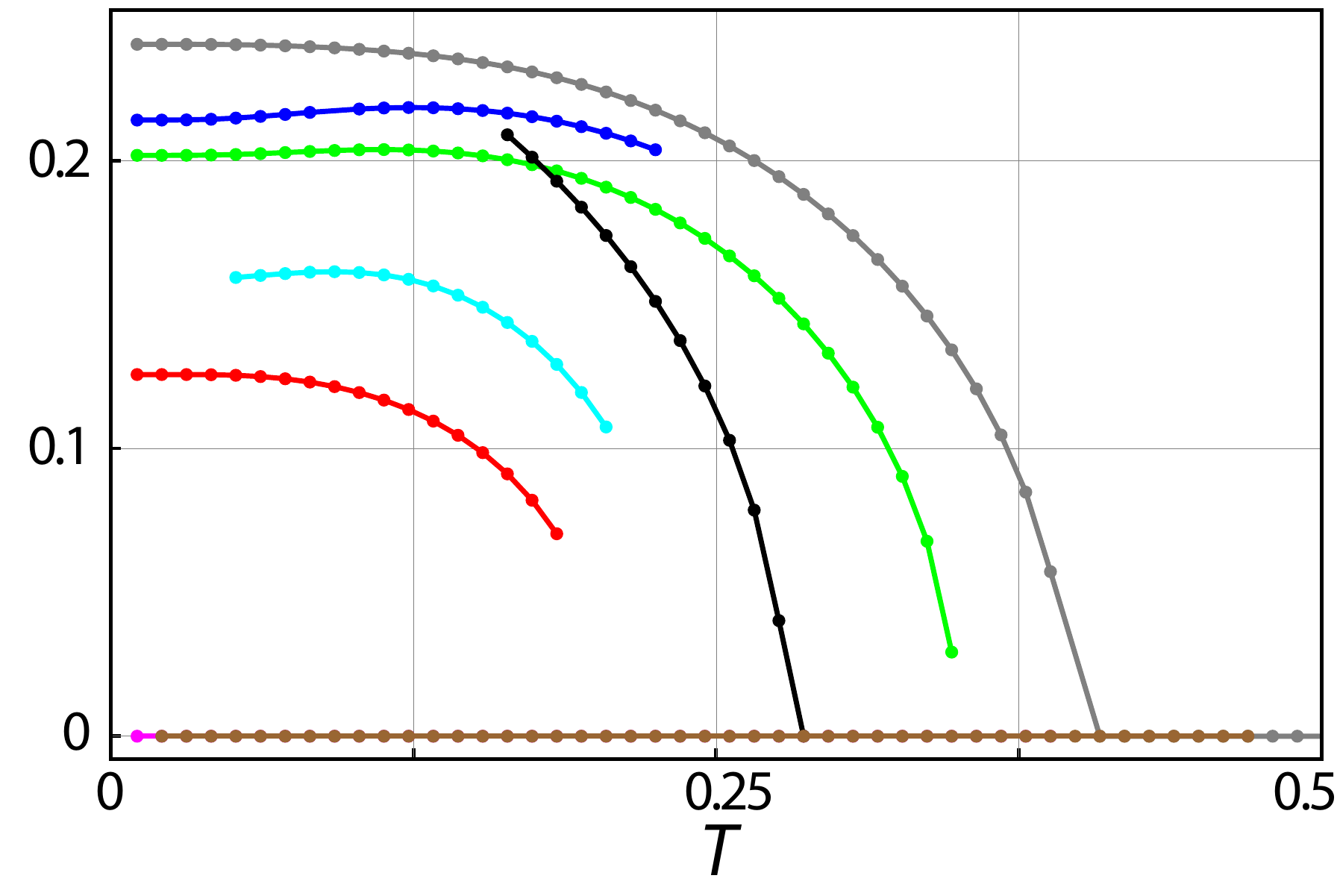}
\put(3,70){\bf(a) $\max_{ij}\Delta_{ij}$}
\end{overpic}\hspace{3mm}
\begin{overpic}
[width=0.65\columnwidth]{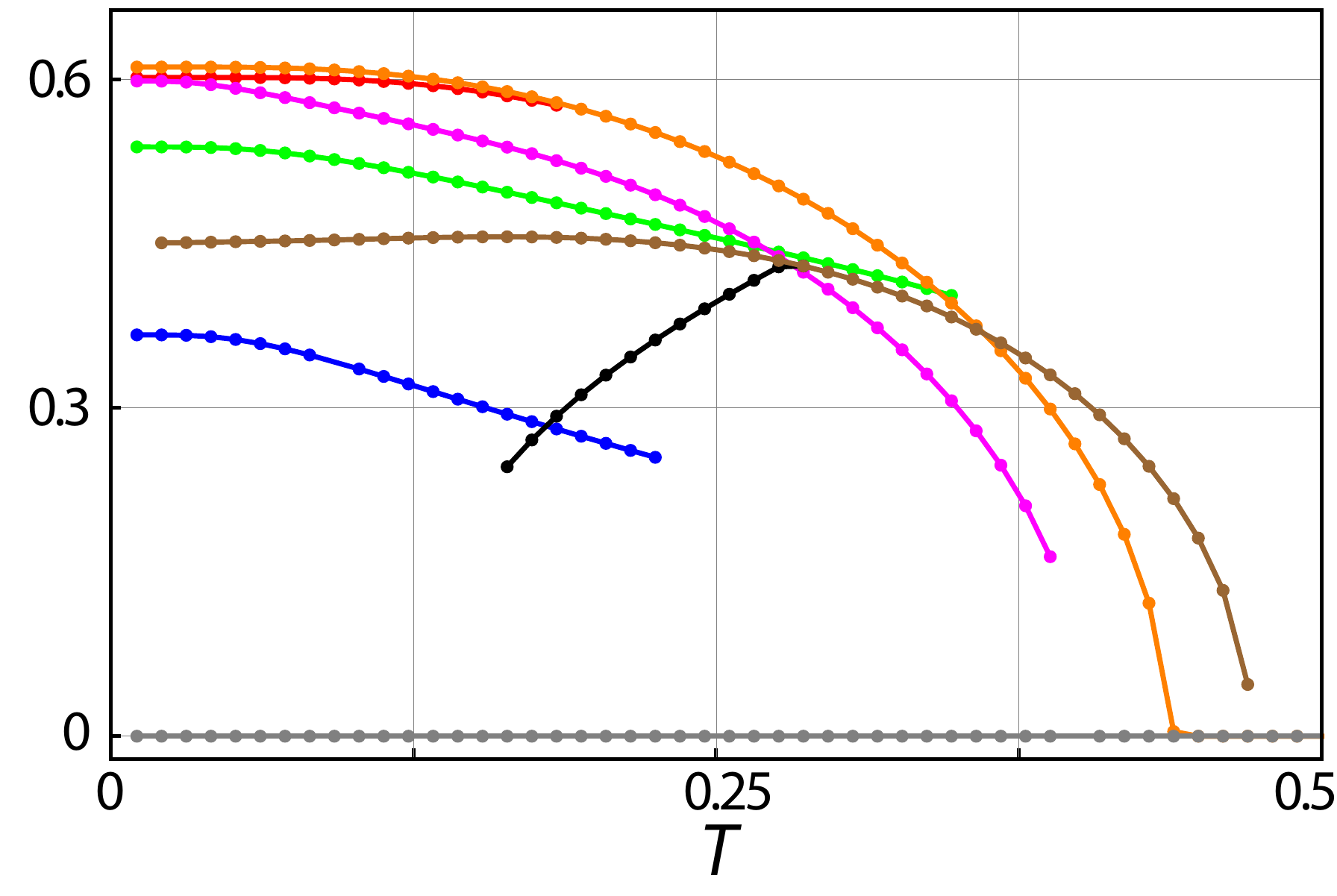}
\put(3,70){\bf(b) $\max_iM_i$}
\end{overpic}\hspace{3mm}
\begin{overpic}
[width=0.65\columnwidth]{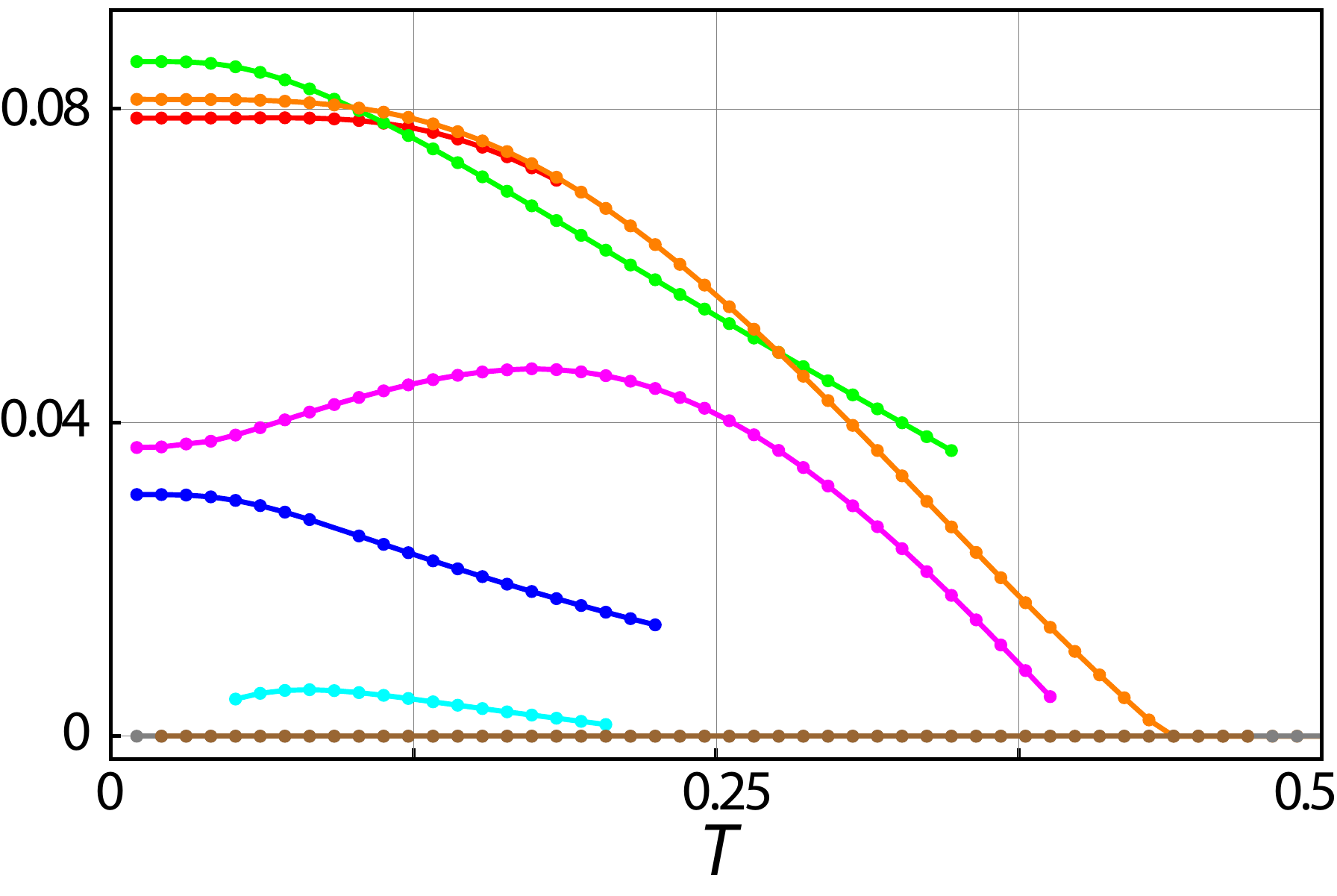}
\put(3,70){\bf(c) $\max_i\delta n_i$}
\end{overpic}
\caption{Upper panel: Selection of (meta) stable solutions of the $U$-model. The groundstate is the $\lambda=8\,a$ AF + mdSC solution. In white fields, the corresponding order is excluded from this solution. Lower panel: Temperature dependence of the maximum SC order parameter (a), the maximum local magnetization (b), and the maximum amplitude of the CDW (c),
characterizing stable solutions of the $U$-model with $U=4.4\,t$ and $V=2\,t$. Some solutions are stable only in specific temperature windows and disappear where the corresponding lines end.}
\label{fig9}
\end{figure*}

Figure~\ref{fig9} shows the maximum values of the SC order parameter $\Delta^{\!s}_{ij}$ (a), the AF order parameter $M_i$ (b), and the maximum amplitude of the CDW $\delta n_i$ (c) as a function of temperature for the $U$-model for $x=1/8$. For better characterization of the various solutions of the $U$-model, we present the corresponding real-space patterns of $m_i$, $n_i$ and $\Delta^{\!d}_i$ in the upper panel.
For the $U$-model we find solutions with stripe order of wavelength $\lambda=8\,a$ and $\lambda=6\,a$, and also homogeneous solutions. The striped solutions exist both in the PDW and the mdSC type, whereas homogeneous antiferromagnetism may coexist with homogeneous $d$-wave superconductivity (dSC). Some solutions disappear abruptly at certain temperatures. More specifically, superconductivity vanishes at temperatures where antiferromagnetism still persists. It is therefore instructive to trace the high-temperature states without superconductivity back to lower temperatures. These special solutions are indicated by white fields in the upper panel of figure~\ref{fig9} where the corresponding order (superconductivity or antiferromagnetism) is suppressed.

Within the $U$-model, the order for the onset of superconductivity and antiferromagnetism is not fixed. Close to $U_{{\rm c}1}$, superconductivity sets in at a higher temperature than antiferromagnetism, while for the parameters used in figure~\ref{fig9}, antiferromagnetism survives to slightly higher temperatures. In this regime, the sudden disappearance of superconductivity is characteristic for the $U$-model, because, as discussed above, superconductivity breaks down, if the SC gap becomes smaller than the AF gap as the temperature is raised. This phenomenon is visible for the $\lambda=8\,a$ AF + PDW and AF + mdSC solutions as well as for the $\lambda=6\,a$ AF + mdSC solution. As the mdSC state vanishes with raising the temperature, the AF stripe order changes spontaneously from a site- to a bond-centered pattern. A remarkable observation is the disappearance of antiferromagnetism in the $\lambda=6\,a$ PDW solution. This state is stable only in a certain temperature window, while for lower temperatures it decays into an AF tartan pattern.

\begin{figure*}[t!]
\centering
\vspace{8mm}
\begin{overpic}
[width=0.65\columnwidth]{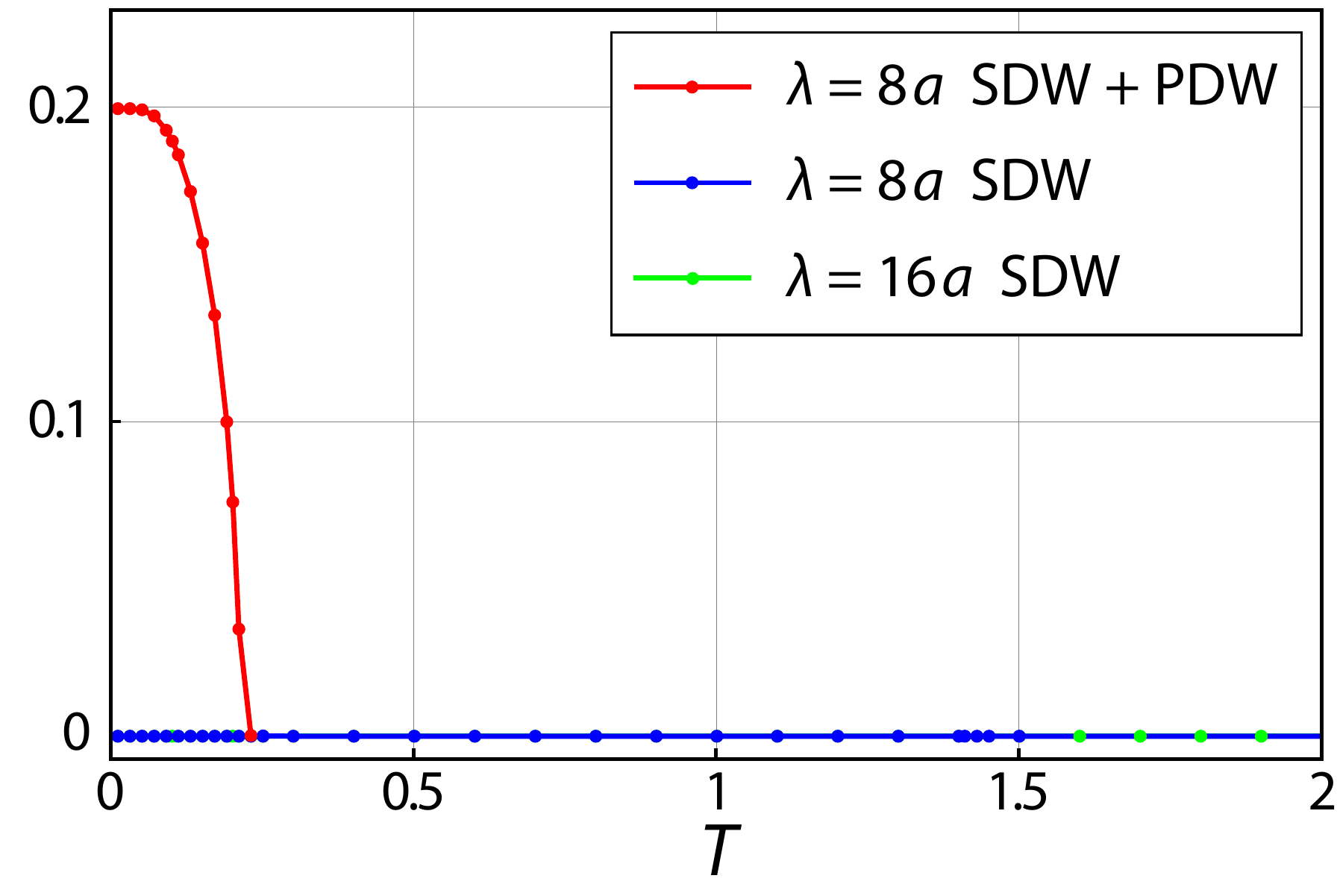}
\put(128,80){\bf $\bm V$-model with $\bm{V=2\,t}$}
\put(3,70){\bf(a) $\max_{ij}\Delta_{ij}$}
\end{overpic}\hspace{3mm}
\begin{overpic}
[width=0.65\columnwidth]{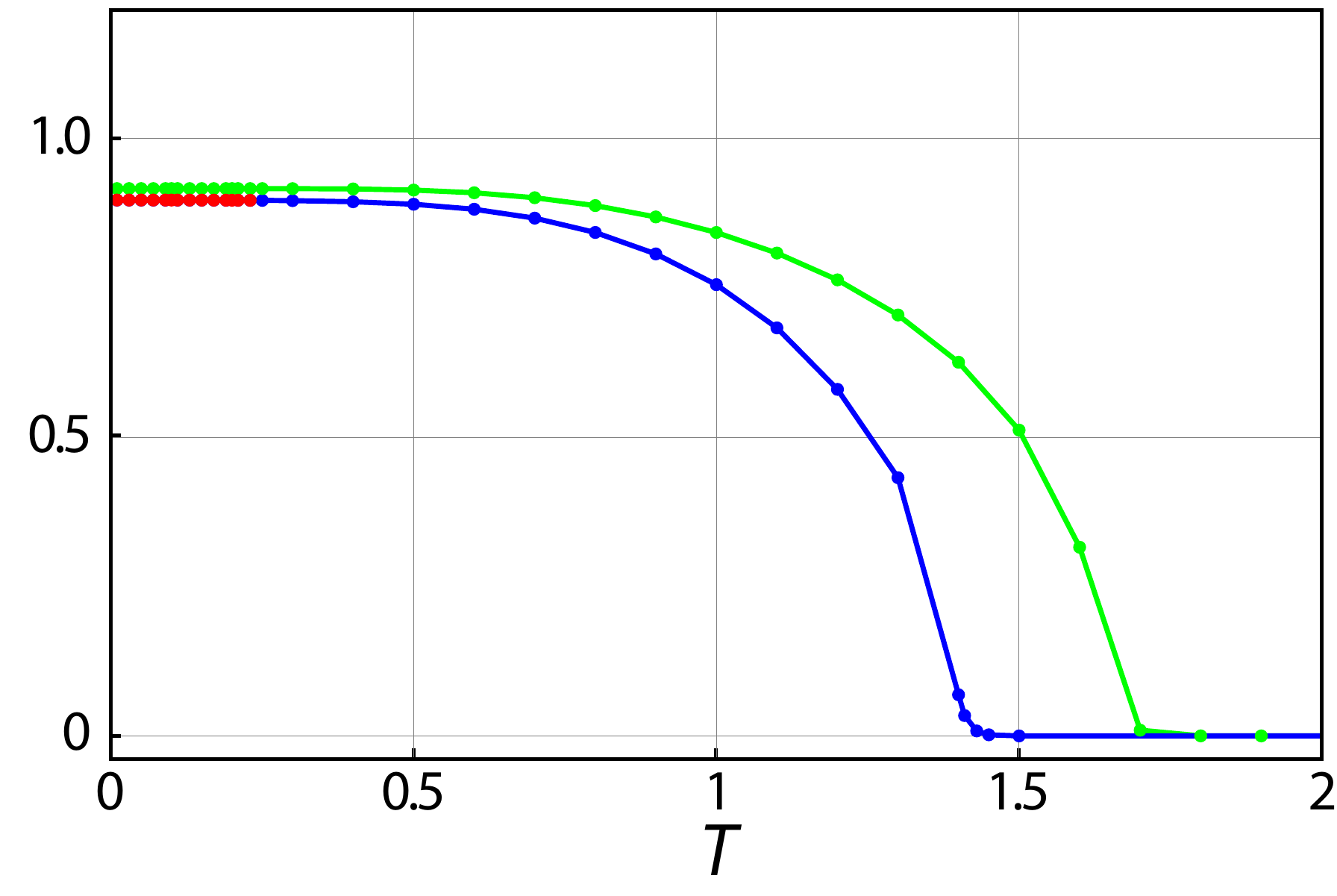}
\put(3,70){\bf(b) $\max_iM_i$}
\end{overpic}\hspace{3mm}
\begin{overpic}
[width=0.65\columnwidth]{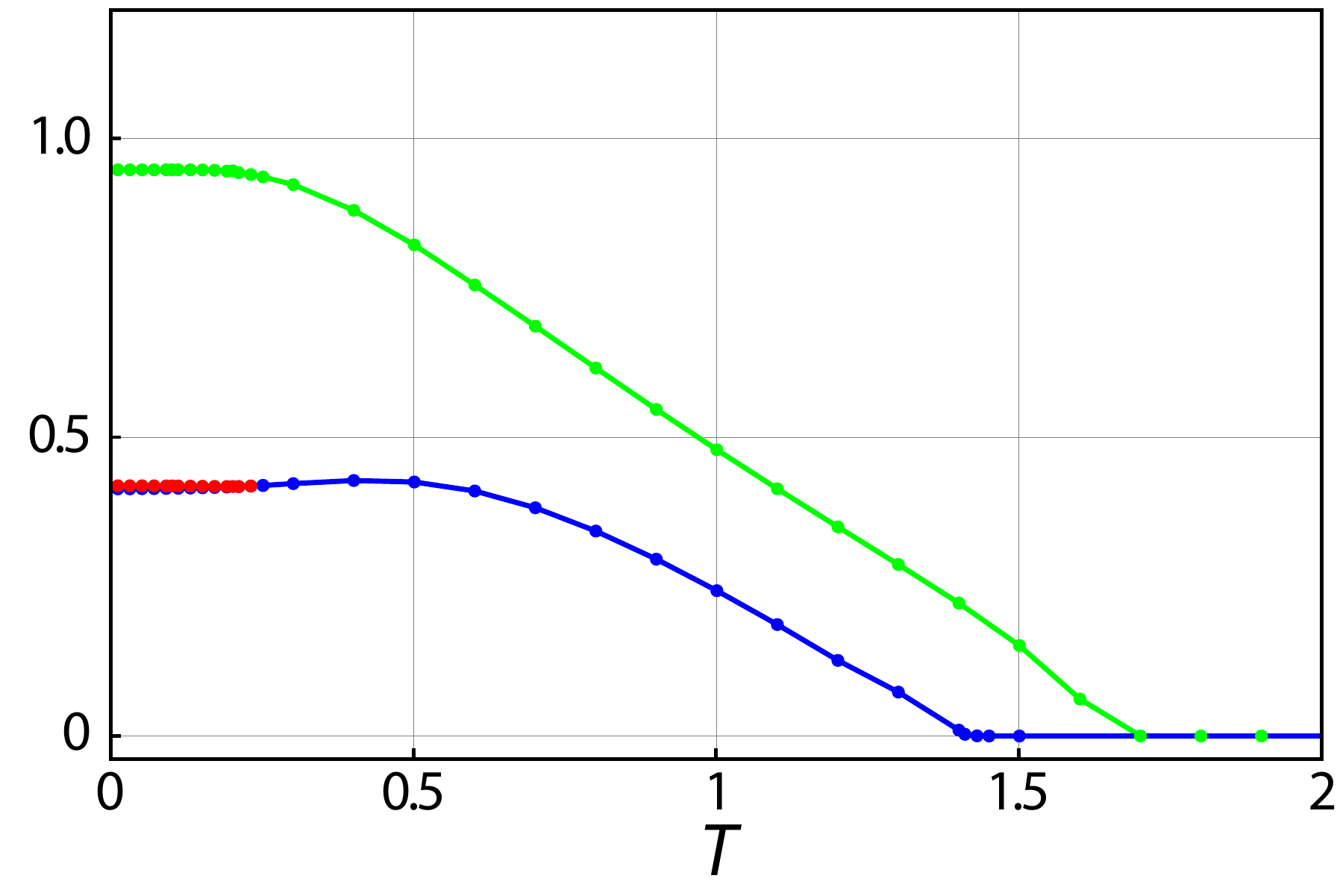}
\put(3,70){\bf(c) $\max_i\delta n_i$}
\end{overpic}
\caption{Temperature dependence of the maximum SC order parameter (a), the maximum local magnetization (b), and the maximum amplitude of the CDW (c),
characterizing stable solutions of the $V$-model with $V=2\,t$.}
\label{fig10}
\end{figure*}

For the $V$-model the magnetic order parameter is dominant and at $x=1/8$ only solutions with
wavelengths of $8\,a$ or $16\,a$ can be stabilized. The temperature dependence of these solutions is shown in figure~\ref{fig10}. Here we also find non-superconducting states with AF stripes ($\lambda=8\,a$ SDW and $\lambda=16\,a$ SDW)
as well as the striped superconducting state with wavelength $\lambda=8\,a$ ($\lambda=8\,a$ SDW+PDW); note that the phase difference of $\Delta_{ij}$ on neighboring stripes is not uniquely determined in the mean-field solution of the $V$-model as discussed in detail in~\cite{loder11}. Since the non-magnetic sites in the $\lambda=16\,a$ SDW state are nearly empty, superconductivity is absent in this solution. A $\lambda=6\,a$ SDW state can be stabilized for higher doping levels, e.g. $x=1/6$. Within the $V$-model, the superconducting $T_{\rm c}$ is typically much smaller than the onset temperature of antiferromagnetism, and the two temperatures become equal, if $V$ is lowered to $V_{{\rm c}1}$. The $\lambda=8\,a$ SDW+PDW solution merges smoothly into the non-SC $\lambda=8\,a$ SDW solution above $T_{\rm c}$, since $M_i$ and $\delta n_i$ are almost temperature independent at $T_{\rm c}$.

The sequence of phase transitions with decreasing temperature is complicated in the striped cuprate materials. It is best investigated in La$_{15/8}$Ba$_{1/8}$CuO$_4$~\cite{LiPRL,Tranquada08}, where a CDW sets in at $T_{\rm CDW}\sim54$$\,$K at the transition into the LTT phase, which is joined by an AF stripe order at a slightly lower temperature. Superconductivity appears only below 10$\,$K, which suggests the dominance of AF order as in the $V$-model. A direct comparison to our mean-field models is however difficult, since these models do not describe fluctuating superconductivity, e.g. non phase-coherent stripes, which is likely to be present at much higher temperature close to $T_{\rm CDW}$.

\section{Conclusions}\label{sec:conc}

In this article, we have explored how far a mean-field approach of the $t$--$J$ model and of closely related models can assist our understanding of static spin- and charge-stripe order observed in underdoped cuprates in coexistence with superconductivity. For hole doping $x=1/8$ the free energy is minimized by an SDW order with wavelength $\lambda=8\,a$ with concomitant CDW order with $\lambda=4\,a$, which agrees with neutron and X-ray scattering experiments for, e.g., La$_{15/8}$Ba$_{1/8}$CuO$_4$ or La$_{15/8-y}$Nd$_y$Sr$_{1/8}$CuO$_4$. This type of order is stable for a wide range of the interaction parameter $\hat J$ in the $t$--$\hat J$ model and in both its descendants for small and large $\hat J$, the $U$-model and the $V$-model, respectively. These latter models represent two limiting cases where a mean-field treatment of the $t$--$\hat J$ model is well justified.

While the energetically favored spin structure is a robust property of the $7/8$ filled $t$--$\hat J$ model, the nature of the SC state coexisting with the spin order changes between different limits. For small values of $\hat J$, an almost homogeneous and isotropic $d$-wave SC state is realized which recedes to one dimensional non-magnetic lines for larger $\hat J$ and also in the $V$-model. For larger $\hat J$, phase coherence perpendicular to the stripes is lost or very fragile, which naturally explains the breakdown of superconductivity in the presence of static spin-stripe order, stabilized by the anisotropic hopping amplitudes in the LTT phase of rare earth doped 214 cuprates~\cite{kampf,Buchner94,Tranquada97,Yang:2009}. If phase coherence is established with decreasing $\hat J$, the favored SC order parameter does not acquire a phase shift between neighboring stripes and therefore has a finite homogeneous $\q=\bm0$ component. It is therefore not of the pure PDW type as suggested in~\cite{Berg09}, but rather similar to the state found in~\cite{Yang:2009}. A definitive answer to the phase relation of the superconducting order parameter on neighboring stripes cannot be given within the presented mean-field description. A phase sensitive Josephson-coupling term is needed, in the spirit of the phenomenological theory of~\cite{Berg09}.

The analysis of the $t$--$\hat J$ model led us to two qualitatively distinct regimes for the coupling constant $\hat J$. Its appropriate value for high $T_{\rm c}$ cuprates is generally believed to be in the range of $\sim0.3$ -- $0.5\,t$. The physics of the cuprates therefore falls somewhere in between the two limiting regimes where the two mean-field approaches are reliably applicable. While the two regimes are not continuously connected on the mean-field level, the model systems may well show a smooth crossover and features of both regimes. For example, ARPES measurements on La$_{15/8}$Ba$_{1/8}$CuO$_4$~\cite{he} show four-fold symmetric Fermi-arc like structures at zero energy, compatible with the \textquotedblleft almost\textquotedblright\ isotropic $d$-wave solution, whereas La$_{15/8-y}$Nd$_y$Sr$_{1/8}$CuO$_4$ with $y=0.6$, which has a much stronger anisotropy in the hopping amplitudes along the $x$- and the $y$-direction, shows a quasi one dimensional Fermi surface with a breach at the Brillouin-zone center~\cite{Zhou:1999}. These features fit well to the solution of the $V$-model in~\cite{loder11}.

A related issue is the influence of disorder and vortices on the phenomenon of stripe formation. Our initial calculations for disordered systems indicate that impurities act as pinning centers for AF stripes by slowing down spin fluctuations and thereby freezing static stripe order. In the $U$-model, the stripe structures flexibly adjust in wavy forms to the impurity pattern, while superconductivity seems only slightly affected. On the other hand, impurities with a moderate scattering potential do not disorder the stripe pattern in the $V$-model at all, although they are detrimental for superconductivity. Since stripes in the cuprates are not perfectly ordered, the $V$-model limit is probably not reached, but rather an intermediate regime. In the presence of antiferromagnetic stripes, the motion of vortices is confined to the stripe direction. Their stable position is at the center of an antiferromagnetic stripe, where superconductivity is weakest. As has been shown within the $U$-model for homogeneous $d$-wave superconductors~\cite{Wang,zhu,Schmid}, antiferromagnetism is enhanced around the vortex core, which, similarly to a strong impurity potential, forces the neighboring superconducting stripes to bend around the vortex.

%
%
%
%
%
%

%\vspace{-13mm}
{\center{\bf Acknowledgements}\\[1mm]} 
\noindent
The authors gratefully acknowledge helpful discussions with Peter Hirschfeld and Raymond Fr\'esard. 
This work was supported by the Deutsche Forschungsgemeinschaft through TRR 80.


\begin{references} %\begin{thebibliography}

\bibitem{Bernordz}
J. G. Bednordz and K. A. M\"uller,
{Z. Phys. B} {\bf 64}, 189 (1986).

\bibitem{TranquadaNickel}
J. M. Tranquada, D. J. Buttrey, V. Sachan,	and J. E. Lorenzo,
{Phys. Rev. Lett.} {\bf 73}, 1003 (1994). 

\bibitem{TranquadaLNdSCO}
J. M. Tranquada, B. J. Sternlieb, J. D. Axe, Y. Nakamura, and S. Uchida,
{Nature} {\bf 375}, 561 (1995).

\bibitem{Buchner94}
B. B\"uchner, M. Breuer, A. Freimuth, and A. P. Kampf,
{Phys. Rev. Lett.} {\bf 73}, 1841 (1994).

\bibitem{TranquadaLSCO}
J. M. Tranquada, H. Woo, T. G. Perring, H. Goka, G. D. Gu, G. Xu, M. Fujita,	and K. Yamada,
{Nature} {\bf 429}, 534 (2004).

\bibitem{fujita}
M. Fujita, H. Goka, K. Yamada, J. M. Tranquada, and L. P. Regnault,
{Phys. Rev. B} {\bf 70}, 104517 (2004).

\bibitem{abbamonte}
P. Abbamonte, A. Rusydi, S. Smadici, G. D. Gu, G. A. Sawatzky, and D. L. Feng,
{Nat. Phys.} {\bf 1}, 155 (2005).

\bibitem{Tranquada97}
J. M. Tranquada, J. D. Axe, N. Ichikawa, A. R. Moodenbaugh, Y. Nakamura, and S. Uchida,
{Phys. Rev. Lett.} {\bf 78}, 338 (1997).

\bibitem{Hucker10}
M. H\"ucker, M. v. Zimmermann, G. D. Gu, Z. J. Xu, J. S. Wen, Guangyong Xu, H. J. Kang, A. Zheludev, and J. M. Tranquada,
{Phys. Rev. B} {\bf 83}, 104506 (2011).

\bibitem{KivelsonRMP}
S. A. Kivelson, E. Fradkin, V. Oganesyan, J. M. Tranquada, A. Kapitulnik and C. Howald,
{Rev. Mod. Phys.} {\bf 75}, 1201 (2003).

\bibitem{berg07}
E. Berg, E. Fradkin, E.-A. Kim, S. A. Kivelson, V. Oganesyan, J. M. Tranquada, and S. C. Zhang,
{Phys. Rev. Lett.} {\bf 99}, 127003 (2007).

\bibitem{Tranquada08}
J. M. Tranquada, G. D. Gu, M. H\"ucker, Q. Jie, H.-J. Kang, R. Klingeler, Q. Li, N. Tristan, J. S. Wen, G. Y. Xu, Z. J. Xu, J. Zhou, and M. v. Zimmermann,
{Phys. Rev. B} {\bf 78}, 174529 (2008).

\bibitem{vojta}
M. Vojta and T. Ulbrecht,
{Phys. Rev. Lett} {\bf 93}, 127002 (2004).

\bibitem{uhrig}
G. S. Uhrig, K. P. Schmidt, and M. Gr\"uninger,
{Phys. Rev. Lett} {\bf 93}, 267003 (2004).

\bibitem{greiter}
M. Greiter and H. Schmidt,
{Phys. Rev. B} {\bf 82}, 144512 (2010).

\bibitem{ZaanenHF}
J. Zaanen and O. Gunnarsson,
{Phys. Rev. B} {\bf 40}, 7391 (1989).

\bibitem{machida}
K. Machida,
{Physica C} {\bf 158}, 192 (1989).

\bibitem{Prelovsek}
P. Prelov\v sek and X. Zotos,
{Phys. Rev. B} {\bf 47}, 5984 (1993).

\bibitem{white}
S. R. White and D. J. Scalapino,
{Phys. Rev. Lett.} {\bf 80}, 1272 (1998);
{Phys. Rev. Lett.} {\bf 81}, 3227 (1998);
{Phys. Rev. B} {\bf 60}, 753(R) (1999).

\bibitem{white09}
S. R. White and D. J. Scalapino,
{Phys. Rev. B} {\bf 79}, 220504 (2009).

\bibitem{himeda}
A. Himeda, T. Kato, and M. Ogata,
{Phys. Rev. Lett.} {\bf 88}, 117001 (2002).

\bibitem{capello}
M. Capello, M. Raczkowski, and D. Poilblanc,
{Phys. Rev. B} {\bf 77}, 224502 (2008).

\bibitem{Agterberg08}
D. Agterberg and H. Tsunetsugu,
{Nat. Phys.} {\bf 4}, 639 (2008).

\bibitem{Baruch:2008}
S. Baruch and D. Orgad.
{Phys. Rev. B} {\bf 77}, 174502 (2008),

\bibitem{Berg09}
E. Berg, E. Fradkin, and S. A. Kivelson,
{Phys. Rev. B} {\bf 79}, 064515 (2009).

\bibitem{Berg:2009}
E. Berg, E. Fradkin, S. A. Kivelson, and J. M. Tranquada,
{New J. Phys.} {\bf 11}, 115004 (2009).

\bibitem{Loder:2010}
F. Loder, A. P. Kampf, and T. Kopp,
{Phys. Rev. B} {\bf 81}, 020511(R) (2010).

\bibitem{fulde}
P. Fulde and R. A. Ferrell,
{Phys. Rev.} {\bf 135}, A550 (1964).

\bibitem{larkin}
A. I. Larkin and Y. N. Ovchinnikov,
{Zh. Eksp. Teor. Fiz.} {\bf 47}, 1136 (1964).

\bibitem{Houzet}
M. Houzet and A. Buzdin,
{Phys. Rev. B} {\bf 63}, 184521 (2001).

\bibitem{Raczkowski07}
M. Raczkowski, M. Capello, D. Poilblanc, R. Fr\'esard, and A. M. Ole\'s,
{Phys. Rev. B} {\bf 76}, 140505(R) (2007).

\bibitem{Yang:2009}
K.-Y. Yang, W. Q. Chen, T. M. Rice, M. Sigrist,  and F.-C. Zhang,
{New J. Phys.} {\bf 11}, 055053 (2009).

\bibitem{kagan}
M. Yu. Kagan and T. M. Rice,
{J. Phys. Condens. Matter} {\bf 6}, 3771 (1994).

\bibitem{loder11}
F. Loder, S. Graser, A. P. Kampf, and T. Kopp,
preprint arXiv:1101.3402.

\bibitem{emery}
V. J. Emery, S. A. Kivelson, and H. Q. Lin,
{Phys. Rev. Lett.} {\bf 64}, 475 (1990).

\bibitem{kivelson.emery}
S. A. Kivelson and V. J. Emery,
{Synth. Met.} {\bf 80}, 151 (1996).

\bibitem{white00}
S. R. White and D. J. Scalapino,
{Phys. Rev. B} {\bf 61}, 6320 (2000).

\bibitem{unger}
S. Unger and P. Fulde,
{Phys. Rev. B} {\bf 47}, 8947 (1993).

\bibitem{anderson87}
P. W. Anderson,
{Science} {\bf 235}, 1196 (1987).

\bibitem{anderson59}
P. W. Anderson,
{Phys. Rev.} {\bf 115}, 2 (1959).

\bibitem{zhang}
F. C. Zhang and T. M. Rice,
{Phys. Rev. B} {\bf 37}, 3759(R) (1988).

\bibitem{scalapino}
See e.g. D. J. Scalapino,
{Phys. Rep.} {\bf 250}, 329 (1995).

\bibitem{Fancini}
F. Mancini, Theoretical Methods for Strongly Correlated Systems (Springer).

\bibitem{ghosal}
A. Ghosal, C. Kallin, and A. J. Berlinsky,
{Phys Rev. B} {\bf 66}, 214502 (2002).

\bibitem{zhang97}
S.-C. Zhang,
{Science} {\bf 275}, 452 (1997).

\bibitem{Wang}
Y. Wang and A. H. MacDonald,
{Phys Rev. B} {\bf 52}, 3876(R) (1995).

\bibitem{Schmid}
B. M. Andersen, P. J. Hirschfeld, A. P. Kampf, and M. Schmid,
{New J. Phys.} {\bf 12}, 053043 (2010).

\bibitem{zhu}
J.-X. Zhu and C. S. Ting,
{Phys Rev. Lett.} {\bf 87}, 147002 (2001).

\bibitem{andersen07}
M. Schmid, B. M. Andersen, A. P. Kampf, and P. J. Hirschfeld,
{Phys. Rev. Lett.} {\bf 99}, 147002 (2007).

\bibitem{Andersen05}
B. M. Andersen and P. Hedeg{\aa}rd,
{Phys. Rev. Lett.} {\bf 95}, 037002 (2005).

\bibitem{kampf}
A. P. Kampf, D. J. Scalapino, and S. R. White,
{Phys Rev. B} {\bf 64}, 052509 (2001).

\bibitem{LiPRL}
Q. Li,  M. H\"ucker, G. D. Gu, A. M. Tsvelik, and J. M. Tranquada,
{Phys. Rev. Lett.} {\bf 99}, 067001 (2007).

\bibitem{he}
R.-H. He, K. Tanaka, S.-K. Mo, T. Sasagawa, M. Fujita, T. Adachi, N. Mannella, K. Yamada, Y. Koike, Z. Hussain, and Z.-X. Shen,
{Nat. Phys.} {\bf 5}, 119 (2009). 

\bibitem{Zhou:1999}
X. J. Zhou, P. Bogdanov, S. A. Kellar, T. Noda, H. Eisaki, S. Uchida, Z. Hussain, and Z.-X. Shen,
{Science} {\bf 286}, 268 (1999).  

\end{references}
\end{document}